\documentclass[12pt,sort&compress]{elsarticle}

\journal{}%Journal of Computational Physics}

\usepackage{afterpage}
\usepackage{enumitem}
\usepackage{mystyle}
\usepackage{algpseudocode}
\usepackage{algorithm}
\usepackage{amsmath}
\usepackage{amssymb}
\usepackage{subfiles}
\usepackage{subfigmat} % enables manipulation of
\usepackage{mathabx}
\usepackage{relsize}
\usepackage{pst-all}
\usepackage{appendix}

\newcommand\figref[1]{Figure \ref{fig:#1}} %Reference to a figure
\newcommand\tabref[1]{Table \ref{tab:#1}} %Reference to a table
\newcommand\eref[1]{Eq. (\ref{eq:#1})} %Reference to an equation`m

\newcommand\p[1]{\partial{#1}}

\begin{document}

\begin{frontmatter}

\title{Invariant Data-Driven Subgrid Stress Modeling in the\\ Strain-Rate Eigenframe for Large Eddy Simulation}

\author[colorado]{Aviral Prakash \corref{cor1}}
\ead{aviral.prakash@colorado.edu}
\author[colorado]{Kenneth E. Jansen}
\author[colorado]{John~A.~Evans}
\cortext[cor1]{Corresponding author}

\address[colorado]{University of Colorado Boulder, Boulder, CO 80309, USA}

\begin{abstract}
We present a new approach for constructing data-driven subgrid stress models for large eddy simulation of turbulent flows.  The key to our approach is representation of model input and output tensors in the filtered strain rate eigenframe.  Provided inputs and outputs are selected and non-dimensionalized in a suitable manner, this yields a model form that is symmetric, Galilean invariant, rotationally invariant,  reflectionally invariant, and unit invariant.  We use this model form to train a simple and efficient neural network model using only one time step of filtered direct numerical simulation data from a forced homogeneous isotropic turbulence simulation.  We demonstrate the accuracy of this model as well as the model’s ability to generalize to previously unseen filter widths, Reynolds numbers, and flow physics using \textit{a priori} and \textit{a posteriori} tests.

\end{abstract}


\end{frontmatter}

\section{Introduction}
\label{sec:introduction}

Turbulent flows at high Reynolds numbers exhibit a broad spectrum of spatial and temporal scales.  Direct Numerical Simulation (DNS) resolves all spatial and temporal scales, but its cost increases quickly with Reynolds number.  Reynolds Averaged Navier-Stokes (RANS) simulations instead model all spatial and temporal scales, so they are much less computationally expensive than DNS.  However, RANS simulations often yield inaccurate flow predictions in the presence of strong streamline curvature, strong adverse and favorable pressure gradients, and flow separation.  Large Eddy Simulation (LES) is an alternative approach to DNS and RANS that resolves the largest scales of motion and models the effect of the unresolved scales of motion on the resolved scales.  As a consequence, LES is less expensive than DNS and, in general, yields more accurate flow predictions than RANS simulations.

LES involves the numerical solution of the filtered Navier-Stokes equations.  These equations involve an unclosed term, the subgrid stress (SGS) tensor, that must be modeled.  Most SGS models fall into one of two categories: functional models and structural models.  Functional models such as the static Smagorinsky model \citep{Smagorinsky1963} and the dynamic Smagorinsky model \citep{Germano1991} aim to accurately predict the SGS energy dissipation, that is, the transfer of energy from filtered scales to subfilter scales.  Structural models such as Clark’s gradient model \citep{Clark1979} and Bardina’s scale-similarity model \citep{Bardina1980} aim to accurately predict the SGS tensor itself.  Functional models are the most popular SGS models in practice today, and they have been used with success in modeling wall-bounded turbulent flows, mixed laminar-turbulent flows, and transitional flows.  However, the SGS tensors predicted by functional models are usually in poor correlation with the exact SGS tensor \citep{Clark1979, Borue1998}, and as a consequence, they perform poorly for certain flows such as those exhibiting significant shearing \cite{Baggett1997}.  Conversely, the SGS tensors predicted by structural models are often in strong correlation with the exact SGS tensor, but structural models typically underpredict the transfer of energy from filtered scales to subfilter scales \citep{Vreman1997}.  This not only can lead to inaccurate flow predictions but also numerical instability \citep{Vreman1996}.

The data-driven approach to SGS closure is an alternative modeling paradigm that has recently gained in popularity \cite{Duraisamy2019}.  This approach leverages advances in machine learning and the availability of high-fidelity DNS data to build improved SGS models.  Below, we provide a non-comprehensive overview of data-driven SGS models in the literature to provide context for the particular data-driven SGS modeling approach introduced in this paper.  One of the first papers employing the data-driven approach to SGS closure used a neural network based model to compute the eddy viscosity in a mixed functional/structural model \citep{Sarghini2003}.  More recent papers have focused instead on learning entirely new SGS representations.  King \textit{et al.} learned a truncated Volterra series model of the SGS tensor using filtered DNS data from a forced homogeneous isotropic turbulence (HIT) simulation and demonstrated the learned model yielded considerably improved predictions of the SGS tensor and energy transfer as compared with the dynamic Smagorinsky model \cite{King2016}.  However, King \textit{et al.}’s model has 5,995 model inputs and is constructed for one unique filter width.  Gamahara and Hattori learned neural network models of the SGS tensor using filtered DNS data from a turbulent channel flow simulation and showed the learned models yielded a similar or higher level of correlation between predicted and exact SGS stresses when compared with Clark’s gradient model and Bardina’s scale-similarity model \cite{Gamahara2017}.  Gamahara and Hattori also demonstrated their models’ ability to generalize to higher Reynolds numbers than the ones they were trained on.  However, Gamahara and Hattori’s models do not depend on filter width and are neither rotationally invariant, reflectionally invariant, nor unit invariant, so they are unable to generalize to filter widths or geometries other than the ones they were trained on.

Wang \textit{et al.} learned a neural network model of the SGS tensor from a forced HIT simulation and demonstrated their model outperforms both the static and dynamic Smagorinsky models in \textit{a priori} and \textit{a posteriori} tests \cite{Wang2018}.  However, like Gamahara and Hattori’s models, Wang \textit{et al.}’s model does not depend on filter width and is neither rotationally invariant, reflectionally invariant, nor unit invariant.  Moreover, Wang \textit{et al.} only examined the performance of their model for forced HIT, and it is unknown whether their model will perform well for other flow physics.  Zhou \textit{et al.} followed up on Wang \textit{et al.}’s work by incorporating filter width as a model input in order to construct a neural network model that better generalizes to previously unseen filter widths \cite{Zhou2019}, though the resulting model is still neither rotationally invariant, reflectionally invariant, nor unit invariant.  To arrive at a model with embedded invariance properties, Xie \textit{et al.} learned a tensor basis neural network model of the SGS tensor \cite{Xie2020a}.  Tensor basis neural network models are very popular for Reynolds stress closure \cite{Ling2016, Parmar2020} and are rotationally and reflectionally invariant by construction.  However, as further discussed in Subsection \ref{subsec:invariant}, tensor basis neural network models rely on the use of minimal tensor and invariant integrity bases for model representation, and these bases increase in size exponentially fast with the number of prescribed tensor inputs.  Subel \textit{et al.} recently pursued transfer learning as an alternative approach to improve the ability of data-driven SGS models to generalize \cite{subel2021data}, though Subel \textit{et al.}'s study focuses on one-dimensional Burger's turbulence.  It should be finally noted a number of authors have pursued the construction of data-driven approximate deconvolution SGS models \citep{Maulik2017, Xie2020b}, but such models employ nonlocal model inputs and a complex neural network architecture for SGS tensor representation. From a cursory view of the literature, it is apparent that many proposed data-driven SGS models have limited applicability and are computationally expensive to train and evaluate.  Moreover, to the best of our knowledge, no data-driven SGS model has yet demonstrated an ability to generalize to previously unseen filter widths, Reynolds numbers, and flow physics in both \textit{a priori} and \textit{a posteriori} tests.

In this paper, we present a new approach for constructing simple and efficient data-driven SGS models with improved generalizability characteristics versus the state-of-the-art.  The key to our approach is representation of model input and output tensors in a flow specific coordinate system, namely the filtered strain rate eigenframe.  This yields a model form that is rotationally and reflectionally invariant.  Additionally, through an appropriate selection and non-dimensionalization of model inputs and outputs, one also attains a model form that is symmetric, Galilean invariant, and unit invariant.  We use this model form to construct a particularly simple and efficient neural network model using a limited amount of training data.  In particular, we train a neural network model with only four inputs and a single hidden layer of twenty neurons using filtered DNS data from just a single time step of a forced HIT simulation at $Re_{\lambda} = 418$.  Moreover, we train this model using a single filter width.  We demonstrate the accuracy of this model as well as the model’s ability to generalize to previously unseen filter widths, Reynolds numbers, and flow physics using a series of \textit{a priori} and \textit{a posteriori} tests.  In particular, the model outperforms both the dynamic Smagorinsky model and Clark’s gradient model in \textit{a posteriori} tests of forced HIT at $Re_{\lambda} = 165$, forced HIT at $Re_{\lambda} = \infty$, decay of HIT, and Taylor-Green vortex flow at $Re = 1,600$. The focus of this paper is periodic turbulence and non-anisotropic filters, but in future work, we plan to extend our approach to wall-bounded turbulent flows and anisotropic filters.

An outline of this paper is as follows.  In Section \ref{sec:FNS}, we recall the filtered Navier-Stokes equations and the SGS tensor, and in Section \ref{sec:SGSPhysics}, we provide a review of functional and structural models of the SGS tensor.  In Section \ref{sec:SGSdata}, we present our new approach for data-driven modeling of the SGS tensor, and in Section \ref{sec:Simple}, we use our approach to train a simple and efficient neural network model of the SGS tensor using filtered DNS data from a forced HIT simulation.  In Section \ref{sec:Results}, we examine the performance of the neural network model constructed in Section \ref{sec:Simple} using \textit{a priori} and \textit{a posteriori} tests.  Finally, in Section \ref{sec:conclusions}, we provide concluding remarks.

\section{The Filtered Navier-Stokes Equations and Subgrid Stress Tensor}
\label{sec:FNS}

LES is predicated on the numerical solution of the filtered Navier-Stokes equations.  Thus, to begin our discussion, we first recall the Navier-Stokes equations for an incompressible flow:

\begin{equation}
\frac{\partial u_i}{\p t} + \frac{\partial}{\p x_j} (u_i u_j)    = - \frac{1}{\rho} \frac{\partial p}{\p x_i} +\frac{\partial }{\p x_j} \left( \nu \left(\frac{\partial u_i}{\partial x_j} + \frac{\partial u_j}{\partial x_i}\right) \right) + f_i,
\end{equation}

\begin{equation}
\frac{\partial u_i}{\p x_i}    = 0.
\end{equation}

\noindent Above, $u_i$ is the $i^\text{th}$ component of the velocity field $\bm{u}$, $p$ is the pressure field, $\rho$ is the density, $\nu$ is the kinematic viscosity, and $f_i$ is the  $i^\text{th}$ component of the body force  $\bm{f}$.  The filtered Navier-Stokes equations are attained by applying a filtering operator

\begin{equation}
    \bar{\phi}(\bm{x}) = \int_{\mathbb{R}^3 } G(\bm{x},\bm{x}') \phi(\bm{x}') d^3 \bm{x}'
\end{equation}

\noindent to the Navier-Stokes equations where $G$ is a filter kernel satisfying

\begin{equation}
    \int_{\mathbb{R}^3 } G(\bm{x},\bm{x}') d^3 \bm{x}' = 1.
\end{equation}

\noindent If the filtering operator commutes with differentiation, the resulting equations are

\begin{equation}
\frac{\partial \bar{u}_i}{\p t} + \frac{\partial}{\p x_j} (\bar{u}_i \bar{u}_j)    = - \frac{1}{\rho} \frac{\partial \bar{p}}{\p x_i} +\frac{\partial }{\p x_j} \left( \nu \left(\frac{\partial \bar{u}_i}{\partial x_j} + \frac{\partial \bar{u}_j}{\partial x_i}\right) \right) - \frac{\p \tau_{ij}}{\p x_j} + \bar{f}_i, \label{eq:FNS-momentum}
\end{equation}

\begin{equation}
    \frac{\partial \bar{u}_i}{\partial x_i} = 0, \label{eq:FNS-mass}
\end{equation}

\noindent where $\bar{u}_i$ is the $i^\text{th}$ component of the filtered velocity field $\bar{\bm{u}}$, $\bar{p}$ is the filtered pressure field, and $\tau_{ij} = \overline{u_i u_j} - \bar{u}_i \bar{u}_j$ is the ${ij}^\text{th}$ component of the SGS tensor $\bm{\tau}$.  The SGS tensor cannot be expressed directly in terms of the filtered velocity and pressure fields, so in practice, it must be modeled.  This process is typically referred to as SGS closure.

By construction, the filtering operator is linear and preserves constant fields, and it commutes with differentiation if the filter kernel is homogeneous.  Isotropic filter kernels are also utilized frequently in practice.  Homogeneous filter kernels take the form
\begin{equation}
    G(\bm{x},\bm{x}') = G_\text{homogeneous}( \bm{x} - \bm{x}'),
\end{equation}
while isotropic filter kernels take the form
\begin{equation}
    G(\bm{x},\bm{x}') = G_\text{isotropic}( |\bm{x} - \bm{x}'| ).
\end{equation}
We typically associate filter kernels with a particular filter width $\Delta$.  For instance, the tensor-product box filter kernel associated with filter width $\Delta$ is equal to
\begin{equation}
    G(\bm{x},\bm{x}') = \left\{ 
\begin{array}{cc}
\frac{1}{\Delta} & \text{ if } | x_i - x'_i | \varleq \frac{\Delta}{2} \text{ for } i = 1, 2, 3 \\
0 & \text{ otherwise. }
\end{array}
 \right.
 \label{eq:boxkernel}
\end{equation}
Isotropic box filter kernels can be defined in a similar manner.  Other common filter kernels include Gaussian filter kernels and sharp spectral filter kernels.  For more details on filtering operators, the reader is pointed to \cite[Chapter 2]{Sagaut2006}.

The SGS tensor admits several important properties which should ideally be preserved under SGS closure.  For instance, the SGS tensor is symmetric, Galilean invariant (that is, its components do not change under Galilean transformations), and unit invariant (that is, its components scale appropriately under a change of units), and if the filter kernel is isotropic, the SGS tensor is also rotationally invariant (that is, it maps as a tensor under a coordinate rotation) and reflectionally invariant (that is, it maps as a tensor under a coordinate reflection).  The SGS tensor additionally admits the decomposition
\begin{equation}
    \tau_{ij} = \tau^d_{ij} + \frac{1}{3} \tau_{kk} \delta_{ij}
\end{equation}
where
$\tau^d_{ij} := \tau_{ij} - \frac{1}{3} \tau_{kk} \delta_{ij}$ is the ${ij}^\text{th}$ component of the deviatoric part of the SGS tensor and $\frac{1}{3} \tau_{kk} \delta_{ij}$ is the ${ij}^\text{th}$ component of the dilational part of the SGS tensor.  Note that the deviatoric part of the SGS tensor is traceless while the trace of the dilational part of the SGS tensor is equal to the trace of the full SGS tensor.  If we define the modified pressure field
\begin{equation}
    P = \bar{p} + \frac{1}{3} \rho \tau_{kk},
\end{equation}
then we can alternately write \eqref{eq:FNS-momentum} as
\begin{equation}
\frac{\partial \bar{u}_i}{\p t} + \frac{\partial}{\p x_j} (\bar{u}_i \bar{u}_j)    = - \frac{1}{\rho} \frac{\partial P}{\p x_i} + \frac{\partial }{\p x_j} \left( \nu \left(\frac{\partial \bar{u}_i}{\partial x_j} + \frac{\partial \bar{u}_j}{\partial x_i}\right) \right) - \frac{\p \tau^d_{ij}}{\p x_j} + \bar{f}_i. \label{eq:FNS-momentum-alt}
\end{equation}
If we elect to solve \eqref{eq:FNS-momentum-alt} and \eqref{eq:FNS-mass} for the filtered velocity field and modified pressure field rather than \eqref{eq:FNS-momentum} and \eqref{eq:FNS-mass} for the filtered velocity field and filtered pressure field, we only need to model the deviatoric part of the SGS tensor rather than the full SGS tensor.  This is convenient as the deviatoric part of the SGS tensor has five independent components while the full SGS tensor has six independent components.  Like the full SGS tensor, the deviatoric part of the SGS tensor is symmetric, Galilean invariant, and unit invariant, and it is also rotationally and reflectionally invariant if the filter kernel is isotropic.

\section{Functional and Structural SGS models}
\label{sec:SGSPhysics}

Most SGS models are characterized as either functional models or structural models.  Functional models aim to accurately predict the SGS energy dissipation $\Pi = - \tau^d_{ij} S_{ij}$ while structural models aim to accurately predict the entire SGS tensor or the deviatoric part of the SGS tensor.  Most functional models invoke the gradient diffusion hypothesis to model SGS energy dissipation.  For instance, the first SGS model proposed, the static Smagorinsky SGS model \cite{Smagorinsky1963}, assumes that the deviatoric part of the SGS tensor satisfies
\begin{equation}
    \tau^d_{ij} \approx -2 \nu_t S_{ij},
\end{equation}
where
\begin{equation}
    \nu_t = (C_s \Delta)^2 \vert \bm{S} \vert
\end{equation}
is the turbulent viscosity, $C_s$ is the so-called Smagorinsky constant, and $S_{ij} = \frac{1}{2}(\partial \bar{u}_i/\partial x_j + \partial \bar{u}_j/\partial x_i)$ is the ${ij}^\text{th}$ component of the filtered strain-rate tensor $\bm{S}$.  In practice, the Smagorinsky constant must be carefully tuned so the SGS energy dissipation is accurately predicted.  For instance, the constant could be chosen as 0.17 for forced HIT \cite{Lilly1967}, but this value does not work well for other flow cases \cite{Germano1991}.  For this reason, Germano proposed a procedure for dynamically computing the Smagorinsky constant by applying two separate filters, a grid filter and a test filter, to the Navier-Stokes equations \cite{Germano1991}.  This procedure gives rise to the so-called dynamic Smagorinsky SGS model.  The dynamic Smagorinsky SGS model typically outperforms the static Smagorinsky SGS model in practice, especially for wall-bounded turbulent flows, mixed laminar-turbulent flows, and transitional flows \cite{Germano1991}.  However, the SGS tensors predicted by both the static and dynamic Smagorinsky model are usually in poor correlation with the exact SGS tensor \cite{Borue1998,Clark1979}, and the same is true of other functional SGS models such as the WALE SGS model \cite{Nicoud1999}, the Vreman SGS model \cite{Vreman2004}, and the Sigma SGS model \cite{Nicoud2011}.  Hence, functional SGS models are said to suffer from poor structural performance.  The fundamental reason for the poor structural performance of functional SGS models is they assume alignment between the deviatoric part of the SGS tensor and the filtered strain-rate tensor, but these tensors are rarely aligned in practice.  As a result, functional SGS models perform poorly for certain classes of flows, such as those exhibiting significant shearing \cite{Baggett1997}.  We will later consider both the static and dynamic Smagorinsky models as representative functional models in our later numerical tests.

Structural models employ powerful mathematical techniques such as formal series expansion to accurately predict the SGS tensor.  For instance, Clark’s gradient model, which approximates the SGS tensor as
\begin{equation}
    \tau_{ij} \approx \frac{1}{12} \Delta^2 \frac{\partial \bar{u}_i}{\partial x_k} \frac{\partial \bar{u}_j}{\partial x_k}, \label{eq:clark}
\end{equation}
may be constructed directly from a truncated Taylor series expansion of the SGS tensor \cite{Clark1979}.  Likewise, Bardina’s self-similarity model invokes a self-similarity hypothesis to arrive at an approximation of the SGS tensor \cite{Bardina1980}, and approximate deconvolution methods use procedures such as van Cittert iteration to arrive at an approximation of the unfiltered velocity field given the filtered velocity field \cite{Stolz2001}.  The SGS tensors predicted by structural models are usually in strong correlation with the exact SGS tensor in \textit{a priori} tests \cite{Borue1998}, but structural models often underperform functional models in \textit{a posteriori} tests \cite{Vreman1997}.  This is due to the fact that structural models typically underpredict the SGS energy dissipation.  As such, structural SGS models are said to suffer from poor functional performance.  In fact, unlike functional models, structural models allow for backscatter of energy from subfilter to filtered scales, and excessive backscatter of energy can lead to numerical instability.  To improve functional performance, the SGS stresses that are predicted by a structural model are often clipped \cite{Liu1994,Prakash2021, Prakash2022} or augmented by an eddy viscosity term \cite{Zang1993,Horiuti1997}.

\section{Data-Driven Modeling of the SGS Tensor}
\label{sec:SGSdata}

The data-driven approach to SGS closure is an attractive alternative to the functional and structural approaches to SGS closure. In the data-driven approach to SGS closure, high-fidelity simulation data is employed in conjunction with machine learning algorithms to build improved SGS models. While a number of data-driven SGS closure strategies have been proposed in the literature, they typically yield SGS models that are significantly more expensive to evaluate than state-of-the-art functional and structural SGS models (due, for instance, to the use of a large number of model inputs or a complex neural network architecture for SGS tensor representation) or do not generalize to arbitrary turbulent flow scenarios or filter widths (due in part to the fact they are often trained on one turbulent flow scenario and one filter width). To address these issues, we propose a novel approach to data-driven SGS closure. The key to our approach is the intelligent selection of a model form that is symmetric, Galilean invariant, rotationally invariant, reflectionally invariant, and unit invariant by construction and that also recovers a simple representation of the first terms of a Taylor series expansion of the exact SGS tensor. This in turn requires an intelligent selection of model inputs and outputs. We demonstrate later that our data-driven approach is capable of generating computationally efficient SGS models that generalize far beyond the specific turbulent flow cases on which they were trained. Moreover, we are able to train these SGS models using a very limited amount of training data. 

In this section, we first detail the construction of a model form for data-driven SGS closure that satisfies the aforementioned desired invariance properties.  Then, we show the first terms of a Taylor series expansion of the exact SGS tensor admit a simple representation using this model form. Finally, we give details on the use of neural networks as a mapping between the chosen set of inputs and outputs.  It should be noted that the procedure outlined here can be used to construct a data-driven model of either the full SGS tensor or deviatoric part of the SGS tensor.

\subsection{Construction of an Invariant Model Form}
\label{subsec:invariant}

To begin the process of building a data-driven model for the SGS tensor, one must first select a particular model form. We describe here a systematic process for constructing a model form that is symmetric, Galilean invariant, rotationally invariant,  reflectionally invariant, and unit invariant. The first step in this process is the selection of model inputs that the SGS tensor is assumed to depend on. We require that model inputs satisfy the following properties:\\

\noindent \textit{Requirement \#1:} The model inputs must be local in space and time.\\
\noindent \textit{Requirement \#2:} The model inputs must be Galilean invariant.\\
\noindent \textit{Requirement \#3:} The model inputs must be able to distinguish different local flow states.\\
\noindent \textit{Requirement \#4:} The model inputs must be able to distinguish different filter widths.\\

\noindent Requirement \#1 is motivated by the fact that we are interested in an SGS model that is computationally efficient. While accurate non-local SGS models exist \cite{Bardina1980,Germano1991,Maulik2017,Stolz2001}, these are typically more expensive than local SGS models. It is also difficult to enforce rotational and reflectional invariance if non-local inputs are allowed.  Requirement \#2 is motivated by the fact that an SGS model is not Galilean invariant if it depends on model inputs that are themselves not Galilean invariant. Note this indicates a Galilean invariant SGS model cannot depend on the local filtered velocity field, but it can depend on spatial derivatives of the filtered velocity field. Requirement \#3 is motivated by the fact that different values of the local derivatives of the filtered velocity field are correlated with different values of the SGS tensor. Finally, Requirement \#4 is motivated by the fact that different values of filter width are correlated with different values of the SGS tensor. In particular, the components of the SGS tensor go to zero as the filter width goes to zero, that is, as the flow is fully resolved, regardless of the values of the local derivatives of the filtered velocity field. 

The minimal set of model inputs that satisfy Requirements \#1 - \#4 is comprised of the local filtered velocity strain-rate tensor $\bm{S}$, the local filtered rotation-rate tensor $\bm{\Omega}$, and the filter width $\Delta$, resulting in SGS models of the form:

\begin{equation}
\label{eq:original_form}
    \tau = \tau^{\text{\text{model}}} (\bm{S}, \bm{\Omega}, \Delta).
\end{equation}

\noindent We focus our attention on models of the above form in this paper. However, the systematic process described here for the model form construction can also accommodate other model inputs such as higher spatial derivatives of the filtered velocity field, the kinematic viscosity, and, for wall-bounded turbulent flows, the distance to the wall. It also supports the filtered pressure gradient as a model input, but we advise against this since the addition of a conservative body force to the body force vector modifies the filtered pressure gradient but not the SGS tensor.  This property is related to the concept of pressure robustness in the numerical solution of the Navier-Stokes equations \cite{Linke2016}.  Spalart has also suggested against the use of pressure gradients in turbulence modeling, and he has even gone so far to list ``Acceleration or pressure gradient is a valid entry in a model'' as a hard fallacy in his recent paper \cite{Spalart2015}.

To arrive at a rotationally and reflectionally invariant model form, we need to employ model inputs and outputs that are themselves rotationally and reflectionally invariant.  A few different strategies have been proposed in the literature for this purpose.  One strategy is to express the SGS tensor in terms of an infinite polynomial expansion of the prescribed tensor inputs.  One can then employ the Cayley-Hamilton Theorem to convert this infinite polynomial expansion to a finite polynomial expansion in terms of a so-called minimal tensor integrity basis.  The coefficients of this expansion in turn can be expressed as functions of the minimal integrity basis of polynomial invariants of the prescribed input tensors \cite{Pope1975}.  If one utilizes the minimal integrity basis of polynomial invariants as a set of model inputs and the expansion coefficients as a set of model outputs, one arrives at a rotationally and reflectionally invariant model form.  While this strategy has been employed for both data-driven Reynolds stress closure modeling \cite{Ling2016, Parmar2020} and data-driven SGS closure modeling \cite{Xie2020a, Reissmann2021, Doronina2020}, it suffers from the issue that sizes of the minimal tensor and invariant integrity bases grow exponentially fast with the number of prescribed tensor inputs.  A second strategy is to model the eigenstructure of the SGS tensor, that is, its eigenvalues and eigenvectors, as functions of invariant model inputs.  This strategy has been used with success for data-driven Reynolds stress closure modeling \cite{Wang2017}, but to the best of our knowledge, this strategy has not been used for data-driven SGS closure to date.  Likewise, we have not had success constructing a simple or efficient data-driven model for the SGS tensor using this approach.  This is largely due to the fact that to arrive at a rotationally and reflectionally invariant model, one must not directly model the eigenvector components in a global frame but rather the rotation tensor from a flow-dependent reference frame to the SGS tensor eigenframe.  Such a rotation tensor can be represented using either Euler angles or quarternions \cite{Wu2019}, but we have found these quantities are  nonlinear, discontinuous, and quite difficult to learn using state-of-the-art supervised machine learning strategies and regression techniques.

In this article, we employ a different strategy to arrive at a rotationally and reflectionally invariant model form.  Namely, we elect to represent the components of the model output and model input tensors in a flow specific coordinate system, namely the filtered strain-rate tensor eigenframe or $S$-frame. Such a representation yields a rotationally and reflectionally invariant model form as the $S$-frame components of a tensor remain unchanged under a coordinate rotation or reflection.  Peters \textit{et al.} first proposed to represent model output tensors in a flow specific coordinate system in the context of data-driven Reynolds stress closure \cite{Peters2020}, but to the best of our knowledge, representation of model input tensors in a flow specific coordinate system has not been considered previously.  Selection of the $S$-frame for model input representation results in a particularly convenient set of model inputs. The components of the filtered strain-rate tensor in the $S$-frame form a diagonal matrix,

\begin{equation}
    \begin{bmatrix}
    \mathrm{S}_{ij}^S
    \end{bmatrix} = 
    \begin{bmatrix}
    \lambda_1^S & 0 & 0 \\
    0 & \lambda_2^S & 0 \\
    0 & 0 & \lambda_3^S 
    \end{bmatrix},
\end{equation}

\noindent where $\lambda_1^S$, $\lambda_2^S$ and $\lambda_3^S$ are the three eigenvalues of the filtered strain-rate tensor, and the components of the filtered rotation-rate tensor in the $S$-frame form an anti-symmetric matrix,

\begin{equation}
    \begin{bmatrix}
    \mathrm{\Omega}_{ij}^S
    \end{bmatrix} = 
    \frac{1}{2}\begin{bmatrix}
    0 & \omega_3^S & -\omega_2^S \\
    -\omega_3^S & 0 & \omega_1^S \\
    \omega_2^S & -\omega_1^S & 0
    \end{bmatrix}, \label{eq:Omega_Sframe}
\end{equation}

\noindent where $\omega_1^S$, $\omega_2^S$ and $\omega_3^S$ are the components of the filtered vorticity in the $S$-frame, that is,

\begin{equation}
 \omega^S_1 = \boldsymbol{\omega} \cdot \bm{v}^S_1, \quad
 \omega^S_2 = \boldsymbol{\omega} \cdot \bm{v}^S_2, \quad
 \omega^S_3 = \boldsymbol{\omega} \cdot \bm{v}^S_3,
\end{equation}

\noindent where $\bm{\omega} = \nabla \times \bar{\bm{u}}$ is the filtered vorticity\footnote{We have assumed a right handed global coordinate system to arrive at \eqref{eq:Omega_Sframe}.  The same result holds for a left handed global coordinate system if we replace the definition $\bm{\omega} = \nabla \times \bar{\bm{u}}$ with $\bm{\omega} = -\nabla \times \bar{\bm{u}}$.  This change of definition is required as filtered vorticity is a pseudovector rather than a true vector.} and $\bm{v}_1^S$, $\bm{v}_2^S$ and $\bm{v}_3^S$ are the three eigenvectors of the filtered strain-rate tensor corresponding to the eigenvalues $\lambda_1^S$, $\lambda_2^S$ and $\lambda_3^S$ respectively. Thus, if we initially assume the SGS tensor depends on the filtered strain-rate tensor $\bm{S}$, the filtered rotation-rate tensor $\bm{\Omega}$, and the filter width $\Delta$, then representation of tensor model inputs and outputs in the $S$-frame yields SGS models of the form

\begin{equation}
\tau_{ij}^S = \tau_{ij}^{S,\text{model}} ( \lambda_1^S, \lambda_2^S, \lambda_3^S, \omega_1^S, \omega_2^S, \omega_3^S, \Delta),
\end{equation}

\noindent where $\tau_{ij}^S$ is the $ij^{th}$ component of the SGS tensor in the $S$-frame. Note that since, in general, the filtered strain-rate tensor has six independent components and the filtered rotation-rate tensor has three independent components, we have reduced the number of model inputs from ten to seven by expressing model input tensors in the $S$-frame. We can further reduce the number of model inputs by recognizing

\begin{equation}
    \lambda_2^S = - \lambda_1^S - \lambda_3^S
    \label{eq:incomp_cond}
\end{equation}

\noindent since the filtered velocity field is divergence-free. We also elect to replace one of the model inputs, $\lambda_1^S$, by the velocity gradient magnitude:

\begin{equation}
\begin{split}
    G & = \Big( S_{ij} S_{ij} + \Omega_{ij} \Omega_{ij} \Big)^{1/2} = \Big( S^S_{ij} S^S_{ij} + \Omega^S_{ij} \Omega^S_{ij} \Big)^{1/2} \\
    & = \Bigg( (\lambda_1^S)^2 + (\lambda_2^S)^2 + (\lambda_3^S)^2 + \frac{1}{2}\Big( (\omega_1^S)^2 + (\omega_2^S)^2 + (\omega_3^S)^2 \Big) \Bigg)^{1/2}.
\end{split}    
\label{eq:G_def}
\end{equation}

\noindent Removal of $\lambda_2^S$ as a model input and substitution of $G$ for $\lambda_1^S$ as a model input yields the simplified model form

\begin{equation}
\tau_{ij}^S = \tau_{ij}^{S,\text{model}} (\lambda_3^S, \omega_1^S, \omega_2^S, \omega_3^S, G, \Delta).
\end{equation}

\noindent SGS models of the above form are necessarily Galilean, rotationally, and reflectionally invariant, and they are symmetric provided $\tau_{ij}^{S,\text{model}} = \tau_{ji}^{S,\text{model}}$. However, they are not necessarily unit invariant.

To arrive at a unit invariant model form, we invoke the Buckingham Pi theorem. Namely, if an SGS model of the form
\begin{equation}
\tau_{ij}^S = \tau_{ij}^{S,\text{model}} (\lambda_3^S, \omega_1^S, \omega_2^S, \omega_3^S, G, \Delta)
\end{equation}

\noindent is unit invariant, then as the model inputs and outputs involve two independent physical units (length and time), the SGS model must admit the dimensionless form
\begin{equation}
\Pi_5 = \hat{\tau}_{ij}^{S,\text{model}} (\Pi_1, \Pi_2, \Pi_3, \Pi_4),
\end{equation}

\noindent where $\Pi_1$, $\Pi_2$, $\Pi_3$, $\Pi_4$ and $\Pi_5$ are suitable chosen dimensionless parameters. We select
\begin{equation}
    \Pi_1 = \hat{\lambda}_3^S = \frac{\lambda^S_3}{G}
\end{equation}
\begin{equation}
    \Pi_2 = \hat{\omega}_1^S = \frac{\omega^S_1}{G}
\end{equation}
\begin{equation}
    \Pi_3 = \hat{\omega}_2^S = \frac{\omega^S_2}{G}
\end{equation}
\begin{equation}
    \Pi_4 = \hat{\omega}_3^S = \frac{\omega^S_3}{G}
\end{equation}
\begin{equation}
    \Pi_5 = \hat{\tau}_{ij}^S = \frac{\tau_{ij}^S}{\Delta^2 G^2},
\end{equation}

\noindent yielding the final model form

\begin{equation}
\tau_{ij}^S = \Delta^2 G^2 \hat{\tau}_{ij}^{S,\text{model}} (\hat{\lambda}_3^S, \hat{\omega}_1^S, \hat{\omega}_2^S, \hat{\omega}_3^S). \label{eq:final_model_form}
\end{equation}

\noindent Note that while the original model form given by \eqref{eq:original_form} has ten model inputs, the final model form given above has only four model inputs.  This reduction was made possible through representation of model input tensors in the $S$-frame, exploitation of the fact that the trace of the filtered strain-rate tensor is zero, and a judicious use of the Buckingham Pi theorem.  It is readily seen that SGS models of the form given by \eqref{eq:final_model_form} are necessarily Galilean, rotationally, reflectionally, and unit invariant, and all Galilean, rotationally, reflectionally, and unit invariant SGS models that depend on the filtered strain-rate tensor, filtered rotation-rate tensor, and filter width must admit the form given by \eqref{eq:final_model_form}. Moreover, models of the form given by \eqref{eq:final_model_form} can be made symmetric by enforcing that $\hat{\tau}_{ij}^{S,\text{model}} = \hat{\tau}_{ji}^{S,\text{model}} $.  It should also be mentioned again that while we have confined our presentation to SGS models that depend only on the filtered strain-rate tensor, filtered rotation-rate tensor, and filter width, the above systematic procedure can also be employed to construct symmetric, Galilean invariant, rotationally invariant, reflectionally invariant, and unit invariant model forms with additional model inputs provided, of course, the additional model inputs are also Galilean invariant.

The above process can be repeated to arrive at a symmetric, Galilean invariant, rotationally invariant, reflectionally invariant, and unit invariant model form for the deviatoric part of the SGS tensor.  If the deviatoric part of the SGS tensor is assumed to depend on the filtered strain-rate tensor, filtered rotation-rate tensor, and filter width, this yields the model form
\begin{equation}
\tau_{ij}^{d,S} = \Delta^2 G^2 \hat{\tau}_{ij}^{d,S,\text{model}} (\hat{\lambda}_3^S, \hat{\omega}_1^S, \hat{\omega}_2^S, \hat{\omega}_3^S) 
\label{eq:final_model_form_dev}
\end{equation}
where $\tau_{ij}^{d,S}$ is the $ij^{th}$ component of the deviatoric part of the SGS tensor in the $S$-frame and $\hat{\tau}_{ij}^{d,S,\text{model}} = \hat{\tau}_{ji}^{d,S,\text{model}} $ due to symmetry.

Heretofore, we have not yet specified an ordering of filtered strain-rate tensor eigenvalues or on orientation of filtered strain-rate tensor eigenvectors. However, such specifications are critical to the success of any data-driven SGS model constructed using our framework. For instance, a data-driven SGS model constructed using our approach will yield inaccurate results if it takes in as input filtered strain-rate tensor eigenvalues in a different ordering than what the model was trained on, and flipping the orientation of the filtered strain-rate tensor eigenvectors changes the signs of select components of the filtered vorticity vector in the $S$-frame as well as select off diagonal components of the SGS tensor in the $S$-frame. For the remainder of this paper, we assume the filtered strain-rate tensor eigenvalues are ordered such that

\begin{equation}
     \lambda^S_1 \vargeq \lambda^S_2 \vargeq \lambda^S_3,
\end{equation}

\noindent and we orient the filtered strain-rate eigenvectors such that

\begin{equation}
    \bm{v}_1^S \cdot \bm{\omega} \vargeq 0
\end{equation}

\begin{equation}
    \bm{v}_3^S \cdot \bm{\omega} \vargeq 0
\end{equation}

\begin{equation}
    \bm{v}_2^S = \bm{v}_3^S \times \bm{v}_1^S.
\end{equation}

\begin{figure}[t!]
    \input{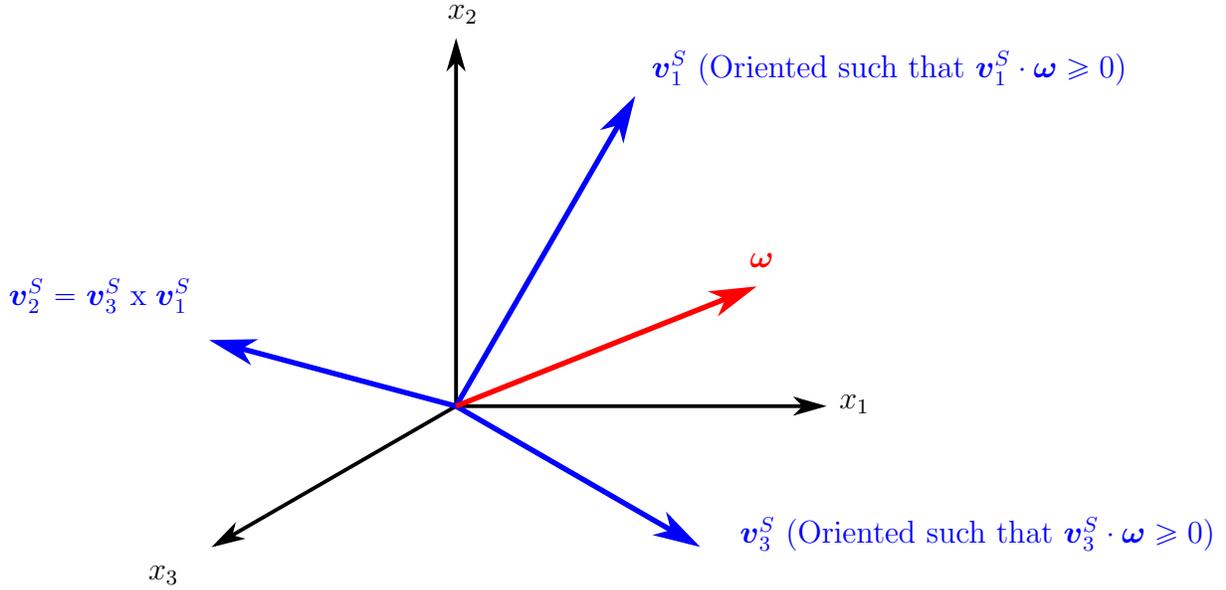}
    \vspace{2pt}
    \caption{Orientation of filtered strain-rate tensor eigenvectors.}
    \label{fig:eigvec_align}
\end{figure}

\noindent A visual depiction of the chosen eigenvector orientation is provided in \figref{eigvec_align}.  This choice of orientation is not unique, and fundamental studies on vorticity alignment with local and nonlocal strain rates could be used to guide alternative choices \cite{Ashurst1987a,Hamlington2008}.  However, the final model forms given in \eqref{eq:final_model_form} and \eqref{eq:final_model_form_dev} hold regardless of the choice of orientation provided it is fixed and consistent, and we have found that our choice of orientation is quite effective in practice.

\subsection{Model Form Representation of the Gradient Model}

For a box filter, if one takes the Fourier transform of the SGS tensor, performs a Taylor series expansion of the transformed quantity with respect to filter width, and then takes the inverse Fourier transform of the resulting expansion (see Chapter 2 of \cite{Aldama1990}), one attains

\begin{equation}
    \tau_{ij} = \frac{1}{12} \Delta^2 \frac{\partial \bar{u}_i}{\partial x_k} \frac{\partial \bar{u}_j}{\partial x_k} + \mathcal{O}(\Delta^4).
    \label{eq:taylor_series}
\end{equation}

\noindent Thus the gradient model corresponds to a Taylor series truncation of the SGS tensor. It is desirable, then, to have an SGS model form that yields a simple representation of the gradient model. The model form given by \eqref{eq:final_model_form} yields such a representation. To see this, note that expressing \eqref{eq:clark} in the $S$-frame results in the equation
\begin{equation}
    \tau^S_{ij} = \frac{1}{12} \Delta^2 \mathrm{G}_{ik}^S \mathrm{G}_{jk}^S
\end{equation}
\noindent where
\begin{equation}
    \begin{bmatrix}
    \mathrm{G}_{ij}^S
    \end{bmatrix} = 
    \begin{bmatrix}
    \mathrm{S}_{ij}^S
    \end{bmatrix} + \begin{bmatrix}
    \mathrm{\Omega}_{ij}^S
    \end{bmatrix} = 
    \begin{bmatrix}
    \lambda_1^S & \omega_3^S /2 & -\omega_2^S /2 \\
    -\omega_3^S /2 & \lambda_2^S & \omega_1^S /2 \\
    \omega_2^S /2 & -\omega_1^S /2 & \lambda_3^S 
    \end{bmatrix}.
\end{equation}
\noindent It follows then that
\begin{equation}
   \tau_{ij}^S = \Delta^2 G^2 \hat{\tau}_{ij}^{S,\text{gradient}} (\hat{\lambda}_3^S,\hat{\omega}_1^S,\hat{\omega}_2^S,\hat{\omega}_3^S)
\end{equation}
where

\begin{equation}
    \hat{\tau}_{11}^{S,\text{gradient}}(\hat{\lambda}_3^S,\hat{\omega}_1^S,\hat{\omega}_2^S,\hat{\omega}_3^S) = \frac{1}{12} \left( (\hat{\lambda}_1^S)^2 + \frac{1}{4} (\hat{\omega}_2^S)^2  + \frac{1}{4} (\hat{\omega}_3^S)^2 \right)
\end{equation}

\begin{equation}
    \hat{\tau}_{22}^{S,\text{gradient}}(\hat{\lambda}_3^S,\hat{\omega}_1^S,\hat{\omega}_2^S,\hat{\omega}_3^S) = \frac{1}{12} \left( (\hat{\lambda}_2^S)^2 + \frac{1}{4} (\hat{\omega}_1^S)^2  + \frac{1}{4} (\hat{\omega}_3^S)^2 \right)
\end{equation}

\begin{equation}
    \hat{\tau}_{33}^{S,\text{gradient}}(\hat{\lambda}_3^S,\hat{\omega}_1^S,\hat{\omega}_2^S,\hat{\omega}_3^S) = \frac{1}{12} \left( (\hat{\lambda}_3^S)^2 + \frac{1}{4} (\hat{\omega}_1^S)^2  + \frac{1}{4} (\hat{\omega}_2^S)^2 \right)
\end{equation}

\begin{equation}
    \hat{\tau}_{12}^{S,\text{gradient}}(\hat{\lambda}_3^S,\hat{\omega}_1^S,\hat{\omega}_2^S,\hat{\omega}_3^S) = \hat{\tau}_{21}^{S,\text{gradient}}(\hat{\lambda}_3^S,\hat{\omega}_1^S,\hat{\omega}_2^S,\hat{\omega}_3^S) = \frac{1}{12} \left( \frac{1}{2} (\hat{\lambda}_2^S - \hat{\lambda}_1^S) \hat{\omega}_3^S - \frac{1}{4} \hat{\omega}_1^S \hat{\omega}_2^S \right)
\end{equation}

\begin{equation}
    \hat{\tau}_{13}^{S,\text{gradient}}(\hat{\lambda}_3^S,\hat{\omega}_1^S,\hat{\omega}_2^S,\hat{\omega}_3^S) = \hat{\tau}_{31}^{S,\text{gradient}}(\hat{\lambda}_3^S,\hat{\omega}_1^S,\hat{\omega}_2^S,\hat{\omega}_3^S) = \frac{1}{12} \left( \frac{1}{2} (\hat{\lambda}_1^S - \hat{\lambda}_3^S) \hat{\omega}_2^S - \frac{1}{4} \hat{\omega}_1^S \hat{\omega}_3^S \right)
\end{equation}

\begin{equation}
    \hat{\tau}_{23}^{S,\text{gradient}}(\hat{\lambda}_3^S,\hat{\omega}_1^S,\hat{\omega}_2^S,\hat{\omega}_3^S) = \hat{\tau}_{32}^{S,\text{gradient}}(\hat{\lambda}_3^S,\hat{\omega}_1^S,\hat{\omega}_2^S,\hat{\omega}_3^S) = \frac{1}{12} \left( \frac{1}{2} (\hat{\lambda}_3^S - \hat{\lambda}_2^S) \hat{\omega}_1^S - \frac{1}{4} \hat{\omega}_2^S \hat{\omega}_3^S \right)
\end{equation}

\noindent and, since $\hat{\lambda}^S_1 + \hat{\lambda}^S_2 + \hat{\lambda}^S_3 = 0$ and $\left( \hat{\lambda}^S_1 \right)^2 + \left( \hat{\lambda}^S_2 \right)^2 + \left( \hat{\lambda}^S_3 \right)^2 + \frac{1}{2} \left( \left( \hat{\omega}^S_1 \right)^2 + \left( \hat{\omega}^S_2 \right)^2 + \left( \hat{\omega}^S_3 \right)^2 \right) = 1$,

\begin{equation}
    \hat{\lambda}_1^S = \frac{1}{2} \Big(- \hat{\lambda}_3^S + \sqrt{2 - 3(\hat{\lambda}_3^S)^2 - (\hat{\omega}_1^S)^2 - (\hat{\omega}_2^S)^2 - (\hat{\omega}_3^S)^2 }\Big)
\end{equation}

\begin{equation}
    \hat{\lambda}_2^S = \frac{1}{2} \Big( - \hat{\lambda}_3^S - \sqrt{2 - 3(\hat{\lambda}_3^S)^2 - (\hat{\omega}_1^S)^2 - (\hat{\omega}_2^S)^2 - (\hat{\omega}_3^S)^2 }\Big).
\end{equation}

\noindent The model form given by \eqref{eq:final_model_form_dev} yields a similar algebraic representation of the deviatoric part of the SGS tensor predicted by the gradient model.  We have conducted numerical tests indicating the algebraic functions $\tau_{ij}^{S,\text{gradient}}$ can be accurately approximated using shallow neural networks, inspiring their use in data-driven SGS closure. It should be noted that an even simpler representation of the gradient model, namely a quadratic polynomial representation, is attained if $\hat{\lambda}^S_1$ and $\hat{\lambda}^S_2$ are retained as model inputs, but we have elected not to do so as $\hat{\lambda}^S_1$ and $\hat{\lambda}^S_2$ may be expressed in terms of the other model inputs and inclusion of $\hat{\lambda}^S_1$ and $\hat{\lambda}^S_2$ as model inputs results in a model with slightly increased computational cost.

\subsection{Functional Mapping Using Artificial Neural Networks}

To arrive at a computable model, an artificial neural network (ANN) may be employed to represent either the functions $\hat{\tau}_{ij}^{S,\text{model}}$ appearing in \eqref{eq:final_model_form} (if we seek a model for the full SGS tensor) or the functions $\hat{\tau}_{ij}^{d,S,\text{model}}$ appearing in \eqref{eq:final_model_form_dev} (if we seek a model for only the deviatoric part of the SGS tensor).  For a fixed ANN architecture, weights and biases can be learned using suitable high-fidelity simulation data (e.g., model input and output data computed using DNS), a suitable loss function (e.g., a loss function  penalizing the difference between the SGS tensor predicted by the ANN model and the exact SGS tensor), and a suitable optimization algorithm (e.g., a stochastic gradient descent algorithm).  ANNs have been used with success in past studies on data-driven Reynolds stress closure \cite{Ling2016, Peters2020} and data-driven SGS closure \cite{Gamahara2017, Maulik2019}, and we have found ANNs yield improved performance over alternative regression techniques such as polynomial expansions in modeling $\hat{\tau}_{ij}^{S,\text{model}}$ or $\hat{\tau}_{ij}^{d,S,\text{model}}$ on an accuracy-versus-cost basis.  In particular, we have had success employing dense ANNs with four inputs (each of $\hat{\lambda}_3^S$, $\hat{\omega}_1^S$, $\hat{\omega}_2^S$, and $\hat{\omega}_3^S$), six outputs (each of $\hat{\tau}_{ij}^{S,\text{model}}$ or $\hat{\tau}_{ij}^{d,S,\text{model}}$), and leaky rectified linear unit (ReLU) activation functions.  For the sake of brevity, we do not review the construction of ANNs in this paper, and we instead refer the reader to \cite{Goodfellow2016} for more details.

\section{A Simple Data-Driven Model for the Deviatoric Part of the SGS Tensor}
\label{sec:Simple}

In the previous section, we presented a model form for data-driven SGS closure that is symmetric, Galilean invariant, rotationally invariant, reflectionally invariant, and unit invariant.  In this section, we demonstrate that this model form can be used to construct a simple data-driven model for the SGS tensor using a limited amount of training data.  In particular, we build a dense feedforward ANN model for the deviatoric part of the SGS tensor with a single hidden layer composed of twenty neurons, each equipped with a leaky ReLU activation function.  In the next section, we examine the accuracy of this model using both \textit{a priori} and \textit{a posteriori} tests, and we also examine the model's capability to generalize to turbulent flow cases outside the training dataset.

To train the ANN model, we use tensor-product box-filtered DNS data from a nondimensional forced HIT simulation at a Taylor Reynolds number of $Re_{\lambda} = 418$.  This data is attained from the Johns Hopkins Turbulence Database (JHTDB) \cite{Li2008}.  While the JHTDB provides DNS data at $1,024^3$ spatial locations for 5,028 time steps after the flow becomes statistically stationary, we train using data at only $196,608$ randomly sampled spatial locations at simulation time $t = 1$ (see Table \ref{tab:Training}). This time corresponds to 0.5 eddy turnover times past the first recorded time step at time $t = 0$ where eddy turnover time is defined as the integral length scale over the root mean square of velocity.  We also train using a single filter width, $\Delta \approx 28.5 \eta$ where $\eta$ is the Kolmogorov length scale.  This filter width lies within the inertial subrange for the considered forced HIT simulation.  Through numerical experimentation, we have discovered that including more spatial locations, time steps, and filter widths in the training dataset does not yield improved accuracy or generalizability.  We hypothesize this is due to the invariance properties strongly embedded in the model form.

The following mean-squared error (MSE) loss function is employed to train the ANN model:
\begin{equation}
    \text{MSE} (\boldsymbol{\hat{W}},\boldsymbol{\hat{b}}) = \frac{1}{n_\text{train}} \sum_{a = 1}^{n_\text{train}} \sum_{i = 1}^{3} \sum_{j = 1}^{3} \left( \hat{\tau}_{ij}^{d,S,\textup{DNS}}(\textbf{x}_a) - \hat{\tau}_{ij}^{d,S,\textup{model}} \left(\hat{\boldsymbol{q}}^\text{DNS}(\textbf{x}_a); \boldsymbol{\hat{W}}, \boldsymbol{\hat{b}}\right) \right)^2 \;,
    \label{eq:MSE}
\end{equation}
where $\hat{\tau}_{ij}^{d,S,\textup{model}}$ denotes the ANN model for the $ij^\text{th}$ component of the deviatoric part of the non-dimensional SGS tensor in the $S$-frame, $\boldsymbol{\hat{W}}$ and $\boldsymbol{\hat{b}}$ denote the weights and biases of the ANN model, $\hat{\tau}_{ij}^{d,S,\textup{DNS}}$ denotes the DNS value of the $ij^\text{th}$ component of the deviatoric part of the non-dimensional SGS tensor in the $S$-frame, $ \hat{\boldsymbol{q}}^{\textup{DNS}}$ denotes the DNS value of the non-dimensional input vector (composed of $\hat{\lambda}_3^S$, $\hat{\omega}_1^S$, $\hat{\omega}_2^S$, and $\hat{\omega}_3^S$), and $\left\{ \textbf{x}_a \right\}_{a=1}^{n_\text{train}}$ denotes the set of training points.  The above loss function penalizes the Frobenius norm of difference between the predicted and exact values of the deviatoric part of the non-dimensional SGS tensor.  The loss function can be further enhanced by including an additional term penalizing the difference between the predicted and exact SGS energy dissipation, but this is not considered in this paper.  We utilize Adam, an adaptive learning rate optimization algorithm designed specifically for training ANNs, in order to find the optimal weights and biases \cite{Adam2015}.

\begin{table}[t!]
    \centering
    \begin{tabular}{ccccc}
        \hline
        \hline
         \textbf{Dataset} & \textbf{No. of Samples} & \textbf{Spatial Locations} & \textbf{Time} & \textbf{Filter Width} \\
         \hline
         Training/Testing & $196,608$/$65,536$ & Randomly Sampled  & $t = 1$ & $\Delta \approx 29 \eta$ \\
        & & From Full Domain & & \\
         \hline
         \hline
    \end{tabular}
    \caption{Training and testing datasets.}
    \label{tab:Training}
\end{table}

\begin{figure}[t!]
    \centering
    \subfigure[\label{fig:Training_MSE}]{\includegraphics[width=0.49\textwidth]{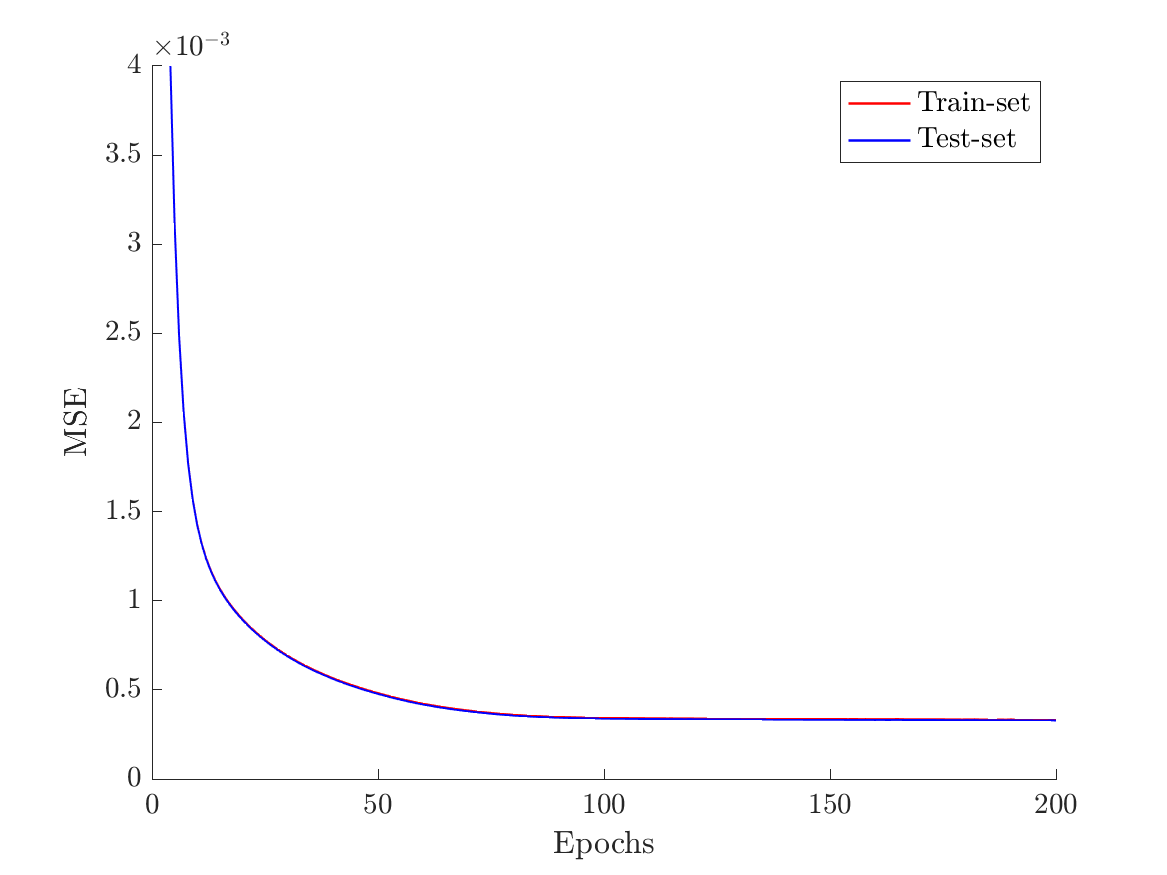}}
    \subfigure[\label{fig:Training_CCs}]{\includegraphics[width=0.49\textwidth]{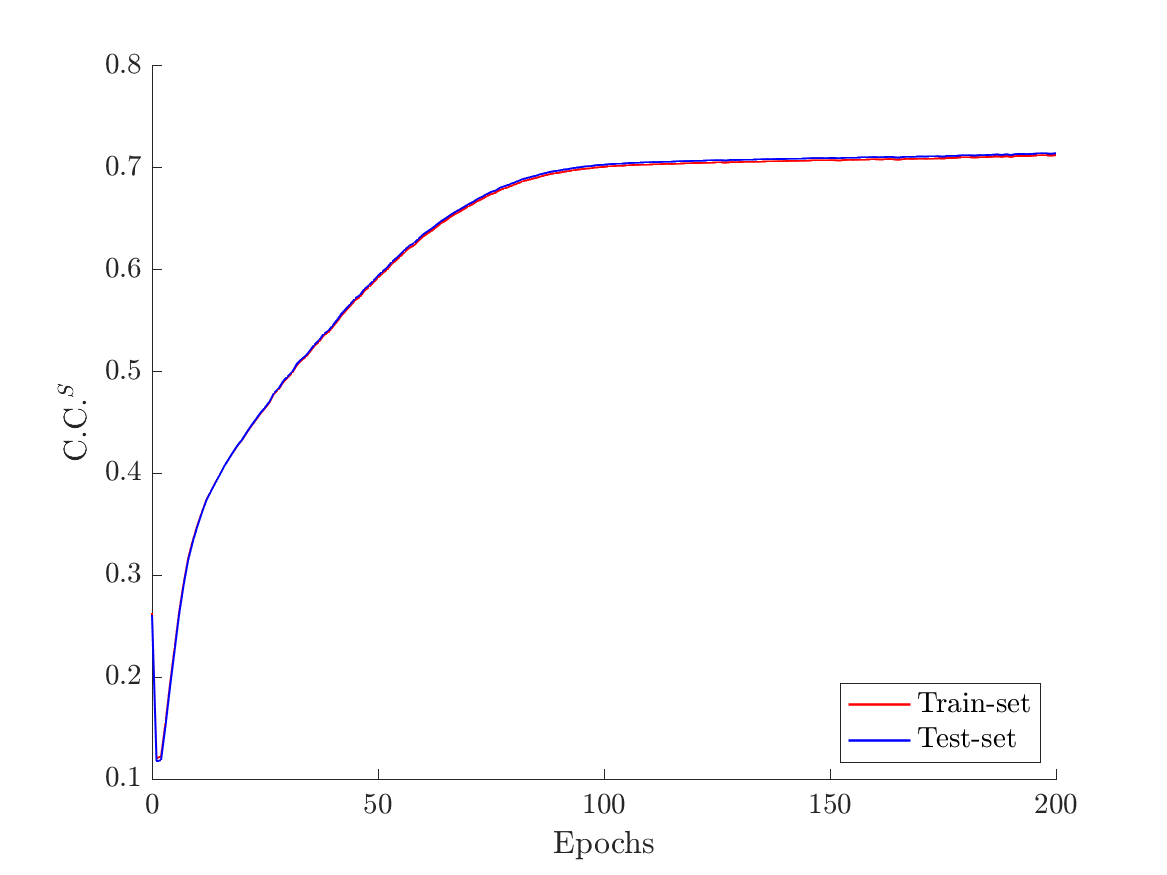}}
    \caption{Neural network training convergence: (a) mean-squared error ($\text{MSE}$) and (b) non-dimensional $S$-frame correlation coefficient ($\text{C.C.}^S$).}
    \label{fig:Training}
\end{figure}

In \figref{Training}(a), the MSE is plotted versus number of epochs for both the training dataset as well as a testing dataset of $65,536$ randomly sampled spatial locations at time $t = 1$.  The MSE quickly drops to a small value for both the training and testing datasets, and the MSE for the testing dataset is virtually identical to that of the training dataset throughout the training process.  In \figref{Training}(b), the non-dimensional $S$-frame correlation coefficient

\begin{equation}
    \text{C.C.}^{S} = \mathlarger{\sum_{i}} \mathlarger{\sum_{j}} \frac{\langle ( \hat{\tau}^{d,S,\text{DNS}}_{ij}- \langle \hat{\tau}^{d,S,\text{DNS}}_{ij} \rangle) ( \hat{\tau}^{d,S,\text{\text{model}}}_{ij} - \langle \hat{\tau}^{d,S,\text{\text{model}}}_{ij} \rangle) \rangle}{(\langle (\hat{\tau}^{d,S,\text{DNS}}_{ij}- \langle \hat{\tau}^{d,S,\text{DNS}}_{ij} \rangle)^2  \rangle)^{1/2}(\langle (\hat{\tau}^{d,S,\text{\text{model}}}_{ij}- \langle \hat{\tau}^{d,S,\text{\text{model}}}_{ij} \rangle)^2  \rangle)^{1/2}}
    \label{eq:CCs}
\end{equation}

\noindent is plotted versus number of epochs for the training and testing sets.  In the above formula, $\langle \cdot \rangle$ indicates taking an average of the considered data samples.  The non-dimensional $S$-frame correlation coefficient quickly converges to a value of approximately 0.7 for both the training and testing datasets.  These results suggest the model training converges without overfitting the training dataset.

\section{Numerical Results}
\label{sec:Results}

In this section, we perform both \textit{a priori} and \textit{a posteriori} tests in order to compare the performance of the data-driven model constructed in the last section to classical SGS models. We use the abbreviations shown in \tabref{abbrev_SGS} to refer to the different SGS models considered in this study. The clipped variants of the gradient and data-driven models locally set each stress component to zero if the local model SGS energy dissipation is negative.  That is, the clipped variants of these models take the form
\begin{equation}
     \tau^{d,M-C}_{ij} = \begin{cases}
     0 & \text{if } \tau^{d,M}_{ij} S_{ij} > 0 \\
     \tau^{d,M}_{ij} & \text{if } \tau^{d,M}_{ij} S_{ij} \varleq 0, \\
     \end{cases}
 \end{equation}
 \noindent where the superscripts ${M}$ and ${M-C}$ indicate unclipped and clipped modeled stresses respectively.  Clipping removes model backscatter and thereby improves numerical stability of \textit{a posteriori} simulations \cite{Liu1994, Vreman1997}. We choose to compare the performance of our trained model with the dynamic Smagorinsky model and the gradient model as these are two popular functional and structural models, but improved results may be attained by other functional, structural, or mixed models.  However, the primary purpose of our \textit{a priori} and \textit{a posteriori} tests is not to compare the performance of our trained model with state-of-the-art SGS models but rather examine our model's ability to generalize to previously unseen filter widths, Reynolds numbers, and flow physics.

\begin{table}[b!]
    \centering
    \begin{tabular}{cc}
        \hline
        \hline
         \textbf{\text{Model}} & \textbf{Abbreviation}  \\
         \hline
         No Model & NM \\
         Static Smagorinsky Model & SM \\
         Dynamic Smagorinsky Model & DSM \\
         Gradient Model & GM \\
         Gradient Model with Clipping & GM-C \\
         Data-Driven Model & DD \\
         Data-Driven Model with Clipping & DD-C \\
         \hline
         \hline
    \end{tabular}
    \caption{List of SGS models for comparison.}
    \label{tab:abbrev_SGS}
\end{table}

\subsection{\textit{A Priori} Testing of the Data-Driven Subgrid Stress \text{Model}}
\label{sec:Apriori}

In \textit{a priori} tests, modeled SGS stresses and energy dissipation are evaluated based on filtered DNS inputs and compared with those from the filtered DNS data. Selecting a subset of data that is different from the training dataset enables us to observe how well the data-driven model performs for previously unseen model inputs and investigate if there is any dependence of the data-driven model on the training data. To facilitate this, in addition to the training and testing datasets, a model validation dataset is used to evaluate the performance of the data-driven model. The model validation dataset is obtained from the same nondimensional forced HIT simulation that was employed for model training and is described in \tabref{Apriori_descp}.
\begin{table}[t!]
    \centering
    \begin{tabular}{ccccc}
        \hline
        \hline
         \textbf{Case} & \textbf{No. of Samples} & \textbf{Spatial Locations} &\textbf{Time} & \textbf{Filter Widths} \\
         \hline
         Case 1 & 262,144 & Randomly Sampled  & $t = 1$ & $\Delta \approx 29 \eta$ \\ 
         & & From Full Domain & & \\
         Case 2 & 512 x 512 x 1 & Uniformly Sampled & $t = 10$ & $\Delta \approx 6.5 \eta - 68 \eta $ \\
         & & From Slice z = $\pi$ & & \\
         \hline
         \hline
    \end{tabular}
    \caption{Description of model validation datasets for \textit{a priori} tests.}
    \label{tab:Apriori_descp}
\end{table}
We use the correlation coefficient (C.C.) and relative error in mean energy flux (R.E.F.) to model performance. C.C. is given as

\begin{equation}
    \text{C.C.} = \mathlarger{\sum_{i}} \mathlarger{\sum_{j}} \frac{\langle ( \tau^{d,\textup{DNS}}_{ij}- \langle \tau^{d,\textup{DNS}}_{ij} \rangle) ( \tau^{d,M}_{ij} - \langle \tau^{d,M}_{ij} \rangle) \rangle}{(\langle (\tau_{ij}^{d,\textup{DNS}}- \langle \tau_{ij}^{d,\textup{DNS}} \rangle)^2  \rangle)^{1/2}(\langle (\tau^{d,M}_{ij}- \langle \tau^{d,M}_{ij} \rangle)^2  \rangle)^{1/2}}
    \label{eq:CC},
\end{equation}

\noindent where $\tau_{ij}^{d,\textup{DNS}}$ denotes the DNS value of the $ij^\text{th}$ component of the deviatoric part of the dimensional SGS tensor in the global frame, $\tau_{ij}^{d,M}$ denotes the model value for the $ij^\text{th}$ component of the deviatoric part of the dimensional SGS tensor in the global frame, and $\langle \cdot \rangle$ indicates taking an average of the considered data samples.  R.E.F. is defined as

\begin{equation}
    \text{R.E.F.} = \frac{\langle \Pi^M \rangle - \langle \Pi^\text{DNS} \rangle }{\langle \Pi^\text{DNS} \rangle}
    \label{eq:REF},
\end{equation}

\noindent where $\Pi^M = - \tau^{d,M}_{ij} S_{ij}$ is the model SGS energy dissipation and $\Pi^\text{DNS} = - \tau^{d,\text{DNS}}_{ij} S_{ij}$ is the exact SGS energy dissipation. C.C. assesses the alignment of the model and exact SGS tensors while R.E.F. assesses the accuracy of the model SGS energy dissipation. It has been suggested in the literature \cite{Baggett1997} that the relative importance of these two metrics in an LES depends on the characteristics of the flow being simulated. For an isotropic flow simulated using a high resolution mesh, correct prediction of the SGS energy dissipation is important for attaining high-quality flow predictions, which corresponds to a low value of R.E.F. \cite{Baggett1997}.  For an anisotropic flow simulated using a low resolution mesh, correct prediction of the SGS tensor itself is important for attaining high-quality flow predictions, which corresponds to a high value of C.C. \cite{Baggett1997}. Thus, it is valuable to assess both the quality of C.C. and R.E.F. in \textit{a priori} tests.

\subsubsection{Case 1}

For Case 1, we consider the model validation dataset to be the union of the training and testing datasets.  Results for this case are shown in \tabref{Apriori_Case1}. The SGS stresses predicted by the gradient model are highly correlated with the exact SGS stresses. The C.C. values shown in \tabref{Apriori_Case1} for the gradient model are consistent with previous studies \cite{Clark1979, Borue1998}. \tabref{Apriori_Case1} also shows a low C.C. value is attained by the static Smagorinsky model. Like the gradient model, the SGS stresses predicted by the data-driven model are highly correlated with the exact SGS stresses. However, the data-driven model gives a much better mean SGS energy dissipation prediction than both the static Smagorinsky and gradient models. The static Smagorinsky model massively overpredicts the mean SGS energy dissipation while the gradient model significantly underpredicts the mean SGS energy dissipation. These observations are consistent with the LES literature. The clipped versions of the gradient and data-driven models provide a higher mean SGS energy dissipation at the expense of reduced correlation of model SGS stresses to the filtered DNS data. Considering both C.C. and R.E.F., the data-driven model seems to exhibit the best performance.

\begin{table}[t!]
    \centering
    \begin{tabular}{ccc}
        \hline
        \hline
         \textbf{\text{Model}} & C.C. & R.E.F.  \\
         \hline
         Static Smagorinsky Model & 0.275 & 1.2619 \\
         Gradient Model & 0.897 & -0.4095 \\
         Gradient Model with Clipping & 0.827 & -0.3411 \\
         Data-Driven Model & 0.891 & 0.0667 \\
         Data-Driven Model with Clipping & 0.860 & 0.0922 \\
         \hline
         \hline
    \end{tabular}
    \caption{\textit{A priori} results for Case 1: (a) Correlation Coefficient (C.C.) and (b) Relative error in mean energy flux (R.E.F).}
    \label{tab:Apriori_Case1}
\end{table}

\subsubsection{Case 2}
\label{sec:apriori_case2}

The results attained in Case 1 indicate that the data-driven model is predictive for inputs that are within the training dataset.  It remains to be seen, however, if the data-driven model is predictive for inputs that are outside the training dataset. To investigate this, we examine the accuracy of the data-driven model on a different spatio-temporal slice of data than the training and testing datasets. The considered spatio-temporal slice occurs roughly 4.5 eddy turnover times after the time of the training dataset. This test is referred to as Case 2 and the corresponding model validation dataset is described in \tabref{Apriori_descp}. The correlation coefficient between the exact SGS tensors for the training and validation datasets is close to zero, indicating the validation dataset is suitable for assessing model accuracy for inputs outside the training dataset. Furthermore, for this case, we investigate a range of filter widths, $ \Delta \approx 6.5\eta - 69 \eta$, rather than just the filter width that was employed to train the data-driven model. This range of filter widths covers both the inertial subrange and dissipation range for the considered flow case. 

\begin{figure}
    \centering
    \subfigure[\label{fig:CC_Apriori_Case2}]{\includegraphics[width=0.49\textwidth]{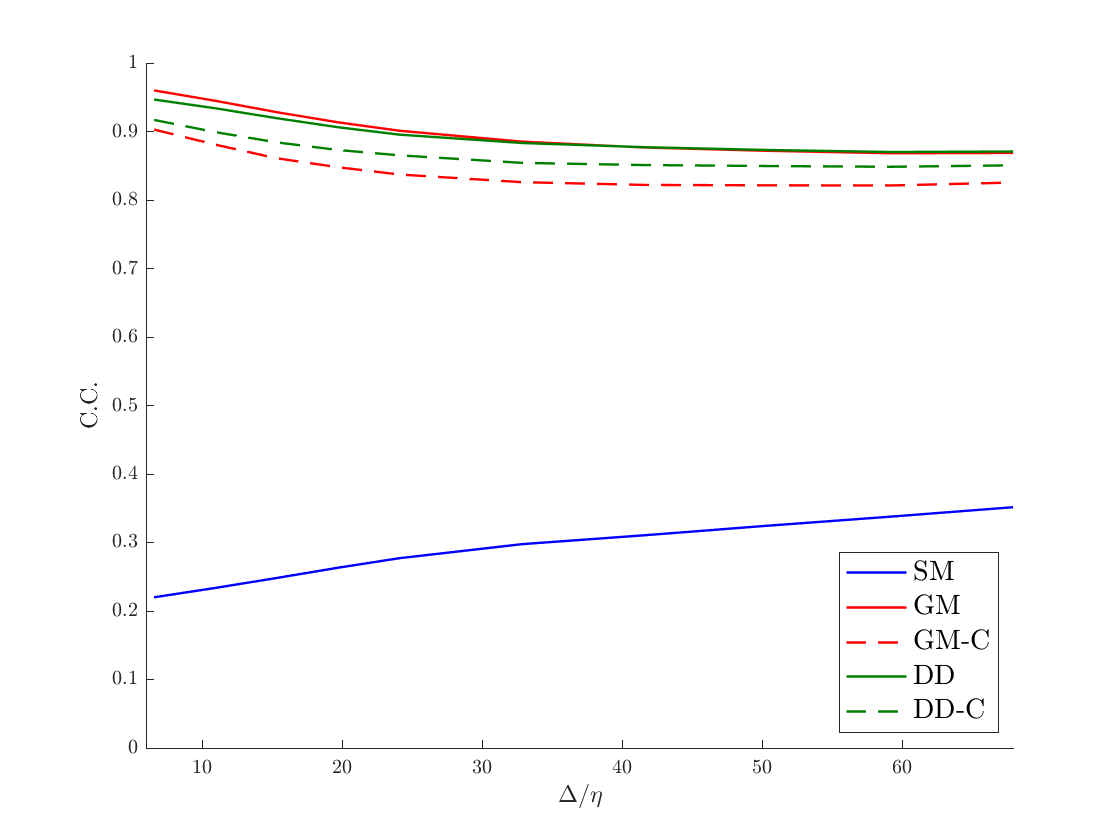}}
    \subfigure[\label{fig:REF_Apriori_Case2}]{\includegraphics[width=0.49\textwidth]{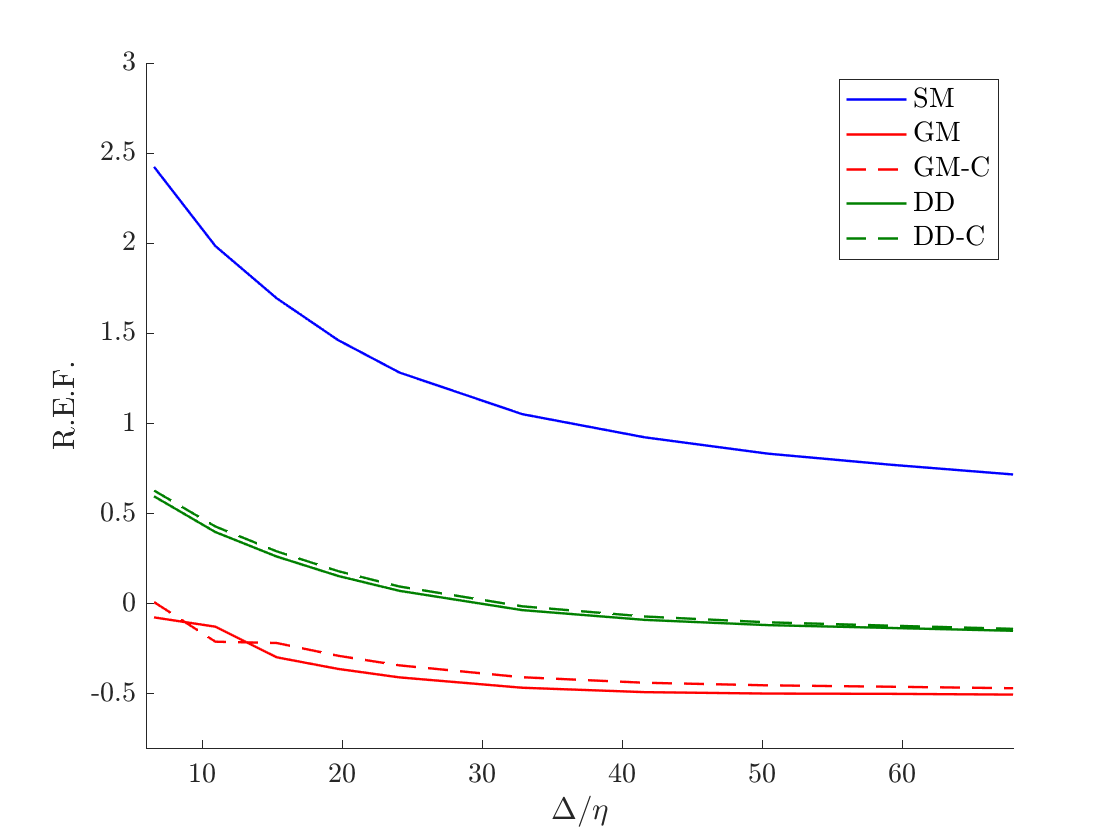}}
    \caption{\textit{A priori} results for Case 2: (a) Correlation Coefficient (C.C.) and (b) Relative error in mean energy flux (R.E.F).}
    \label{fig:Apriori_Case2}
\end{figure}

\figref{Apriori_Case2} shows C.C. and R.E.F. for the different SGS models under consideration. We observe that the model SGS stresses for both the data-driven and gradient models are highly correlated with the exact SGS stresses throughout the range of considered filter widths.  Clipping the model SGS stresses lowers the C.C. for both the data-driven and gradient models, though clipping affects the data-driven model less than the gradient model.  The mean SGS energy dissipation predicted by the gradient model is accurate for filter widths within the dissipation range, but the gradient model significantly underpredicts the mean SGS energy dissipation for filter widths in the inertial subrange.  The static Smagorinsky model significantly overpredicts the mean SGS energy dissipation throughout the range of considered filter widths.  The data-driven model slightly overpredicts the mean SGS energy dissipation for filter widths within the dissipation range, but it accurately predicts the mean SGS energy dissipation for filter widths within the inertial subrange.  This result is highly encouraging as we typically aim to filter in the inertial subrange in LES.  Moreover, viscous stresses dominate the SGS stresses when we filter in the dissipation range, so we anticipate overprediction of mean SGS energy dissipation for filter widths within the dissipation range to have a minimal effect on flow statistics such as energy spectra.

\begin{figure}[t!]
    \centering
    \subfigure[$\Delta \approx 7 \eta$]{\includegraphics[width=0.49\textwidth]{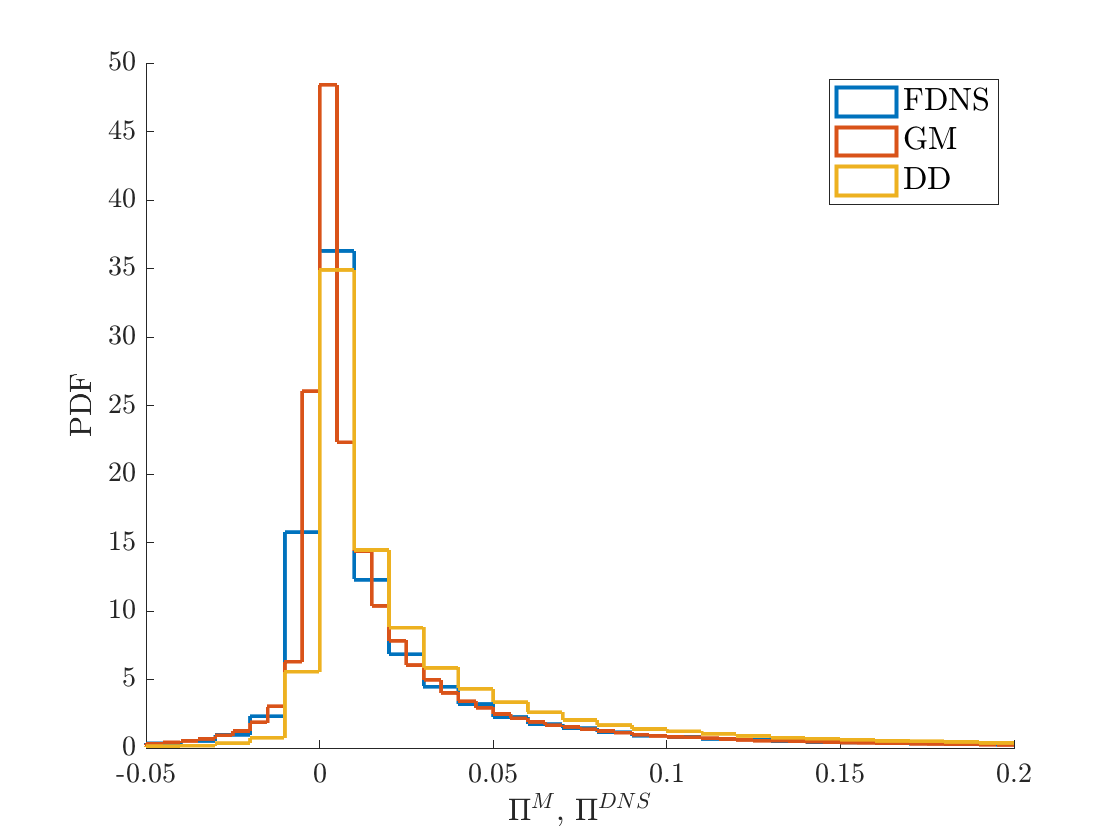}}
    \subfigure[$\Delta \approx 42 \eta$]{\includegraphics[width=0.49\textwidth]{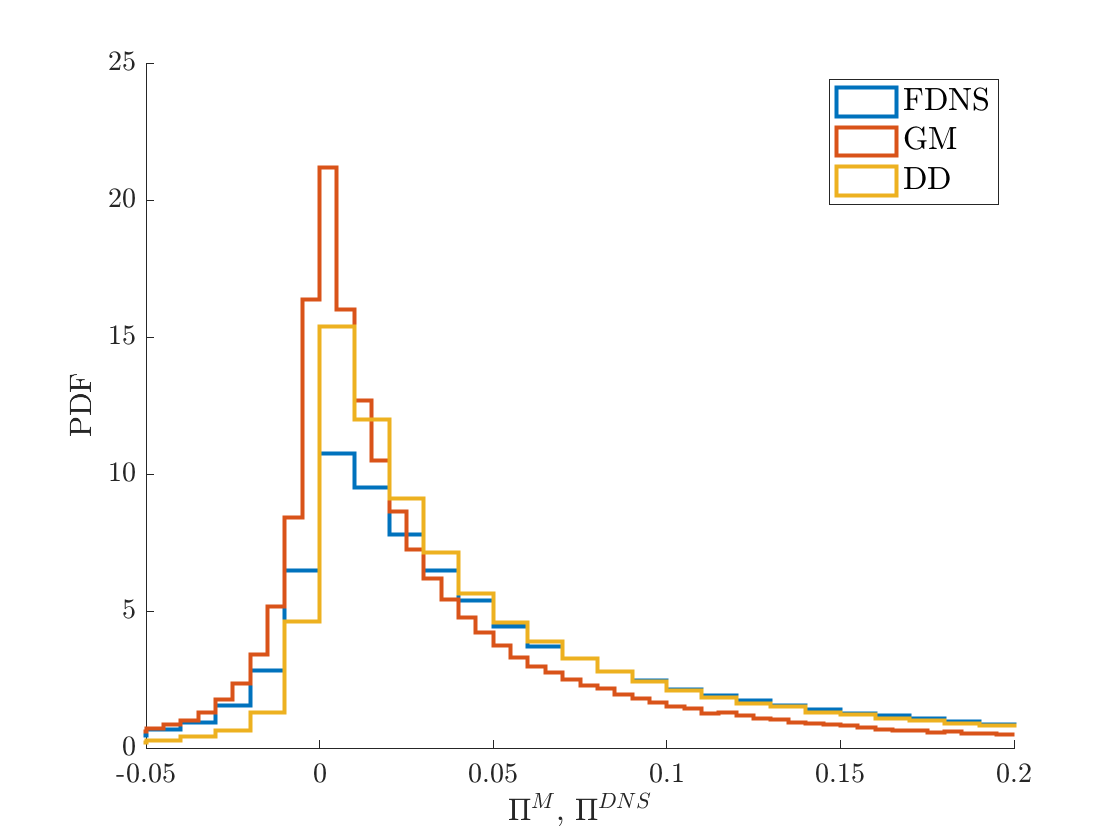}}
    \subfigure[$\Delta \approx 68 \eta$]{\includegraphics[width=0.49\textwidth]{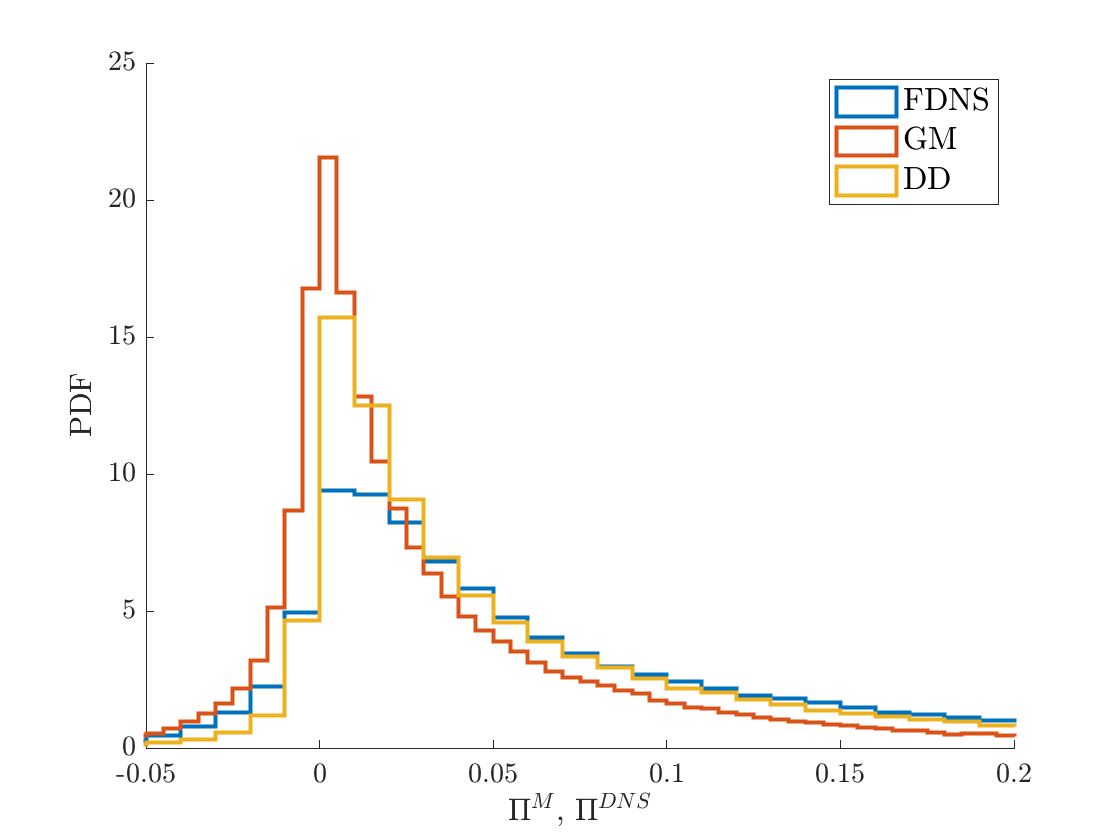}}
    \caption{\textit{A priori} results for Case 2: Probability density functions of model and exact SGS energy dissipation for three different filter widths.}
    \label{fig:PDF_dissip_apriori}
\end{figure}

To better understand the dissipative behavior of both the data-driven model and the gradient model without clipping, we can examine probability density functions (PDFs) of model SGS energy dissipation for these two models and compare these with PDFs of exact SGS dissipation.  Such PDFs are displayed in \figref{PDF_dissip_apriori} for three different filter widths.  The smallest filter width, $\Delta \approx 7 \eta$, is in the dissipation range, and we find that the PDF of model SGS energy dissipation closely matches the PDF of exact SGS dissipation for both models.  The other two filter widths, $\Delta \approx 42 \eta$ and $\Delta \approx 68 \eta$, lie within the inertial subrange, and we find that the PDFs of model SGS energy dissipation deviate from the PDF of exact SGS dissipation for both these filter widths.  However, while \figref{PDF_dissip_apriori} shows the gradient model predicts more large backscatter events than actually occur for $\Delta \approx 42 \eta$ and $\Delta \approx 68 \eta$, it also shows the data-driven model predicts less large backscatter events than actually occur.  This suggests clipping has a less pronounced impact on the data-driven model than the gradient model.  This is in agreement with what we observe in \figref{Apriori_Case2}.  Moreover, the data-driven model dissipation PDFs for $\Delta \approx 42 \eta$ and $\Delta \approx 68 \eta$ closely match those of exact SGS energy dissipation for forward scatter events, while the gradient model dissipation PDFs are below those of exact SGS energy dissipation for such events.  Consequently, the gradient model predicts less large forward scatter events than actually occur, and we might expect the gradient model to be underdissipative in \textit{a posteriori} tests for filter widths within the inertial subrange.  Our later numerical experiments confirm this is indeed the case.

We observe that the data-driven model C.C. and R.E.F. results in the inertial subrange are similar to those attained for Case 1. This is encouraging as we are evaluating the data-driven model using a validation dataset that is not correlated with the training dataset, and we are evaluating the data-driven model using several filter widths rather than just the single filter width used for model training. These observations suggest that the data-driven model is indeed predictive for inputs outside the training dataset, highlighting the fact that a limited amount of the data can be effectively used to develop an accurate data-driven SGS model using our new approach for data-driven SGS modeling.  It should further be highlighted that the data-driven model exhibits both good functional and structural performance even though it was only trained for good structural performance.  We have also investigated the use of other data slices and random data point selection for model training, testing, and validation.  This investigation has revealed trained model results are statistically unaffected by training dataset selection.  These results are not shown in this paper for brevity.

\subsection{\textit{A Posteriori} Testing of the Data-Driven Subgrid Stress \text{Model}}
\label{sec:Aposteriori}

\textit{A priori} tests alone are inadequate to assess the performance of an SGS model. As simulation time progresses, the dissipative characteristics of the model play an important role in the evolution of the turbulence statistics. \textit{A posteriori} tests involve performing a LES run and analyzing the evolved flow statistics. Such tests have shown that the unclipped gradient model exhibits stability issues that can be attributed to the underestimation of SGS energy dissipation \cite{Vreman1996, Vreman1997}. On the other hand, even though the SGS stresses predicted by the dynamic Smagorinsky model exhibit a lower correlation with filtered DNS data as compared with the gradient model, the dissipative nature of the dynamic Smagorinsky model results in much better predictive performance \cite{Vreman1997}. Therefore, the natural next step in model validation is to investigate the performance of the data-driven model through \textit{a posteriori} tests and compare it to other commonly used models, that is the dynamic Smagorinsky model and the gradient model. We investigate model performance for four different test cases: forced HIT at $Re_{\lambda} = 165$, forced HIT at $Re_{\lambda} = \infty$, decaying HIT, and 3-D Taylor-Green vortex flow at $Re = 1,600$. These four separate test cases allows us to investigate model performance for flow regimes that were not a part of the training set and thereby, allows us to comment on the generalization characteristics of the neural network-based data-driven model. Nondimensional simulations are conducted for the forced HIT and Taylor-Green vortex flow test cases, while dimensional simulations matching the experimental setup are conducted for the decaying HIT test case. \textit{A posteriori} simulations are carried out with the SUPG/PSPG/grad-div-stabilized finite element based solver \cite{Whiting2001} named PHASTA \cite{PHASTA}. In \textit{a posteriori} simulations, there is unavoidable interaction between numerical dissipation and model dissipation, but it has been shown that SUPG/PSPG/grad-div-stabilization is minimally dissipative \cite{tejada2005interaction}. Moreover, application of the SUPG/PSPG/grad-div-stabilized finite element method to the filtered Navier-Stokes equations may be viewed as a variational multiscale method wherein the influence of the numerically unresolved filtered scales (that is, the filtered scales that cannot be represented by the finite element discretization) on the numerically resolved filtered scales is explicitly modeled \cite{Bazilevs2007}. More details on the stabilization parameters employed in the studies reported on here can be found in \ref{sec:Appendix1}. The spatial discretization utilizes piecewise trilinear polynomials on a tensor-product mesh, whereas the temporal discretization is based on the generalized-$\alpha$ method \cite{Jansen2000}.  The density is chosen to be one for all simulations.

For the forced HIT and 3-D Taylor-Green vortex flow cases, a periodic box domain of side length $2\pi$ is employed. For the decaying HIT test case, the side length is approximately $11 M$ where $M = 0.00508$ m is the lattice length for the grid that is used for generating turbulence in validation experiments \cite{Comte1971}.  Three hexahedral element topology based grid resolutions, meshes of $32^{3}$ elements, $64^{3}$ elements, and $128^{3}$ elements, respectively, are examined for the forced HIT and decaying HIT cases. For Taylor-Green Vortex flow, we consider meshes of $64^{3}$ elements and $128^{3}$ elements. The filter width input for each of the considered models is based on the relation by Deardoff \cite{Deardorff1970}, $\Delta = (\Delta_x \Delta_y \Delta_z)^{1/3}$ where $\Delta_x$, $\Delta_y$, and $\Delta_z$ are the side-lengths of mesh elements in the $x, y,$ and $z$ directions respectively. The filter width ratio used for explicit filtering in the dynamic Smagorinsky model is taken to be $\sqrt{3}$ following \cite{Tejada2003}. Finally, we do not average the Smagorinsky constant in homogeneous directions in our \textit{a posteriori} tests, though local clipping of the Smagorinsky constant is performed.

\subsubsection{Forced HIT at $Re_{\lambda} = 165$}

As the data-driven model is trained on forced HIT data, it is natural to conduct \textit{a posteriori} tests for this same flow case albeit at a different Reynolds number. Thus, for our first test case, we consider forced HIT at $Re_{\lambda} = 165$ as we also have DNS simulation data to compare to for this Reynolds number \cite{Langford1999}.  In the LES runs, we attempt to solve for this case using a forcing that has same power input as the forcing in the DNS. The forcing is based on the formulation employed in \cite{Bazilevs2007}. The forcing is given as

\begin{equation}
    \bm{f} (\bm{x}) = \sum_{\substack{\bm{k} \\ \vert k_i \vert < k_f \\ \bm{k} \neq 0}} \frac{P_{in}}{2 E_{kf}} \hat{\bm{u}}_{\bm{k}} \exp{(\iota \bm{k}\cdot\bm{x})}, \quad i = 1,2,3,
\end{equation}

\begin{equation}
    E_{kf} = \frac{1}{2} \sum_{\substack{\bm{k} \\ \vert k_i \vert < k_f \\ \bm{k} \neq 0}} \hat{\bm{u}}_{\bm{k}} \cdot \hat{\bm{u}}_{\bm{k}}  , \quad i = 1,2,3,
\end{equation}

\begin{equation}
    \hat{\bm{u}}_{\bm{k}} = \frac{1}{\vert \Omega \vert} \int_{\Omega} \bm{u}^h (\bm{x}) \exp{(-\iota \bm{k}\cdot\bm{x})},
\end{equation}

\noindent where $P_{in} = 62.84$ is the input power to the forcing and $k_f = 3$ \cite{Bazilevs2007}.  This forcing is slightly different than that employed in the DNS simulation, so the LES energy spectra are found to be slightly different than the DNS spectra results at the lowest wavenumbers.  However, the forcing yields good agreement between the LES and DNS spectra at intermediate and high wavenumbers.  The required value of $Re_{\lambda}$ is achieved by setting viscosity to $1/150$. Each simulation is initialized by subsampling the velocity and pressure fields corresponding to the first time step of the forced HIT simulation at $Re_{\lambda} = 418$ in the JHTDB \cite{Li2008}. Each simulation is run until a statistical convergence of energy spectra is observed. In the legend, the DNS and filtered DNS data are denoted as DNS and FDNS respectively. The filtered DNS energy spectra are computed by applying a differential filter approximating the tensor-product box filter \cite{Germano1986, Bull2016} to the DNS spectra as follows:

\begin{equation}
    \bar{E}(k) = \frac{E(k)}{(1 + \alpha^2 k^2)^2}
    \label{eq:diff_filter}
\end{equation}

\noindent where $E(k)$ is the DNS energy spectrum function, $\bar{E}(k)$ is the filtered DNS energy spectrum function, and $\alpha^2 = \Delta^2/40$.

\begin{figure}
    \centering
    \subfigure[No Model\label{fig:Re165_NM}]{\includegraphics[width=0.49\textwidth]{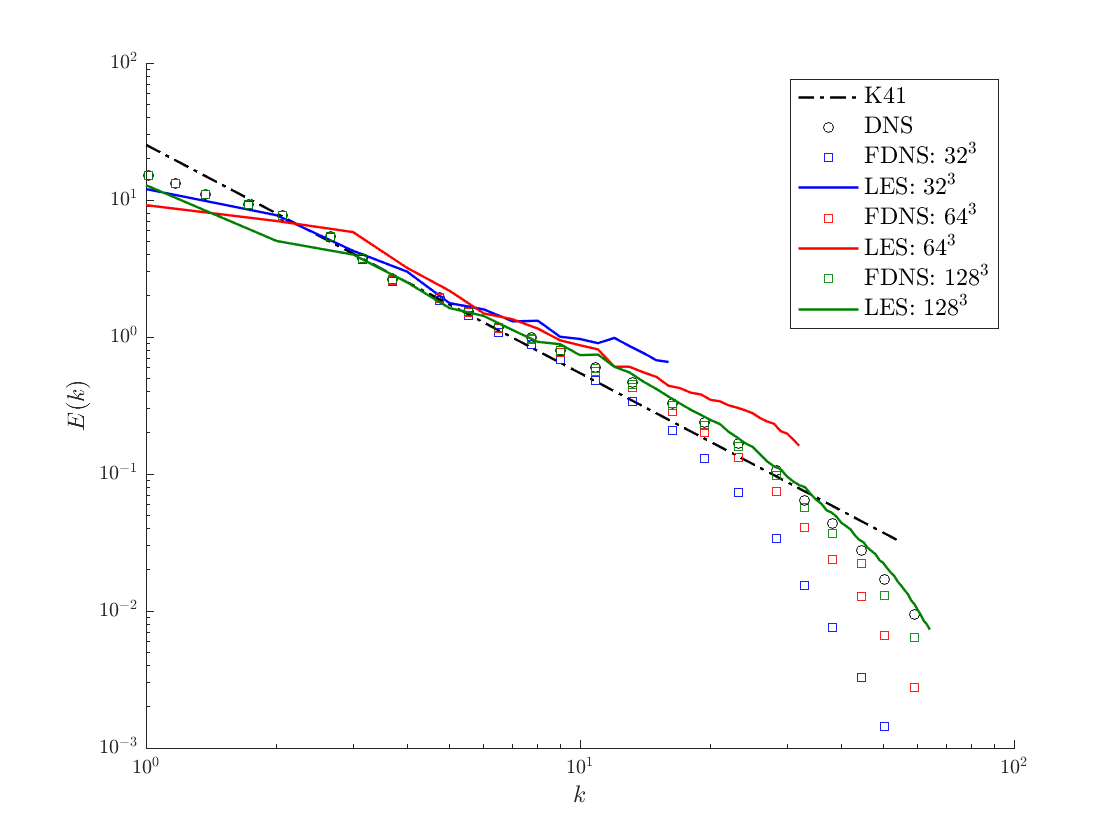}}
    \subfigure[Dynamic Smagorinsky Model\label{fig:Re165_DS}]{\includegraphics[width=0.49\textwidth]{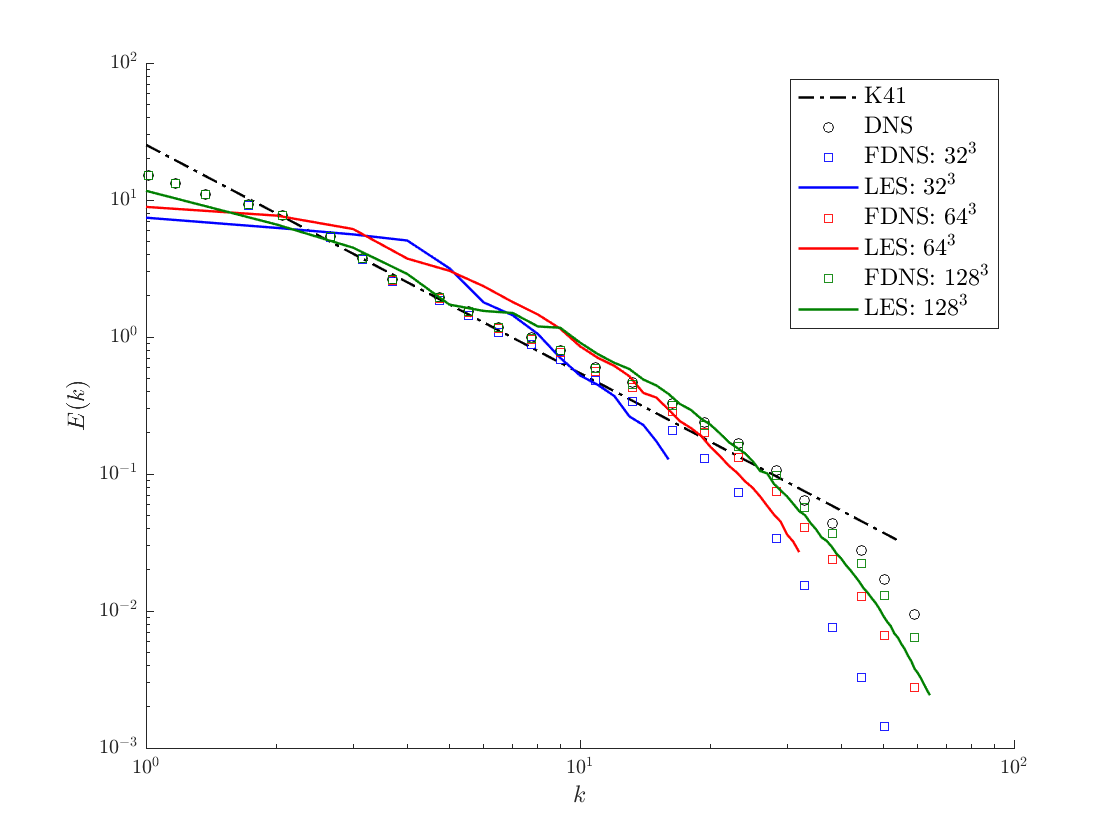}}
    \subfigure[Gradient Model\label{fig:Re165_GM}]{\includegraphics[width=0.49\textwidth]{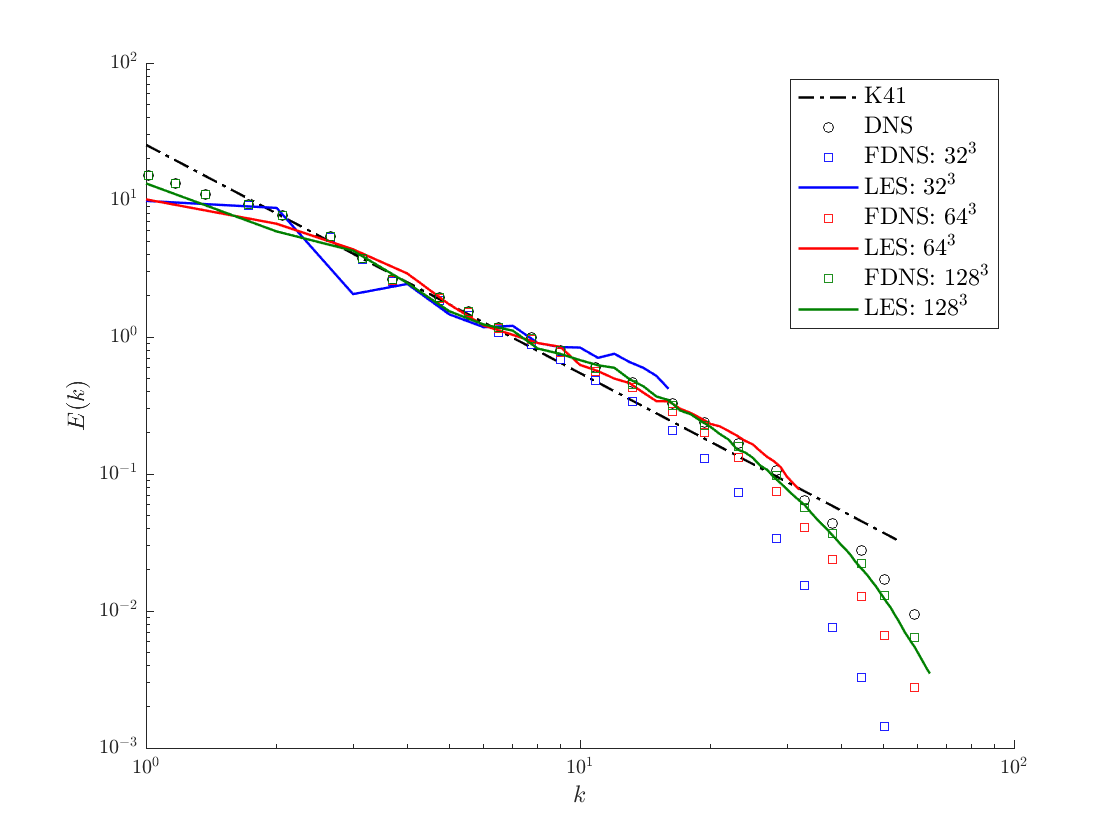}}
    \subfigure[Gradient Model with Clipping\label{fig:Re165_GM_C}]{\includegraphics[width=0.49\textwidth]{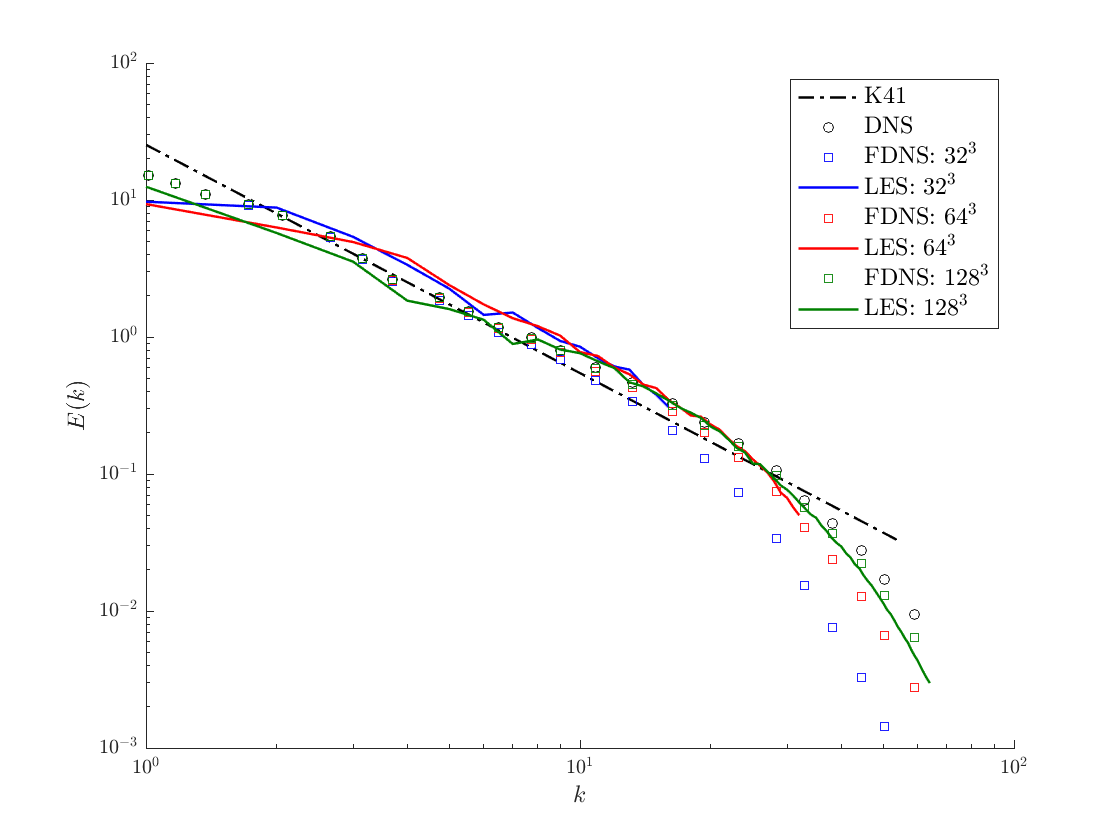}}
    \subfigure[Data-Driven Model\label{fig:Re165_DD}]{\includegraphics[width=0.49\textwidth]{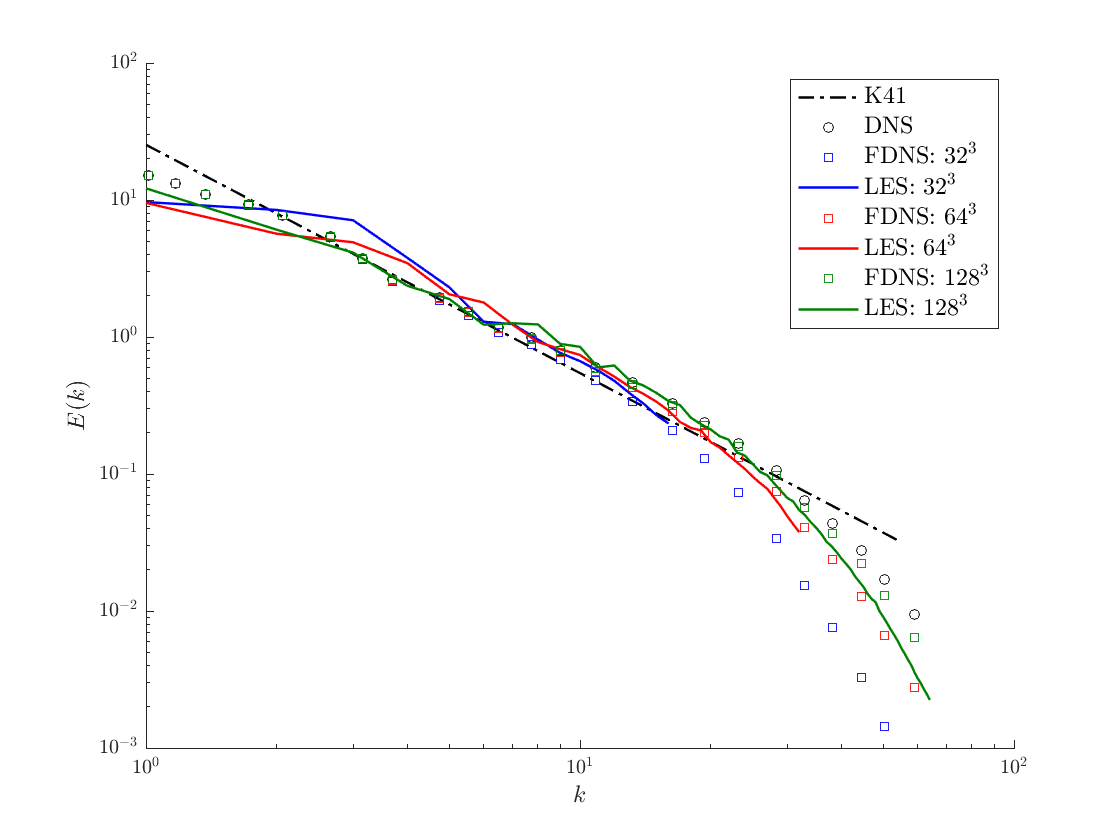}}
    \subfigure[Data-Driven Model with Clipping \label{fig:Re165_DD_C}]{\includegraphics[width=0.49\textwidth]{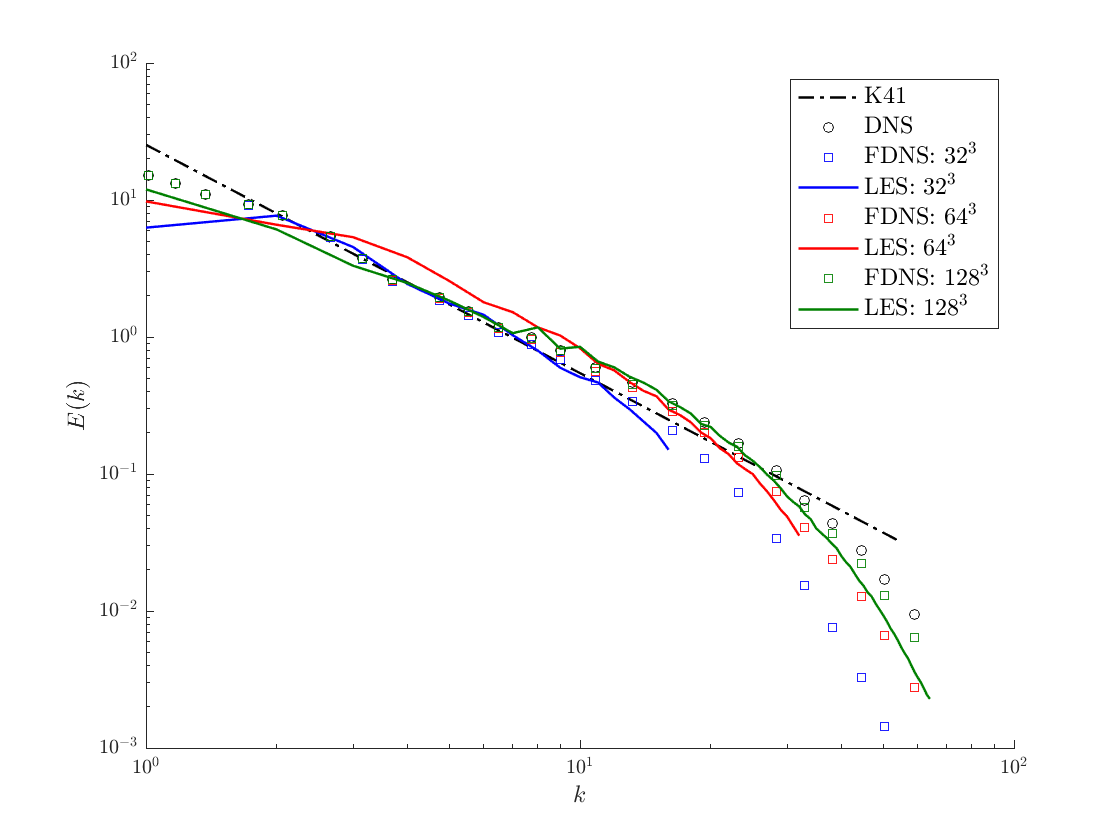}}    
    \caption{Energy spectra for forced HIT at $Re_{\lambda} = 165$ for (a) No Model, (b) Dynamic Smagorinsky Model, (c) Gradient Model, (d) Gradient Model with Clipping, (e) Data-Driven Model, and (f) Data-Driven Model with Clipping.}
    \label{fig:Re165}
\end{figure}

Time-instantaneous energy spectra results for all considered models and meshes are shown in \figref{Re165}. We observe that using no explicit LES model gives an accurate prediction of the entire energy spectra for the $128^3$ element mesh. This is due to the fact the flow is nearly resolved at this mesh resolution. However, using no model leads to energy pile up at higher wavenumbers for the $32^3$ and $64^3$ element meshes. Utilization of any of the considered explicit LES model considerably improves the prediction of the energy spectra for coarser LES runs. As expected, the dynamic Smagorinsky model does not lead to a pile-up of energy. However, it slightly overpredicts energy at intermediate wavenumbers that lie in the inertial subrange for the $32^3$ and $64^3$ element meshes. The gradient model spectra results are fairly close to the filtered DNS spectra results for each of the considered meshes, however, for the $32^3$ and $64^3$ element meshes, we observe a slight pile-up of energy at the highest wavenumbers. On the other hand, the clipped gradient model spectra do not exhibit pile-up and in fact are in better agreement with the filtered DNS results for the $32^3$ and $64^3$ element meshes.  Lastly, we observe that the results obtained with both the unclipped and clipped data-driven models are very close to the filtered DNS results. In fact, the data-driven model without clipping gives the best agreement with filtered DNS out of all considered models for the $32^3$ and $64^3$ element meshes. Remarkably, the data-driven model without clipping does not exhibit pile-up for any of the considered meshes.  This is encouraging as this is consistent with the \textit{a priori} tests in which both the structural and functional performance of the data-driven model was found to be superior to those of the other models. Even though the Taylor Reynolds number for this case is different from that of the training dataset, we observe that the data-driven model still gives the best performance of all the considered models for the $32^3$ and $64^3$ element meshes. However, it should be remarked that the data-driven model is slightly overdissipative for the $128^3$ element mesh, and it is less accurate than the gradient model for this case.  This is due to the fact that the flow is nearly resolved at this mesh resolution, and the data-driven model was trained using a filter width corresponding to the inertial subrange rather than the dissipation range.  Improved results for this mesh resolution can be attained by including viscosity as a model input as discussed in Section \ref{sec:conclusions}.

\subsubsection{Forced HIT at $Re_{\lambda} = \infty$}

We next examine the performance of the data-driven model in the asymptotic limit of $Re_{\lambda} = \infty$. As most industrial flow cases are characterized by high Reynolds number, this test case allows us to comment on the suitability of the considered SGS models for such cases. An infinite Reynolds number is approximately achieved by setting the viscosity to an extremely small value, $10^{-12}$. The forcing and initial conditions are taken to be the same as the last case. There is no reference DNS simulation for such a high Reynolds number. However, as the Reynolds number is so high, the time-instantaneous energy spectra attained by LES runs are compared to the Kolmogorov spectrum,

\begin{equation}
    E(k) = C \epsilon^{2/3} k^{-5/3}.
\end{equation}

\noindent We set $C = 1.6$ in accordance with previous experimental and numerical studies \cite{Pope2000}. As we are considering a statistically stationary equilibrium state, the dissipation, $\epsilon$, is equal to the power input of the forcing. Each simulation is run until statistical convergence of energy spectra is observed.

\begin{figure}
    \centering
    \subfigure[No Model\label{fig:ReInf_NM}]{\includegraphics[width=0.49\textwidth]{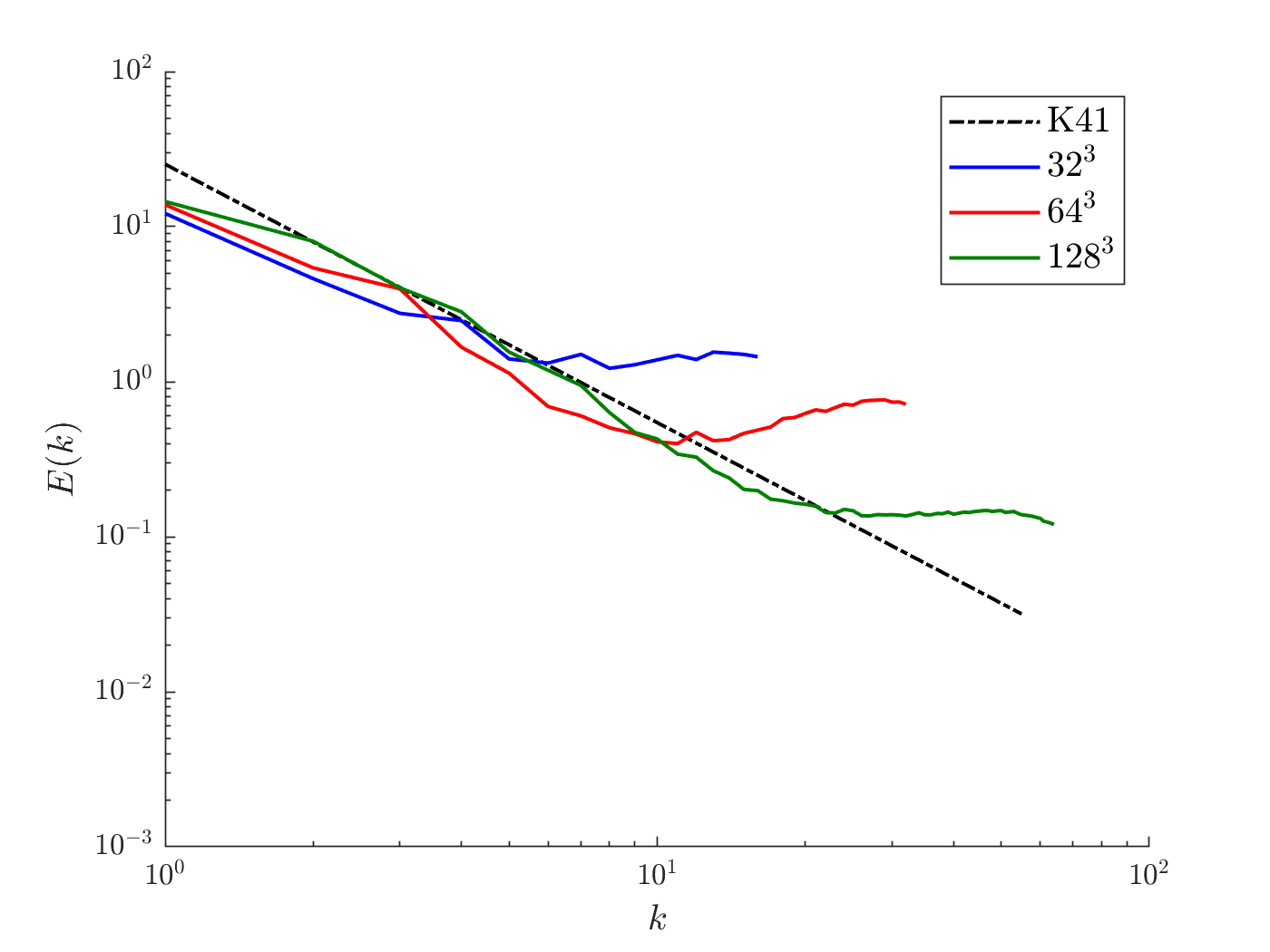}}
    \subfigure[Dynamic Smagorinsky Model\label{fig:ReInf_DS}]{\includegraphics[width=0.49\textwidth]{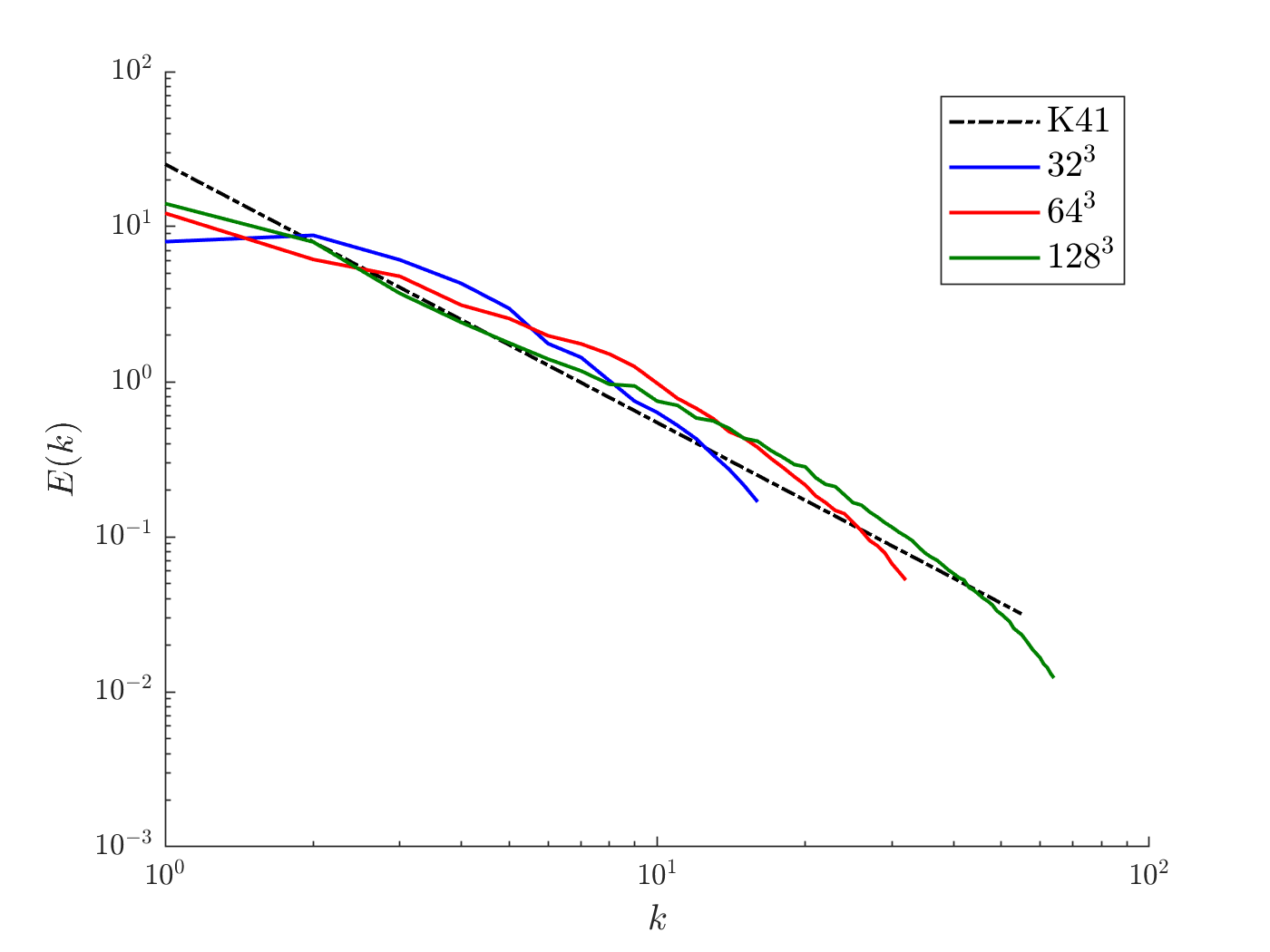}}
    \subfigure[Gradient Model \label{fig:ReInf_GM}]{\includegraphics[width=0.49\textwidth]{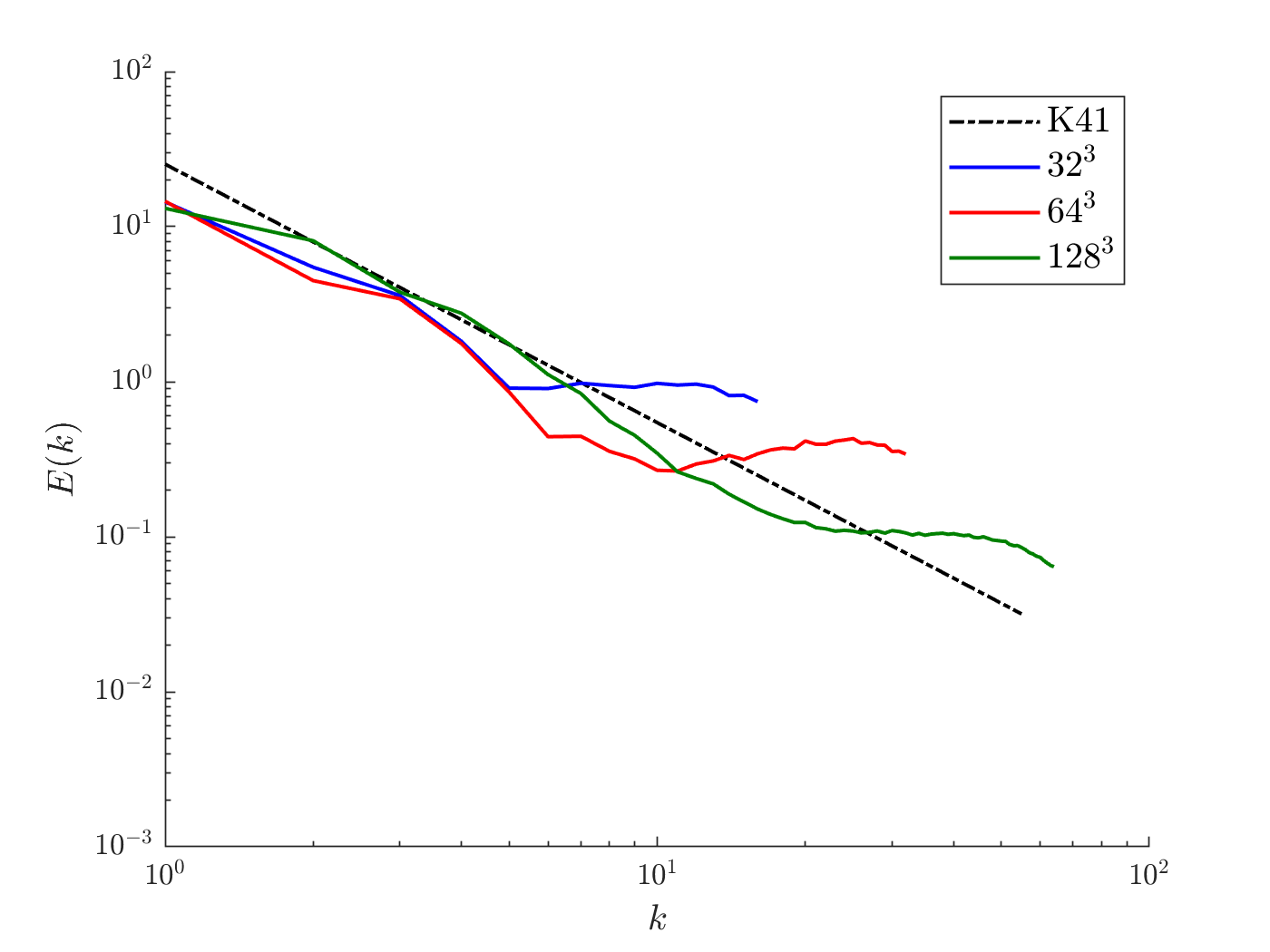}}
    \subfigure[Gradient Model with Clipping\label{fig:ReInf_GM_C}]{\includegraphics[width=0.49\textwidth]{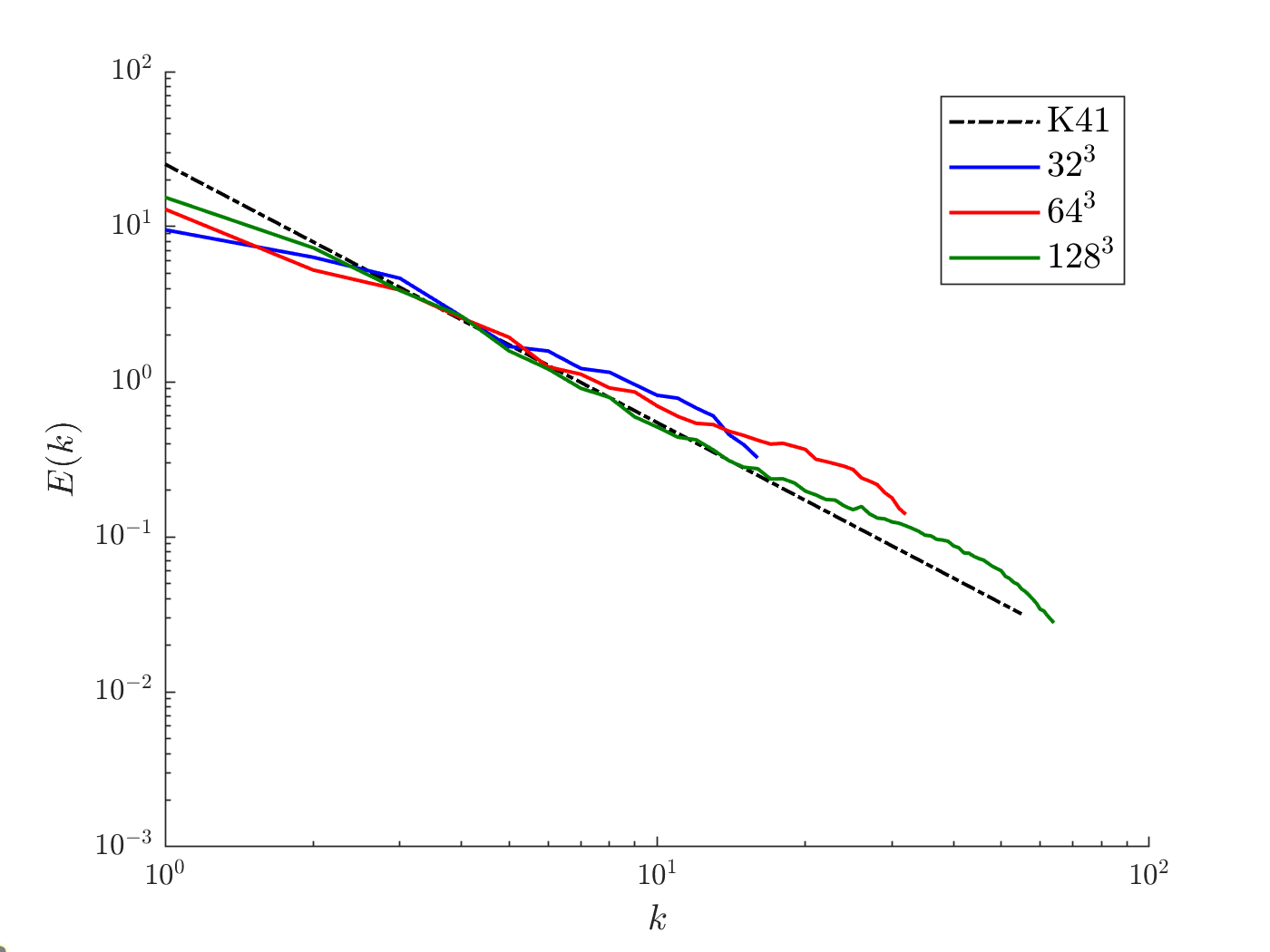}}
    \subfigure[Data-Driven Model\label{fig:ReInf_DD}]{\includegraphics[width=0.49\textwidth]{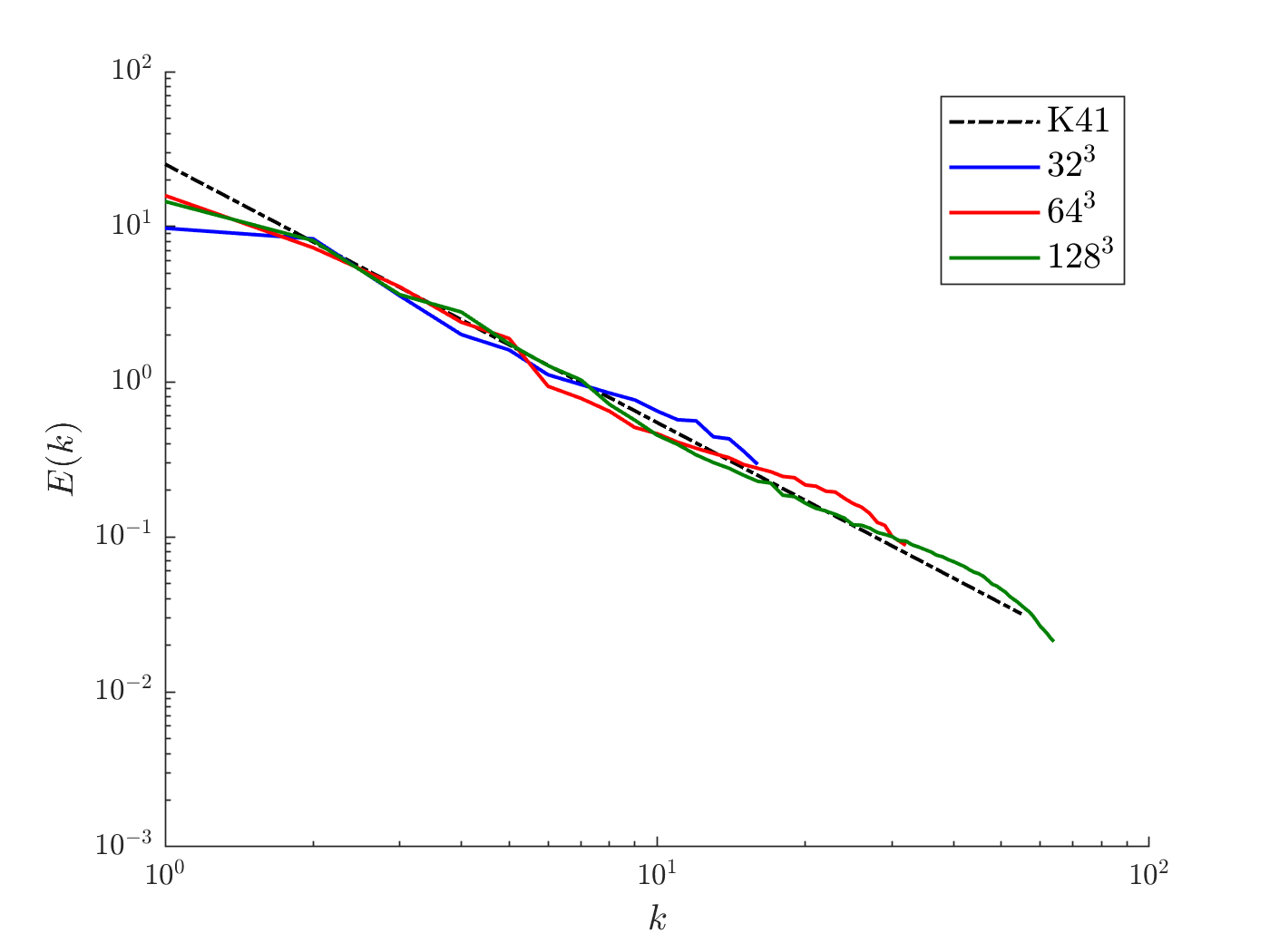}}
    \subfigure[Data-Driven Model with Clipping\label{fig:ReInf_DD_C}]{\includegraphics[width=0.49\textwidth]{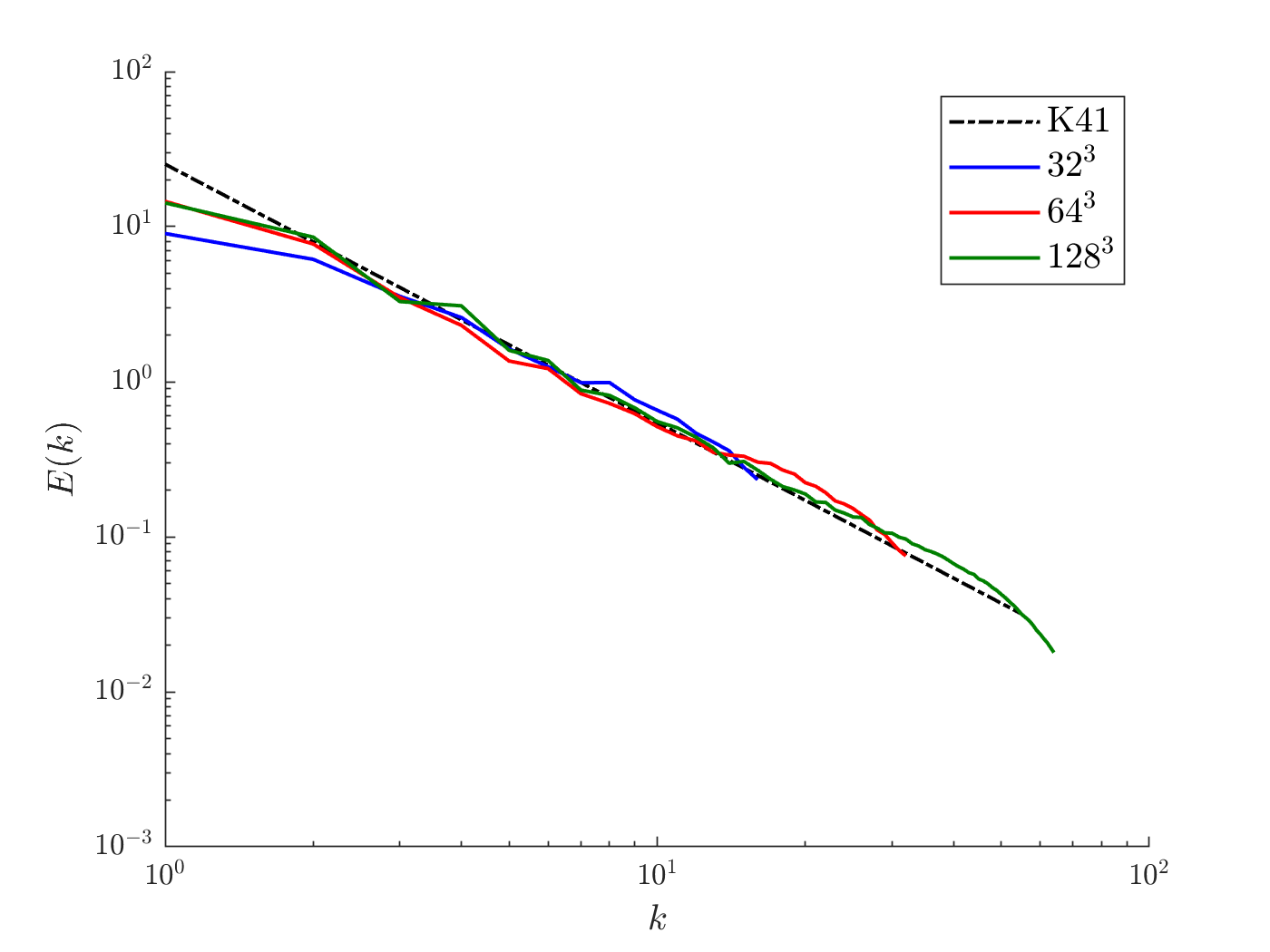}}    
    \caption{Energy spectra for forced HIT at $Re_{\lambda} = \infty$ for (a) No Model, (b) Dynamic Smagorinsky Model, (c) Gradient Model, (d) Gradient Model with Clipping, (e) Data-Driven Model, and (f) Data-Driven Model with Clipping.}
    \label{fig:ReInf}
\end{figure}

In \figref{ReInf}, we compare the time-instantaneous energy spectra obtained using several SGS models. We observe that the use of no explicit LES model leads to a massive pile-up of energy at higher wavenumbers for each of the considered meshes. The use of any of the considered SGS models improves the results. The use of the dynamic Smagorinsky model does not result in energy pile-up but it again leads to overprediction in energy at the intermediate wavenumbers. The gradient model exhibits energy pile-up at the higher wavenumbers for each of the considered meshes, and its accuracy at intermediate wavenumbers is also poor. Clipping significantly improves the results for the gradient model at intermediate wavenumbers, though there is still a slight pile-up of energy at the highest wavenumbers. Lastly, we observe that the data-driven model either with or without clipping gives significantly better results than all the other models. The predicted energy spectra at intermediate wavenumbers agree with the Kolmogorov spectrum and the energy pile up at higher wavenumbers is insignificant.

It is important to highlight the role of clipping for the stability of the structural SGS models. Clipping is shown to be essential for the gradient model as it significantly improves the energy spectra prediction. However, the energy spectra results shown here suggest that the data-driven model performs well even without clipping. This observation is in agreement with results in Section \ref{sec:apriori_case2} where we showed that the data-driven model predicts less large backscatter events than actually occur and the gradient model predicts more large backscatter events.  Thus clipping has a much more significant impact on the gradient model than the data-driven model.

The observations from this test case suggest the data-driven model is capable of yielding high quality results even for flow cases for which it was not trained on as the model was trained using forced HIT data for a much lower Reynolds number than the one considered here. In our experience, training an accurate and generalizable data-driven model using a model form without embedded invariance properties requires a large amount of training data and a complex neural network architecture. As DNS training data is not available for high Reynolds number flows, the resulting model would still likely not perform well for the case considered here. While classical SGS models such as the dynamic Smagorinsky model have previously demonstrated an ability to yield accurate flow predictions at the asymptotic limit of $Re_{\lambda} = \infty$, to the best of our knowledge, this has not been previously shown for a data-driven SGS model.

\subsubsection{Decaying HIT}

\begin{figure}[t!]
    \centering
    \includegraphics[width=0.6\textwidth]{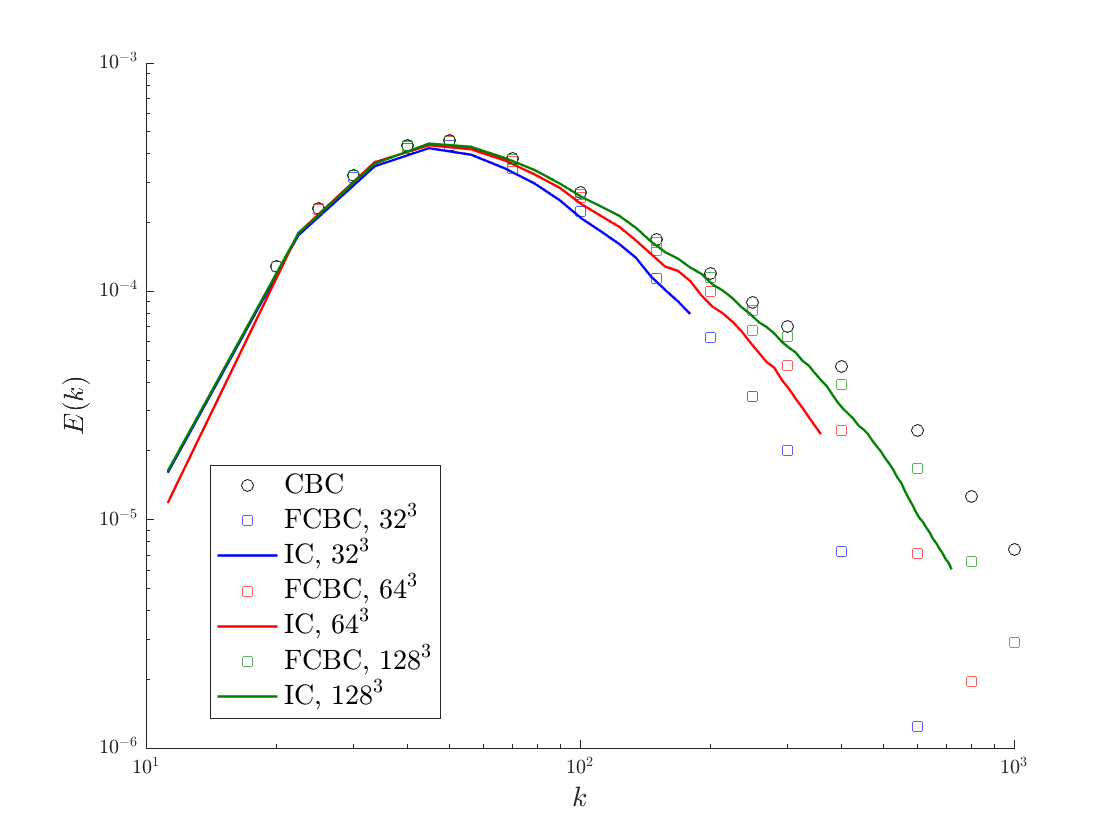}
    \caption{Initial condition for decaying HIT \textit{a posteriori} tests. F-CBC denotes the filtered experimental energy spectra.}
    \label{fig:DHIT_IC}
\end{figure}

Up until now, we have validated the data-driven model using \textit{a posteriori} tests involving flow physics similar to that it was trained on. In particular, the data-driven model was both trained and validated using statistically stationary turbulent flows. As a next step, we investigate the performance of the data-driven model using a turbulent flow case that is not statistically stationary - decaying HIT. Physically, decaying HIT models the decay of turbulence in the absence of mean shear, a phenomenon which is quite common in practice. Furthermore, there exists experimental data to validate our data-driven model for this flow case. In particular, the seminal experimental work reported in \cite{Comte1971} can be used as a reference to assess the quality of \textit{a posteriori} simulation results attained using the data-driven model. These experiments were carried out at a grid mesh Reynolds number, $U_0 M/ \nu$, of 34,000, where $M = 0.00508$ m is the lattice length of the grid and $U_0 = 10$ m/s is the speed on incoming airflow. Energy spectra are provided at three non-dimensional timestamps in \cite{Comte1971}, $t^* = 42, \;98, \; \text{and} \;171 $ where $t^* = t U_0/M$. To replicate their energy spectra, accurate initial conditions are required. In the literature, initial conditions are often synthetically generated for $t^* = 42$ and propagated through the filtered Navier-Stokes equation to compare LES energy spectra with experimental data at $t^* = 98$ and $t^* = 171$. Here, we generate initial conditions at $t^* = 42$ using the methodology proposed in \cite{Kang2003, Tejada2002}.  For the initial velocity field, Fourier coefficient amplitudes are chosen to meet a target filtered energy spectrum at $t^* = 42$ while phases are randomly generated.  To compute a target filtered energy spectra at $t^* = 42$, we apply a differential filter approximating the tensor-product box filter (\eref{diff_filter}) to the unfiltered energy spectra at $t^* = 42$. Then, using this initial condition, a preliminary LES simulation is performed for each SGS model until $t^* = 98$. It is expected that propagation of initial flow-field results in partial physical correlation among the phases \cite{Tejada2002}. However, as the amplitudes of the velocity field Fourier coefficients decay during this propagation, we rescale the Fourier coefficient amplitudes to again match the target filtered energy spectra at $t^* = 42$.  This new flow field is then used as the initial condition for \textit{a posteriori} tests. The energy spectra of the synthetic turbulent initial conditions for the data-driven model are shown in \figref{DHIT_IC} and compared to filtered experimental spectra. Different initial conditions are produced for the other models considered here but they are characterized by the same energy spectra. In the legend, CBC indicates the energy spectra obtained from experiments \citep{Comte1971} and FCBC indicates the filtered energy spectra obtained by applying the differential filter (\eref{diff_filter}) to the experimental energy spectra.

\begin{figure}
    \centering
    \subfigure[No Model\label{fig:DHIT_NM_98}]{\includegraphics[width=0.49\textwidth]{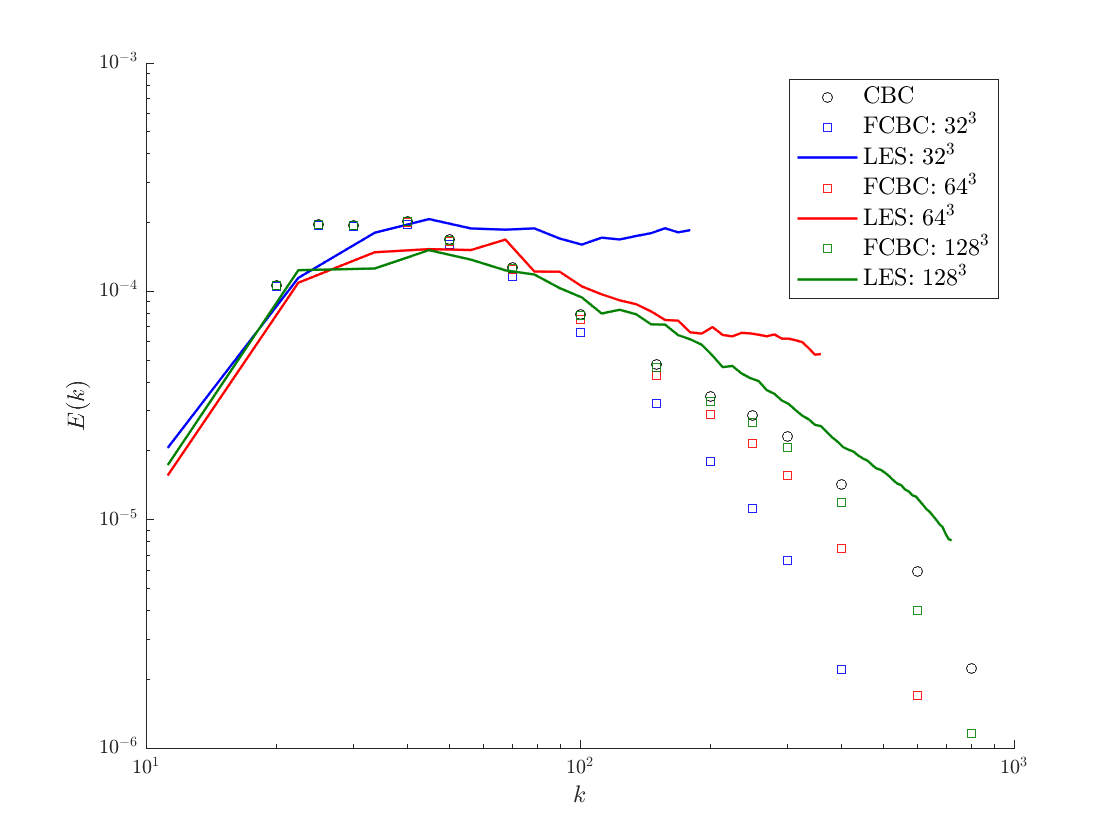}}
    \subfigure[Dynamic Smagorinsky Model\label{fig:DHIT_DS_98}]{\includegraphics[width=0.49\textwidth]{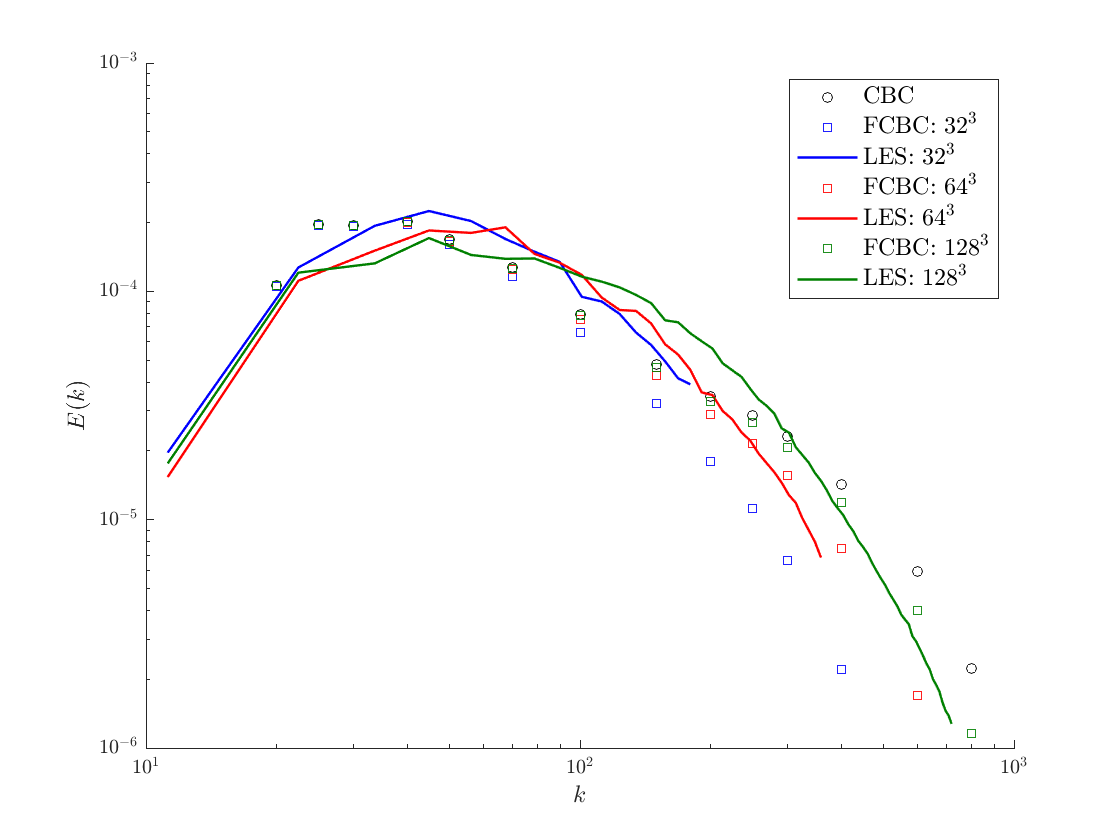}}
    \subfigure[Gradient Model\label{fig:DHIT_GM_98}]{\includegraphics[width=0.49\textwidth]{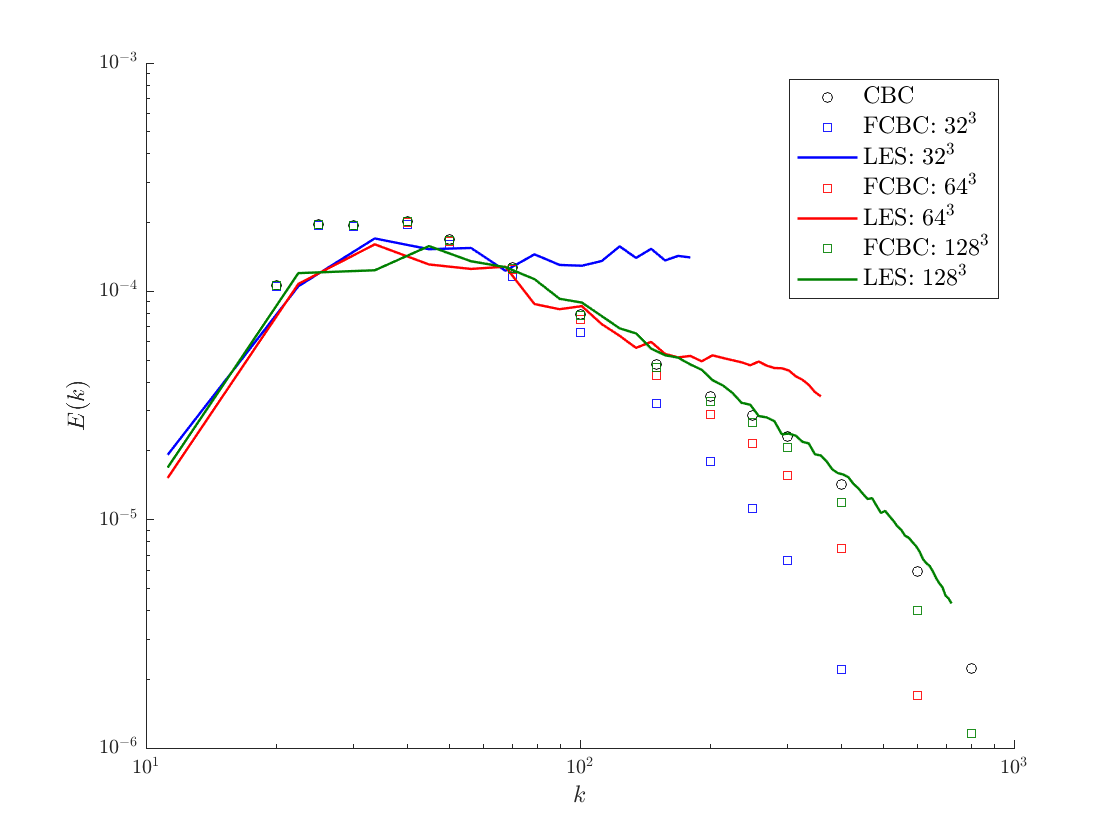}}
    \subfigure[Gradient Model with Clipping\label{fig:DHIT_GM_98_C}]{\includegraphics[width=0.49\textwidth]{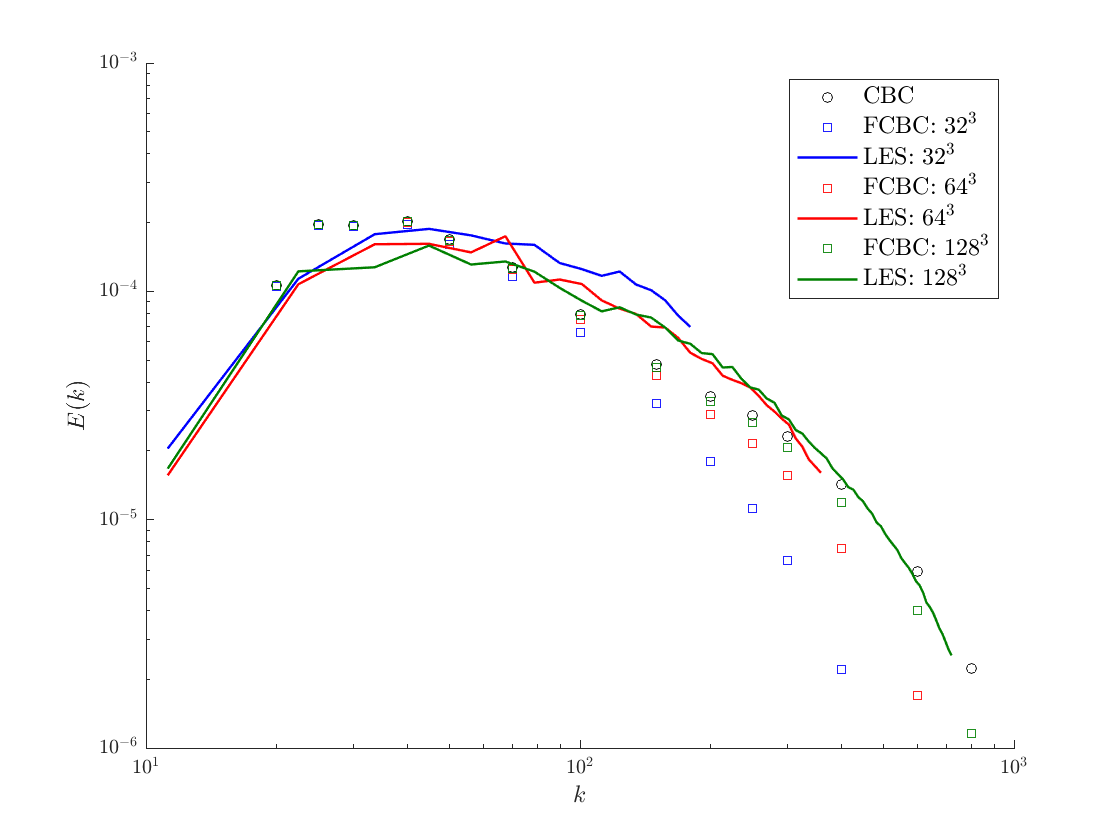}}    
    \subfigure[Data-Driven Model\label{fig:DHIT_DD_98}]{\includegraphics[width=0.49\textwidth]{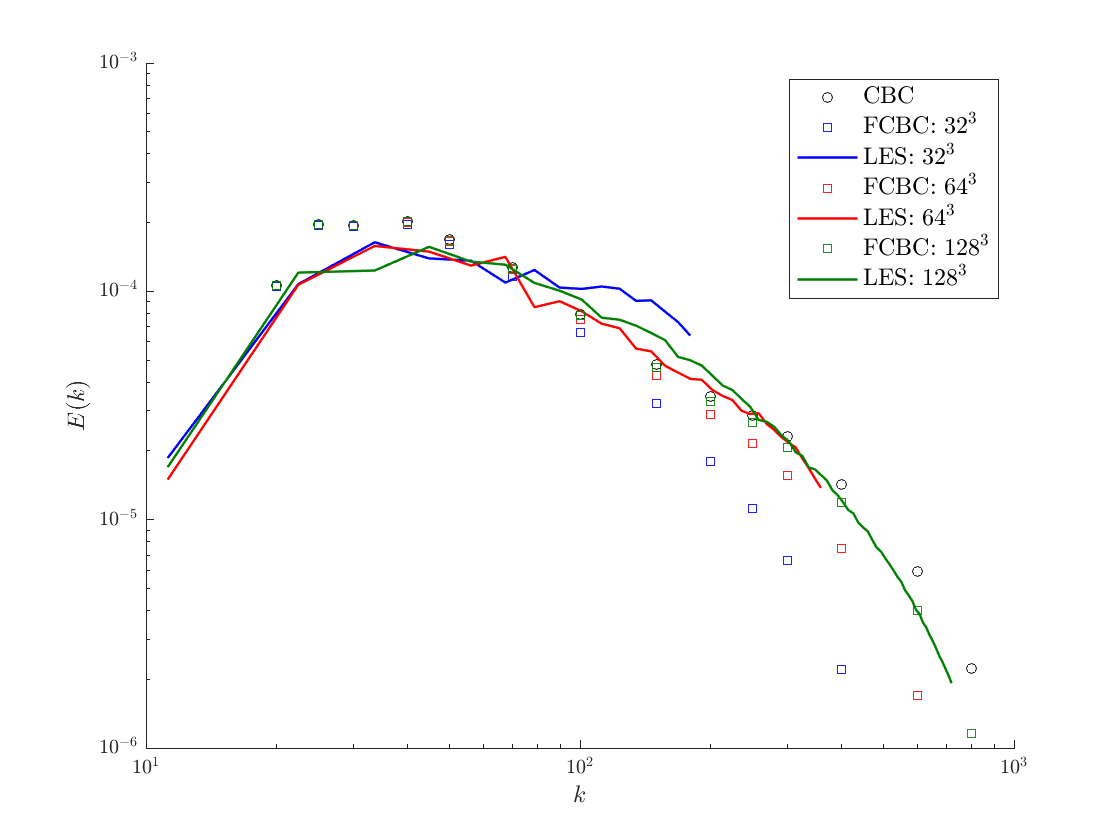}}
    \subfigure[Data-Driven Model with Clipping\label{fig:DHIT_DD_98_C}]{\includegraphics[width=0.49\textwidth]{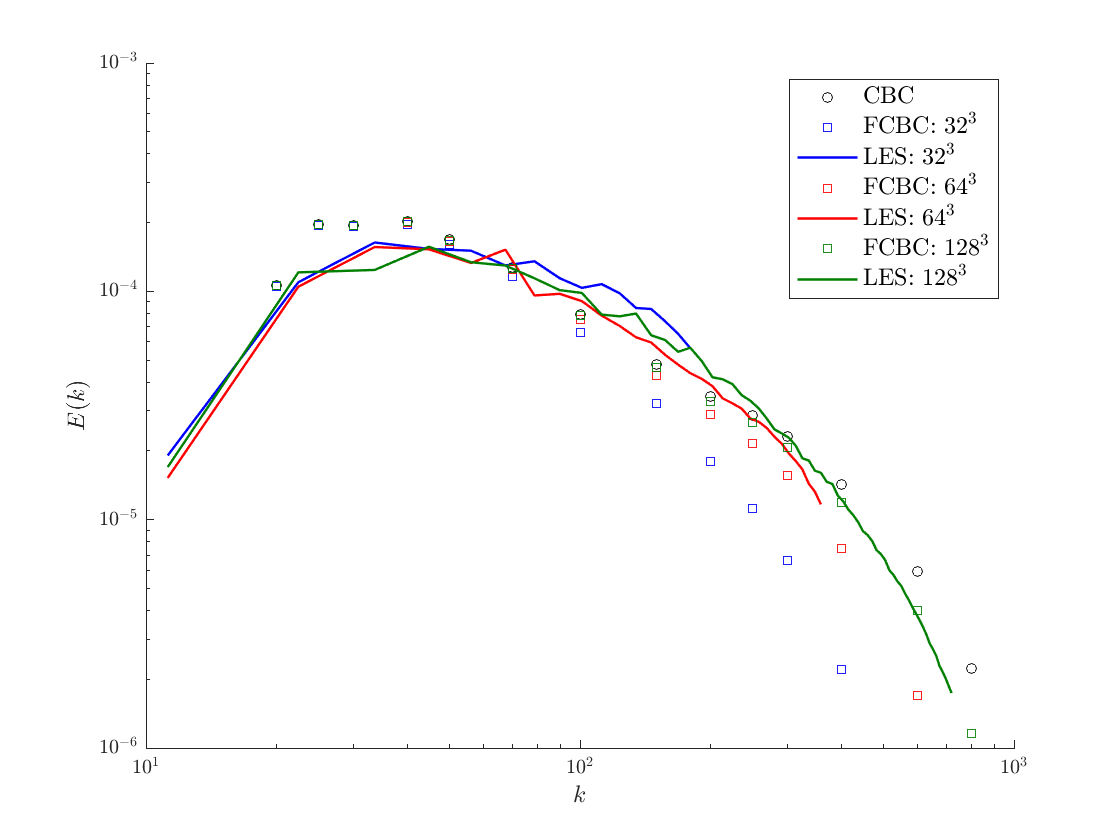}}    
    \caption{Energy spectra for decaying HIT at $t^* = 98$ for (a) No Model, (b) Dynamic Smagorinsky Model, (c) Gradient Model, (d) Gradient Model with Clipping, (e) Data-Driven Model and (f) Data-Driven Model with Clipping.}
    \label{fig:DHIT_98}
\end{figure}

\begin{figure}
    \centering
    \subfigure[No Model\label{fig:DHIT_NM_171}]{\includegraphics[width=0.49\textwidth]{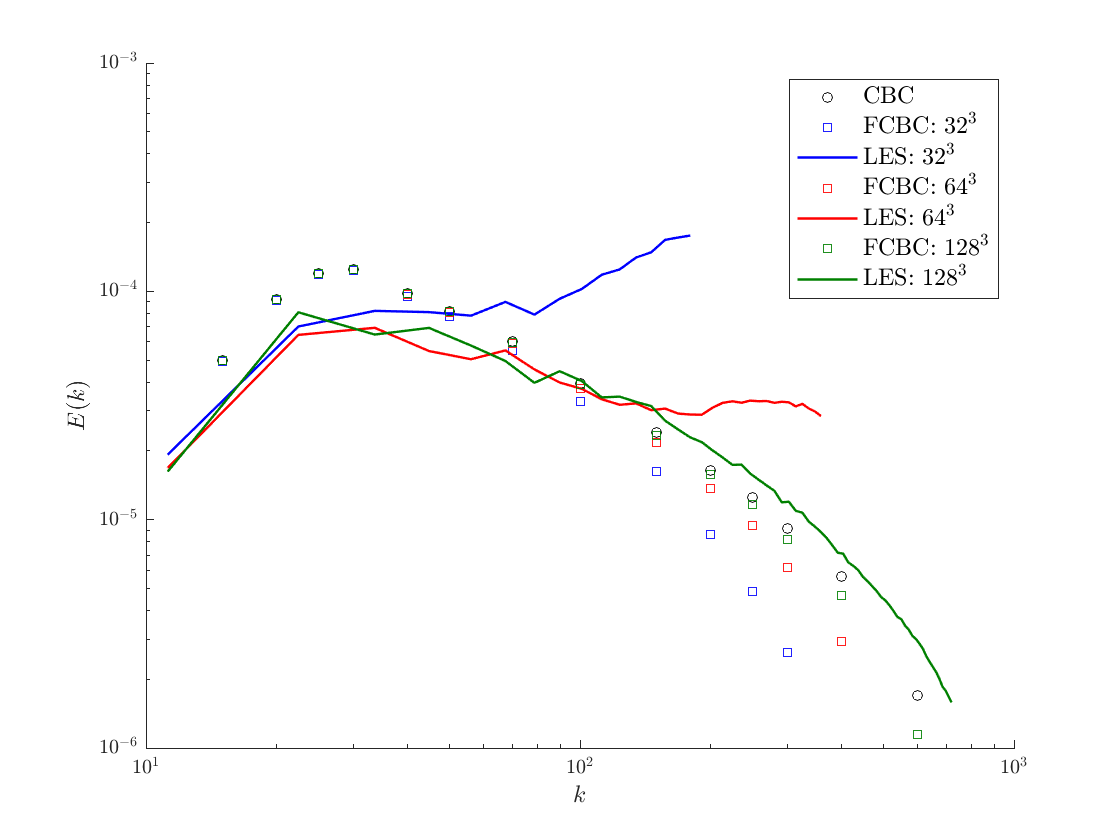}}
    \subfigure[Dynamic Smagorinsky Model\label{fig:DHIT_DS_171}]{\includegraphics[width=0.49\textwidth]{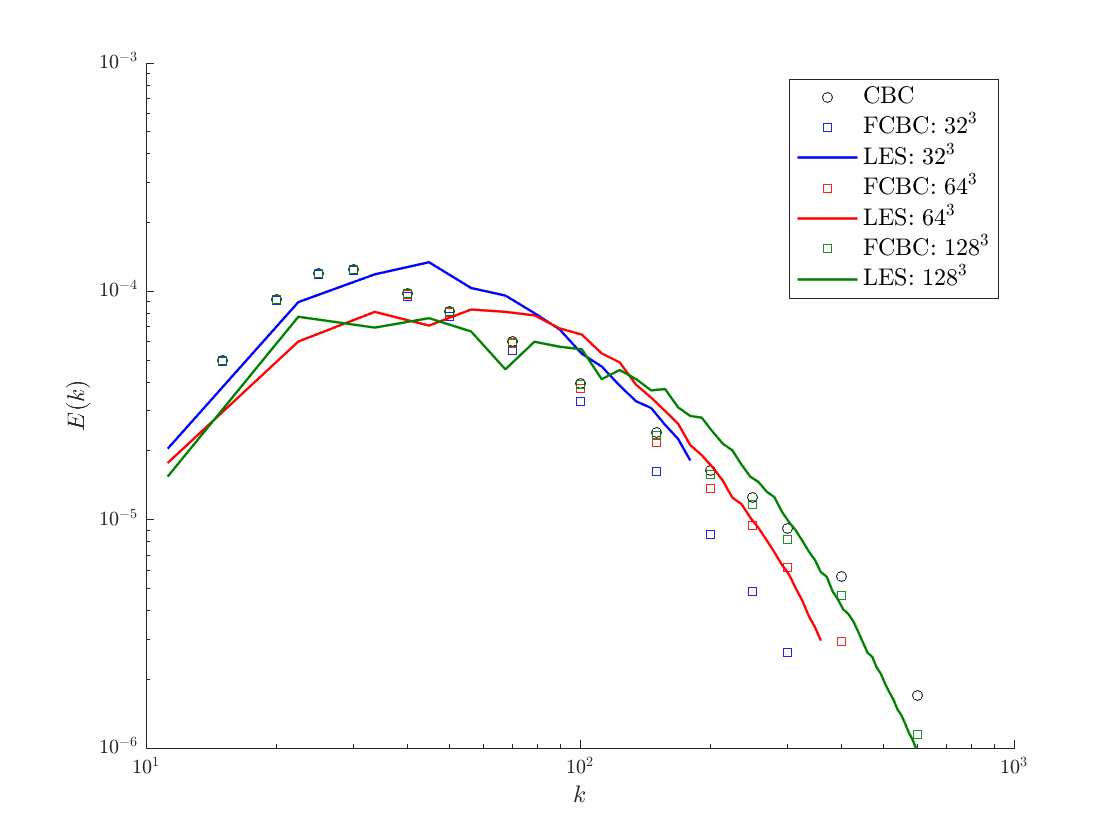}}
    \subfigure[Gradient Model\label{fig:DHIT_GM_171}]{\includegraphics[width=0.49\textwidth]{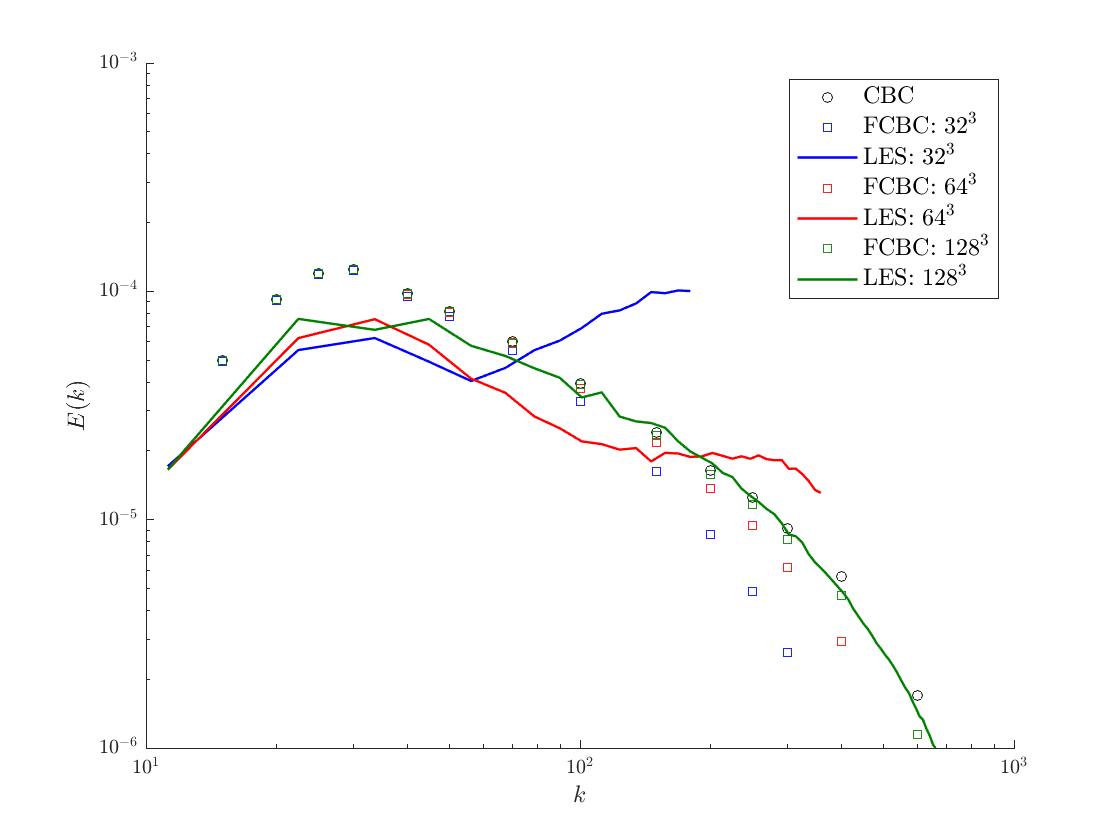}}
    \subfigure[Gradient Model with Clipping\label{fig:DHIT_GM_171_C}]{\includegraphics[width=0.49\textwidth]{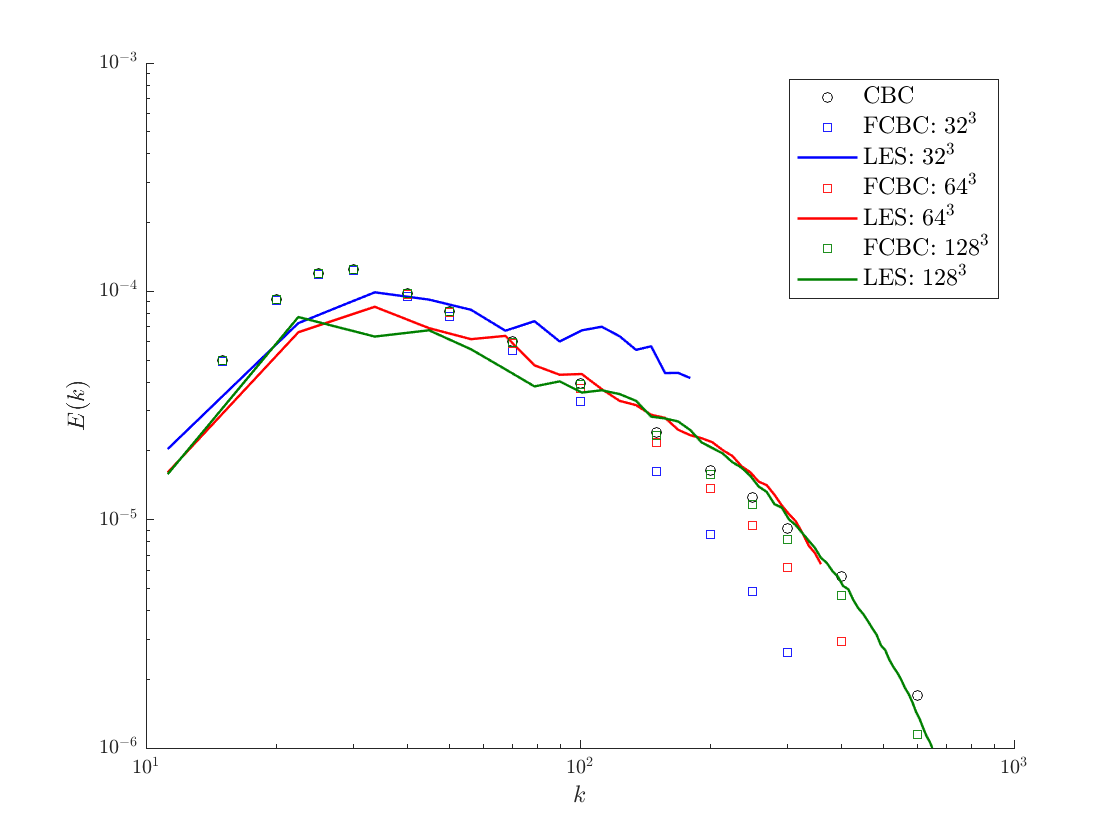}}    
    \subfigure[Data-Driven Model\label{fig:DHIT_DD_171}]{\includegraphics[width=0.49\textwidth]{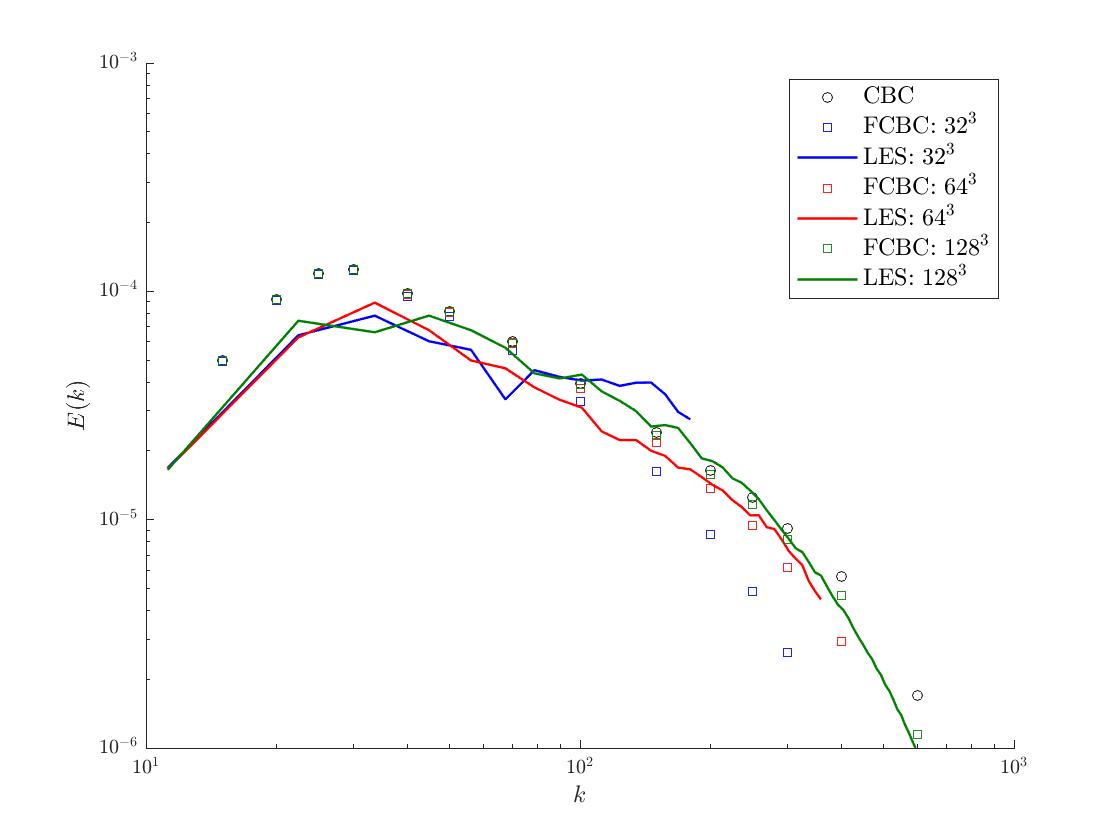}}
    \subfigure[Data-Driven Model with Clipping\label{fig:DHIT_DD_171_C}]{\includegraphics[width=0.49\textwidth]{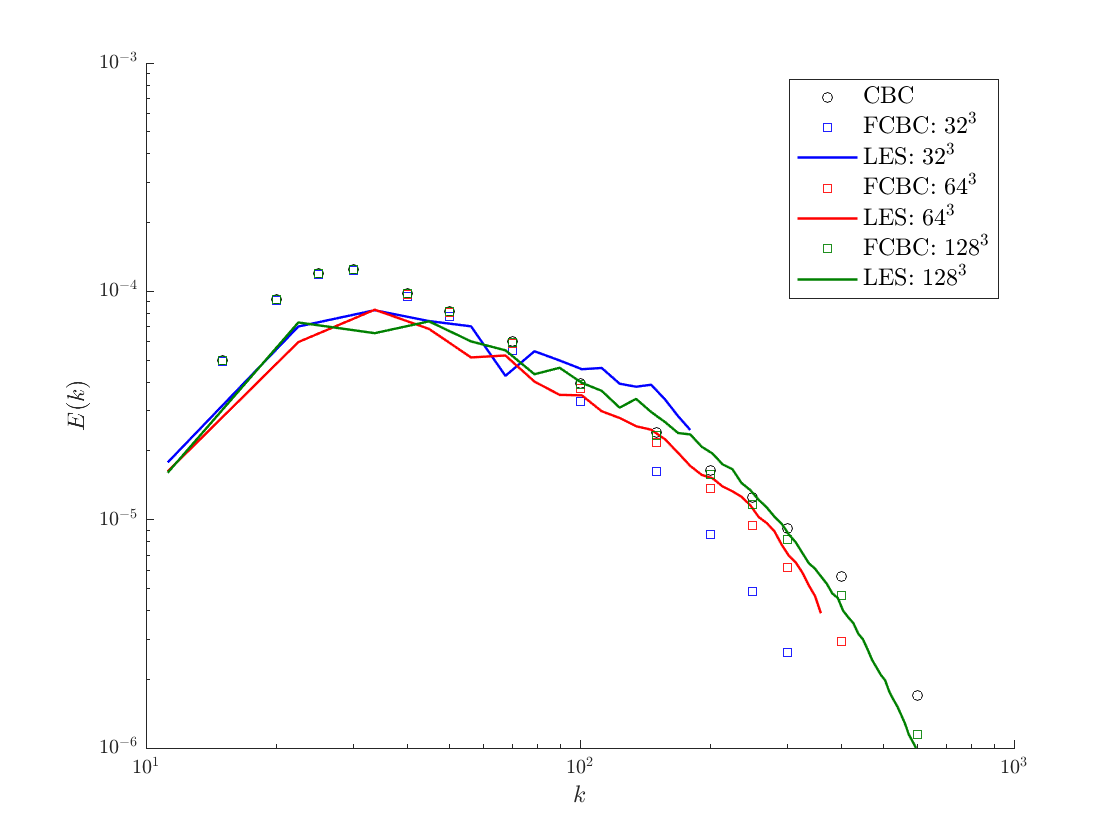}}    
    \caption{Energy spectra for decaying HIT at $t^* = 171$ for (a) No Model, (b) Dynamic Smagorinsky Model, (c) Gradient Model, (d) Gradient Model with Clipping, (e) Data-Driven Model and (f) Data-Driven Model with Clipping.}
    \label{fig:DHIT_171}
\end{figure}

Energy spectra at $t^* = 98$ attained using \textit{a posteriori} tests for each of the considered models are shown in \figref{DHIT_98}. The use of no explicit LES model leads to a significant pile-up of energy at higher wavenumbers for the $32^3$ and $64^3$ element meshes. The dynamic Smagorinsky model yields improved results free of energy pile-up, however, it slightly overpredicts the energy at intermediate wavenumbers. The gradient model also leads to energy pile-up at higher wavenumbers for the $32^3$ and $64^3$ element meshes. The clipped gradient model gives a much better prediction, however, there is a slight over-prediction of energy spectra at intermediate wavenumbers. Energy spectra results attained with the unclipped and clipped data-driven models are in better overall agreement with the filtered experimental energy spectra as compared to the other models with the clipped data-driven model producing the best results for the $64^3$ and $128^3$ element meshes. For the $32^3$ element mesh, the dynamic Smagorinsky model gives a slightly better energy spectra prediction than the clipped data-driven model, though both models overpredict the filtered experimental spectra for this very coarse mesh.

Energy spectra at $t^* = 171$ are shown in \figref{DHIT_171}. Similar model behavior is seen at this later time. There is less agreement between the dynamic Smagorinsky model energy spectra and the filtered experimental energy spectra at the intermediate wavenumbers as compared to $t^* = 98$. The data-driven model exhibits considerably less energy pile-up as compared to the gradient model.  Clipping significantly improves the performance of the gradient model for the $32^3$ and $64^3$ element meshes.  Clipping slightly improves the performance of the data-driven model for the $32^3$ element mesh, but the unclipped and clipped data-driven models exhibit very similar performance for the $64^3$ and $128^3$ element meshes.  

It has long been hypothesized that the turbulent kinetic energy $k(t)$ of a homogeneous isotropic turbulent flow decays like $t^{-n}$ for some $n \in \mathbb{R}^+$ in the absence of external forcing, and theory has suggested that $n = 10/7 \approx 1.43$ (see, e.g., \citep{Kolmogorov1941b, ComteBellot1966}) while experiments have suggested that $n \in [1.2,1.35]$ (see, e.g., \citep{ComteBellot1966}).  Consequently, it is worth comparing the decay rates predicted by each of the considered SGS models with theory and experiment.  We have computed the decay rate $n \in \mathbb{R}^+$ predicted by each of the considered SGS models and each of the considered meshes using the predicted turbulent kinetic energy values at non-dimensional times of $t^\ast_1 = 98$ and $t^\ast_2 = 171$ and displayed these rates in \tabref{DecayRate}.  Note that the decay rates attained using both the clipped and unclipped data-driven models are remarkably consistent with mesh resolution and within 5\% of the theoretical value of $n = 10/7 \approx 1.43$ and within 10\% of the experimental range of $n \in [1.2,1.35]$.  On the other hand, the dynamic Smagorinsky model and both the clipped and unclipped gradient models exhibit significant variation in their decay rate predictions across the different considered mesh resolutions, though their predictions appear to be converging to those predicted by the clipped and unclipped gradient models with increasing mesh resolution.

\begin{table}[t!]
    \centering
    \begin{tabular}{ccccccc}
        \hline
        \hline
         \textbf{Grid Resolution} & \multicolumn{6}{c}{\textbf{Decay Rate ($n$)}} \\
          &\textbf{NM} & \textbf{DS} & \textbf{GM} & \textbf{GM-C} & \textbf{DD} & \textbf{DD-C}\\
         \hline
         $32^3$ & 0.80 & 1.08 & 1.18 & 1.19 & 1.46 & 1.37\\
         $64^3$ & 1.51 & 1.21 & 1.75 & 1.43 & 1.52 & 1.48\\
         $128^3$ & 1.66& 1.45& 1.53& 1.56& 1.46& 1.48\\
         \hline
         \hline
    \end{tabular}
    \caption{Turbulent kinetic energy decay rate for decaying HIT. \label{tab:DecayRate}}
\end{table}

\subsubsection{3-D Taylor-Green Vortex Flow at $Re = 1,600$}

We finally examine the performance of the data-driven model using 3D Taylor-Green vortex flow. Like decaying HIT, 3-D Taylor-Green vortex flow is not statistically stationary. In fact, 3D Taylor-Green vortex flow starts as laminar and then transitions to a turbulent state provided the Reynolds number is sufficiently high. After the flow fully transitions, the turbulence decays. As we already examined the performance of the data-driven model for decaying HIT, we focus here on the performance of the data-driven model during transition. 

The initial velocity profile for 3D Taylor-Green vortex flow is

\begin{equation}
    \textbf{u} = \begin{bmatrix} \sin{(x_1)} \cos{(x_2)} \cos{(x_3)} \\ - \cos{(x_1)} \sin{(x_2)} \cos{(x_3)} \\ 0
    \end{bmatrix}.
\end{equation}

\noindent
We investigate this flow case for a Reynolds number of $Re = 1,600$, where $Re = U L/\nu$ and $U L = 1$. In this article, we consider two quantities of interest: energy spectra at $t = 9$, a time near peak dissipation, and the temporal evolution of resolved dissipation,

\begin{equation}
    \epsilon (t) = 2 \nu \frac{1}{\Omega} \int_{\Omega} \frac{\bm{\omega}(t) \cdot \bm{\omega}(t)}{2} d\Omega.
    \label{eq:dissip_tg}
\end{equation}

\noindent We compare LES results to the DNS results from \cite{Shoraka2017}. We consider $64^3$ and $128^3$ element meshes for this case, corresponding to filter widths of $\Delta = 2\pi/64$ and $\Delta = 2\pi/128$ respectively. Total and resolved dissipation time histories are provided in \cite{Shoraka2017} for filter widths $\Delta = 2\pi/64$ and $\Delta = 2\pi/128$. However, \cite{Shoraka2017} only presents unfiltered energy spectra at $t = 9$ s. We again apply a differential filter approximating the tensor-product box filter to the unfiltered energy spectra (see \eref{diff_filter}) to arrive at filtered spectra data for comparison purposes.  The use of an explicit LES model approximates the filtered Navier-Stokes equations, so statistics attained using a perfect SGS model will be in agreement with the filtered DNS data.  If no model is employed and the flow is fully resolved, statistics are instead expected to be close to the unfiltered DNS data. 

\begin{figure}[t!]
    \centering
    \subfigure[\label{fig:TG_dissip_64}]{\includegraphics[width=0.49\textwidth]{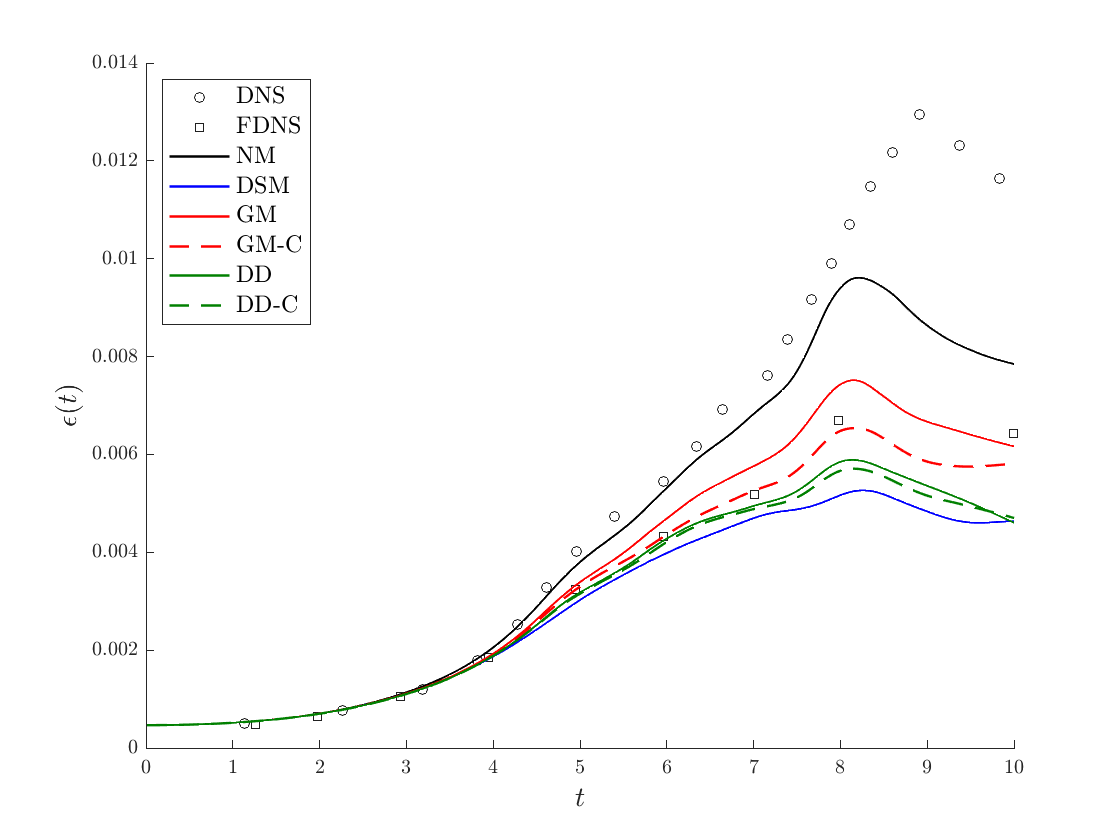}}
    \subfigure[\label{fig:TG_dissip_128}]{\includegraphics[width=0.49\textwidth]{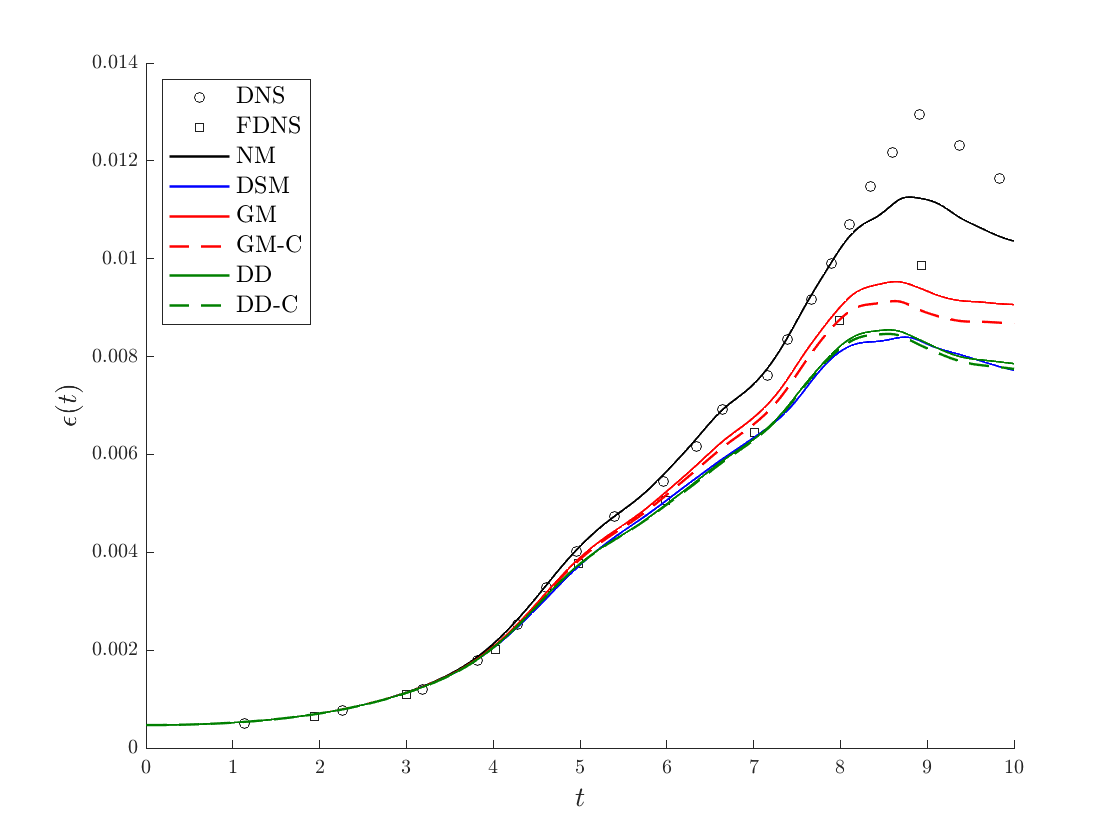}}
    \caption{Temporal evolution of resolved dissipation for 3-D Taylor-Green vortex flow for (a) the $64^3$ element mesh and (b) the $128^3$ element mesh.}
    \label{fig:TG_dissip}
\end{figure}

\figref{TG_dissip} shows the temporal evolution of the resolved dissipation. In the legend, the DNS and filtered DNS data are denoted as DNS and FDNS respectively. We observe that for both the mesh cases, the clipped gradient model gives the best estimate of the resolved dissipation as compared to the filtered DNS value.  The unclipped gradient model slightly overpredicts the resolved dissipation. Both the clipped and unclipped data-driven models give a better prediction of the resolved dissipation than the dynamic Smagorinsky model, though the resolved dissipation is slightly underpredicted for both the meshes. The slight underprediction of the resolved dissipation for the clipped and unclipped data-driven models can attributed to their overdissipative nature at small filter widths as observed in our earlier \textit{a priori} tests. As discussed in Section \ref{sec:conclusions}, this issue can be overcome by including viscosity as a model input at the expense of increased computational cost.

\begin{figure}[t!]
    \centering
    \subfigure[\label{fig:TG_ES_64}]{\includegraphics[width=0.49\textwidth]{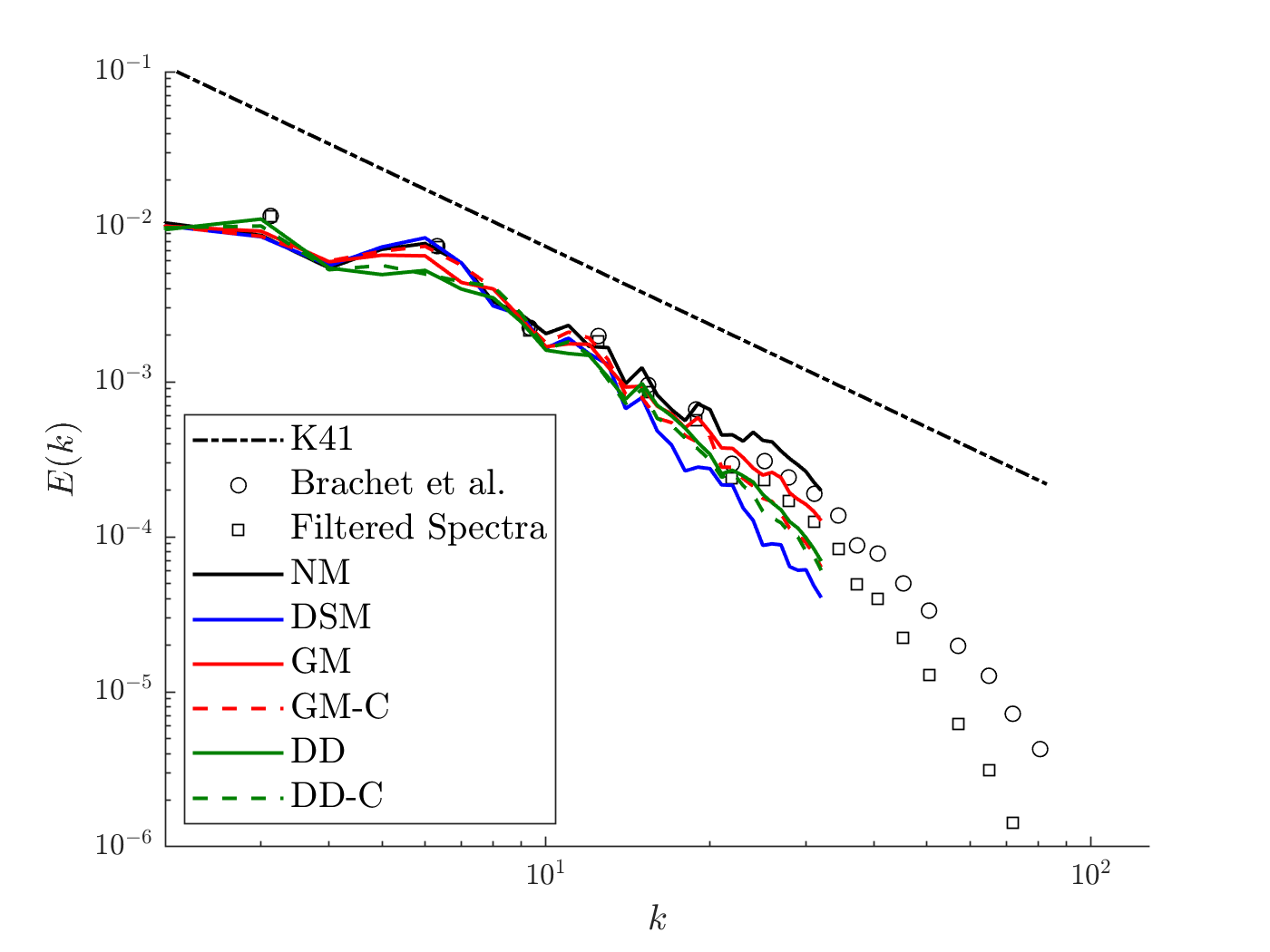}}
    \subfigure[\label{fig:TG_ES_128}]{\includegraphics[width=0.49\textwidth]{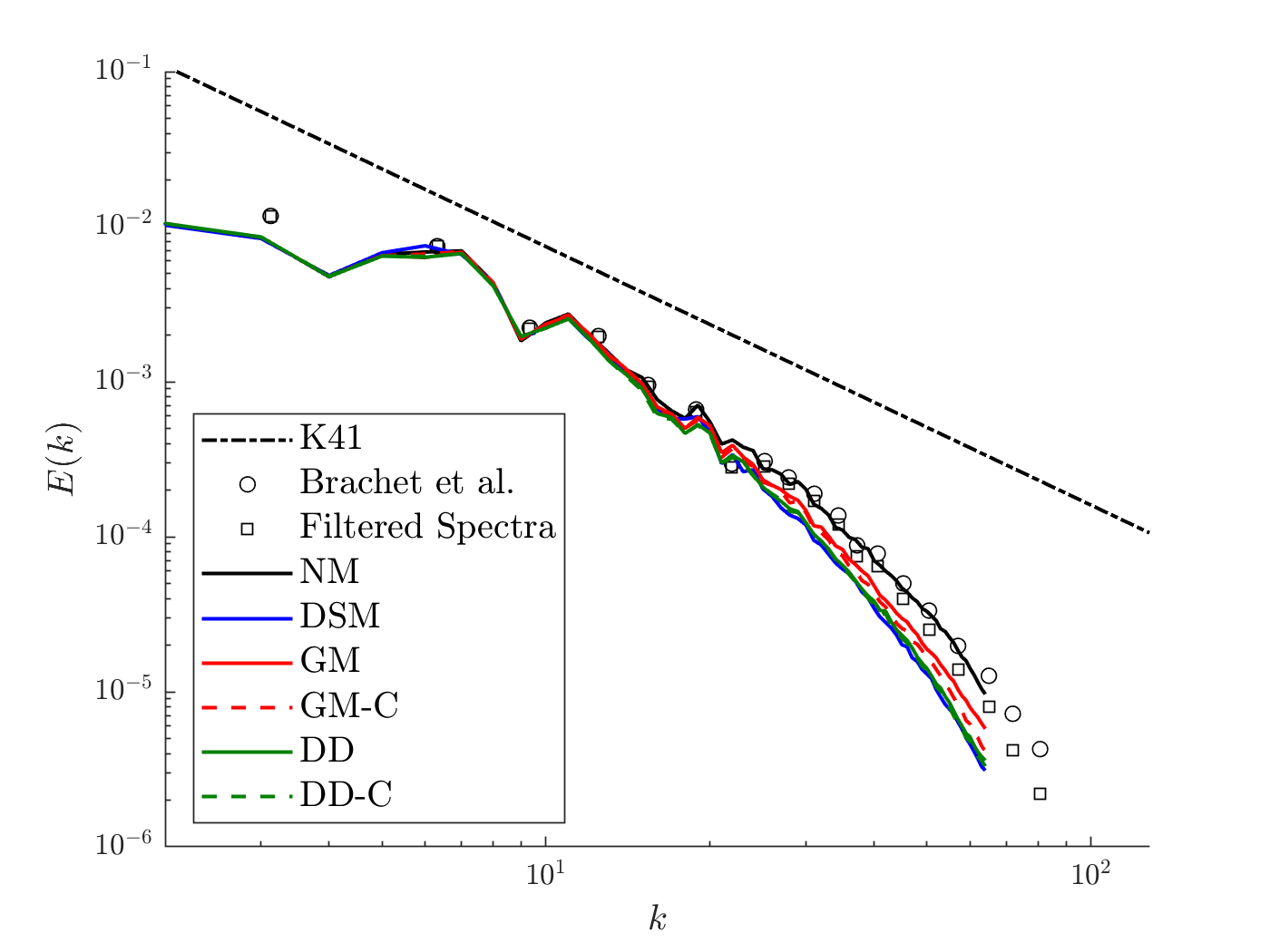}}
    \caption{Energy spectra for 3-D Taylor-Green vortex flow at $t = 9$ for (a) the $64^3$ element mesh and (b) the $128^3$ element mesh.}
    \label{fig:TG_ES}
\end{figure}

Lastly, we compare the energy spectra at $t = 9$ for the considered SGS models in \figref{TG_ES}.  The gradient model without clipping yields the best agreement with the filtered spectra for both mesh resolutions, though it slightly overpredicts the filtered spectra for the coarser mesh resolution and slightly underpredicts the filtered spectra for the finer mesh resolution.  The clipped data-driven model, unclipped data-driven model, and clipped gradient model exhibit similar performance and slightly underpredict the filtered spectra for both mesh resolutions.  The clipped and unclipped data-driven models' underprediction of the filtered spectra can again be attributed to their overdissipative nature at small filter widths.  The dynamic Smagorinsky model significantly underpredicts the filtered spectra for the coarse mesh resolution. For the coarse mesh resolution, the clipped and unclipped data-driven models noticeably underpredict the energy spectrum at wavenumber $k=5$, and the clipped and unclipped gradient models also underpredict the energy spectrum at this wavenumber. Though the precise reason for this discrepancy is not known, we believe that it could be due to non-linear triadic interactions from the dynamically evolving flow field. However, this is just a conjecture, and one cannot expect to have a perfect match of time-instantaneous quantities of interest due to the acute sensitivity of the turbulent flow field to perturbations resulting from model inaccuracies or even finite precision arithmetic.

\section{Conclusions}
\label{sec:conclusions}

State-of-the-art approaches to data-driven SGS modeling typically employ a complex neural network architecture for SGS tensor representation, require a large amount of training data, and do not generalize well to previously unseen filter widths, Reynolds numbers, and flow physics.  In this paper, we presented a new approach for constructing data-driven SGS models with embedded invariance properties to address these issues.  The cornerstone to our new approach is representation of model input and output tensors in the filtered strain rate eigenframe.  Provided one’s chosen model inputs are Galilean invariant and non-dimensionalized using the Buckingham Pi theorem, this yields a final model form that is symmetric, Galilean invariant, rotationally invariant, reflectionally invariant, and unit invariant.  We illustrated our approach for the specific case when the filtered strain-rate tensor, filtered rotation-rate tensor, and filter width are chosen as model inputs, and for this case, we attained a final model form with only four model inputs.  Moreover, we recovered a simple representation of the first terms of a Taylor series expansion of the exact SGS tensor using this final model form, indicating its potential for data-driven SGS modeling.  We then used this final model form to construct a remarkably simple and efficient data-driven SGS model using a limited amount of training data.  In particular, we trained a neural network model with only a single hidden layer composed of twenty neurons using just one time step of filtered direct numerical simulation data from a forced homogeneous isotropic turbulence simulation.  We also only employed one filter width for model training.  \textit{A priori} tests revealed that this model not only accurately predicts the exact SGS tensor and SGS energy dissipation for inputs within the training set but also inputs outside the training set.  \textit{A posteriori} tests revealed the model outperforms classical SGS models such as the dynamic Smagorinsky model and Clark’s gradient model, even when applied to test cases that involve filter widths, Reynolds numbers, and flow physics that are different from the training set.

We believe that the success of our new approach is due to two factors.  First, as invariance properties are directly embedded into SGS models using our approach, these invariance properties do not need to be learned during the training process.  A large amount of training data would be required to learn invariance properties for a model form without embedded invariance properties, and a complex neural network architecture would also be required to ensure the final neural network model actually can satisfy such invariance properties.  Second, as our approach yields model forms with simple representations of the first terms of a Taylor series expansion of the exact SGS tensor, our approach necessarily can yield an accurate SGS model even for a simple neural network architecture.

\begin{figure}[b!]
    \centering
    \subfigure[\label{fig:TG_dissip_64_BN}]{\includegraphics[width=0.49\textwidth]{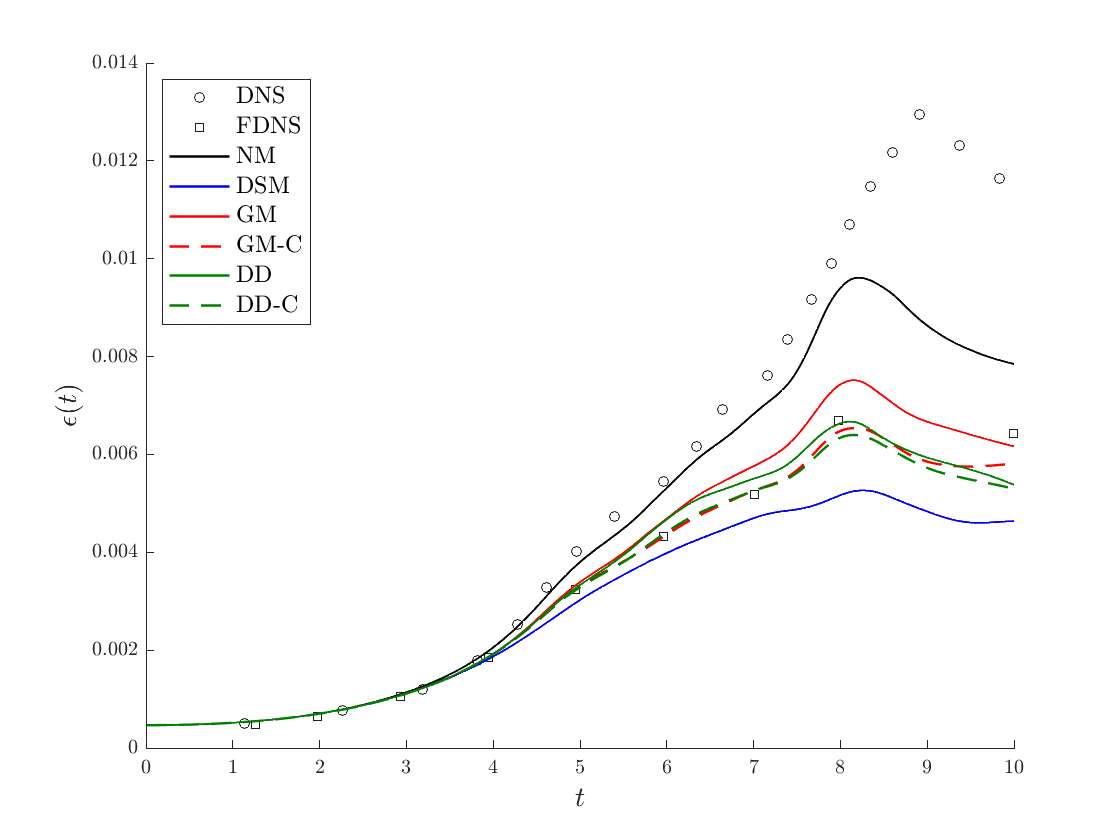}}
    \subfigure[\label{fig:TG_dissip_128_BN}]{\includegraphics[width=0.49\textwidth]{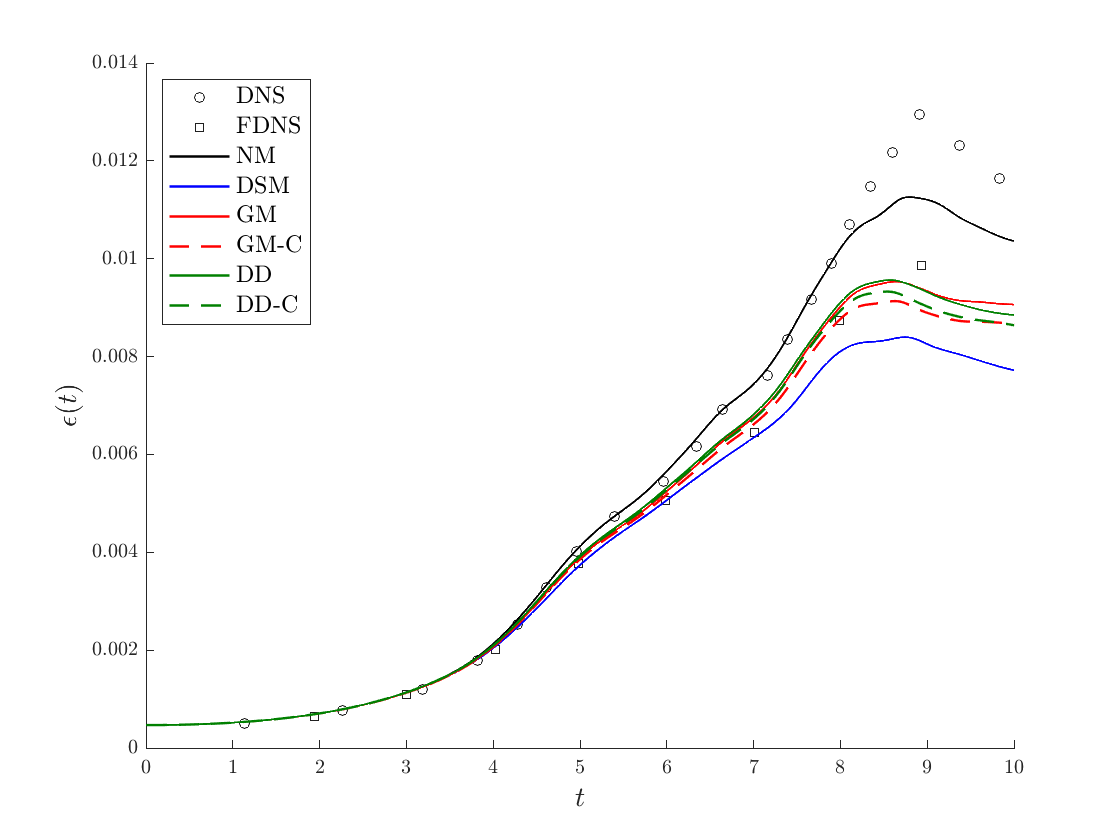}}
    \caption{Temporal evolution of resolved dissipation for 3-D Taylor-Green vortex flow using the neural network model with $\hat{\nu} = \frac{\nu}{G \Delta^2}$ as an additional input parameter for (a) the $64^3$ element mesh and (b) the $128^3$ element mesh.}
    \label{fig:TG_dissip_BN}
\end{figure}

\begin{figure}[t!]
    \centering
    \subfigure[\label{fig:TG_ES_64_BN}]{\includegraphics[width=0.49\textwidth]{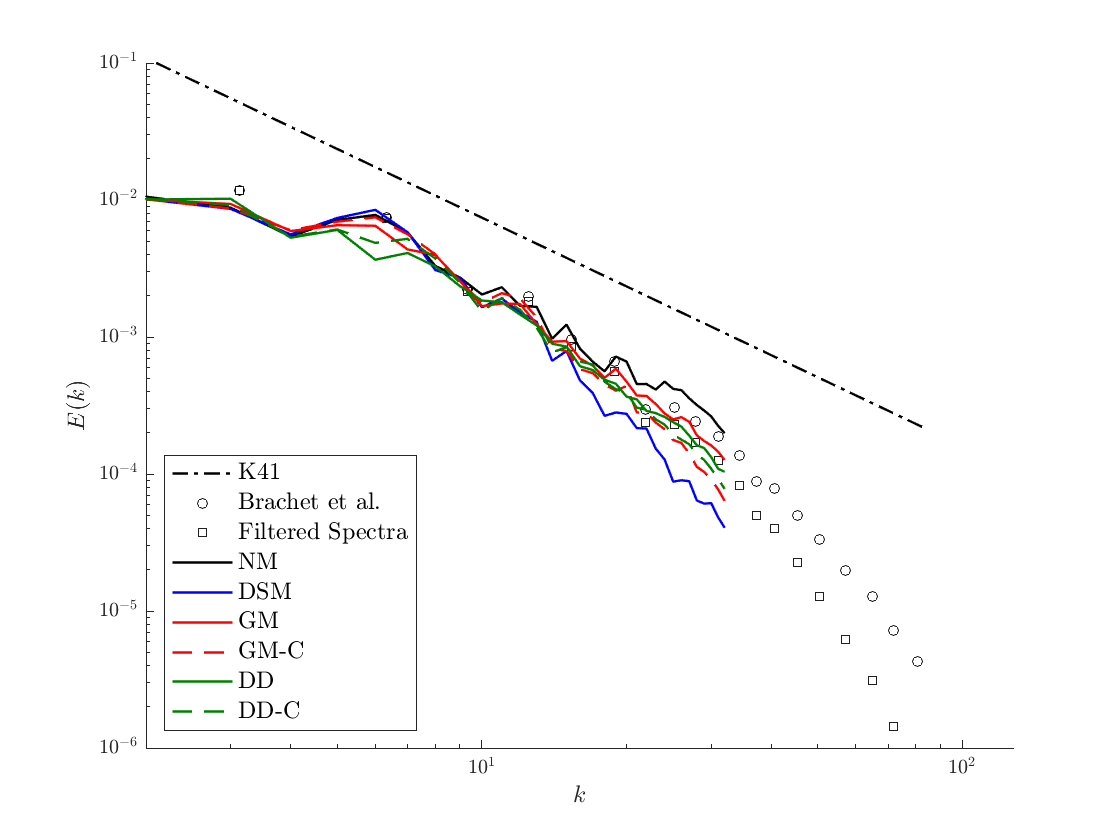}}
    \subfigure[\label{fig:TG_ES_128_NM}]{\includegraphics[width=0.49\textwidth]{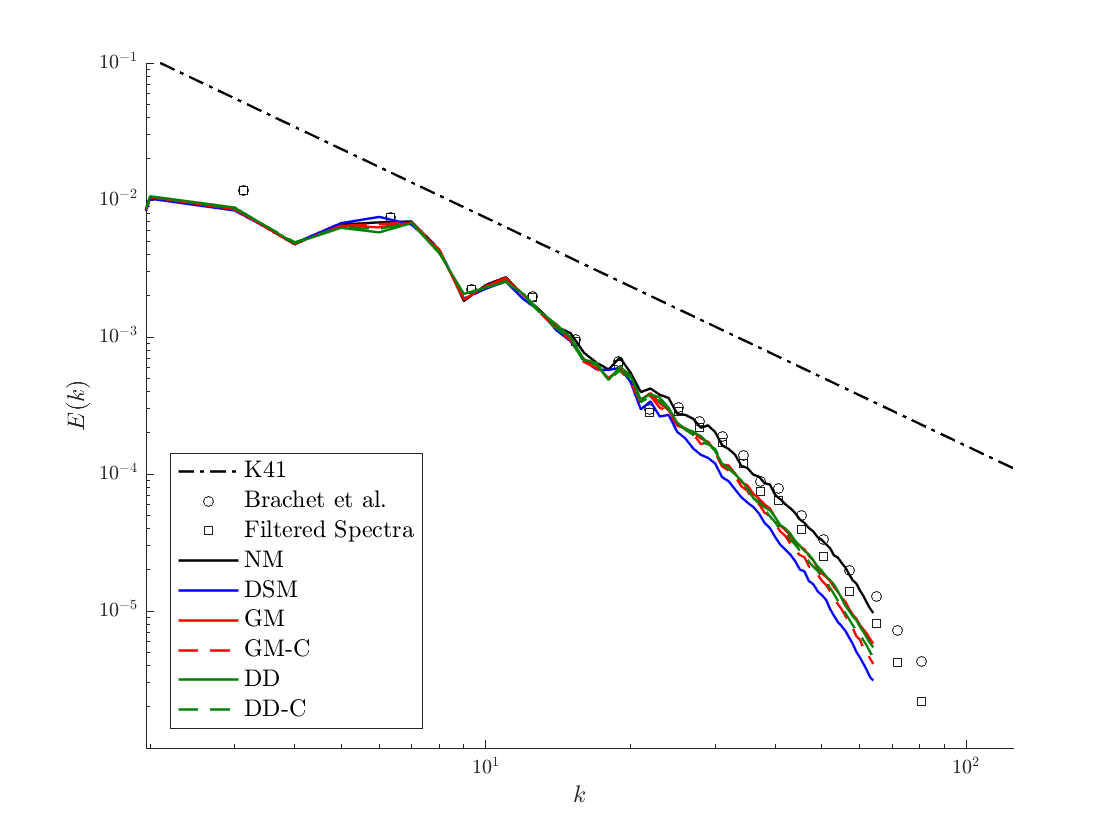}}
    \caption{Energy spectra for 3-D Taylor-Green vortex flow at $t = 9$ using the neural network model with $\hat{\nu} = \frac{\nu}{G \Delta^2}$ as an additional input parameter for (a) the $64^3$ element mesh and (b) the $128^3$ element mesh.}
    \label{fig:TG_ES_BN}
\end{figure}

While we have demonstrated our new approach is capable of yielding accurate, efficient, and generalizable SGS models for periodic flows, performing LES for wall-bounded flows necessitates the use of anisotropic and nonuniform grids.  Thus, in future work, we plan to extend our approach to accommodate anisotropic and nonuniform filters.  We also plan to explore the use of additional model inputs beyond just the filtered strain-rate tensor, filtered rotation-rate tensor, and filter width.  We have already begun to explore the use of viscosity as an additional model input.  This yields the alternative final model form
\begin{equation}
\tau_{ij}^S = \Delta^2 G^2 \hat{\tau}_{ij}^{S,\text{model}} (\hat{\lambda}_3^S, \hat{\omega}_1^S, \hat{\omega}_2^S, \hat{\omega}_3^S,\hat{\nu}) \label{eq:alt_final_model_form}
\end{equation}
where $\hat{\nu} = \frac{\nu}{G \Delta^2}$ is a non-dimensional viscosity input parameter.  We have used this model form to train an alternative neural network model composed of two hidden layers with one hundred neurons each using the $Re_{\lambda} = 418$ forced HIT simulation dataset from the JHTDB.  This alternative neural network model exhibits similar performance to the neural network model constructed earlier in this paper for filter widths in the inertial subrange, but it exhibits superior performance for filter widths in the dissipation range.  This is not surprising, as the non-dimensional viscosity input parameter $\hat{\nu}$ enables the alternative neural network model to distinguish between filter widths in the inertial subrange and filter widths in the dissipation range.  As shown in \figref{TG_dissip_BN} and \figref{TG_ES_BN}, the alternative neural network model also exhibits superior performance for the $Re = 1,600$ 3-D Taylor-Green vortex flow case.  However, the alternative neural network model has a considerably more complex network architecture than the neural network model constructed earlier in this paper, so it is more expensive to train and evaluate.  This underscores an inevitable tradeoff between accuracy and computational cost.

\section{Acknowledgements}
The authors would like to acknowledge the Computational and Data-Enabled Science and Engineering (CDS\&E) program of the National Science Foundation (NSF) CBET-1710670, as well as the Transformational Tools and Technologies Project of the National Aeronautics and Space Administration (NASA) 80NSSC18M0147 for funding of this work. Moreover, they thank the Argonne Leadership Computing Facility (ALCF) for the resources on which the simulations and post-processing were performed.  Finally, they thank Eric Peters, Basu Parmar, Riccardo Balin, and Alireza Doostan for conversations that contributed to this, as well as the anonymous reviewers whose thoughtful feedback improved the quality of this manuscript.

\bibliographystyle{unsrt}
\bibliography{main}

\newpage
\appendix
\section{Numerical Method for \textit{A Posteriori} Simulations}
\label{sec:Appendix1}

All \textit{a posteriori} simulations in this paper are carried out using a SUPG/PSPG/grad-div-stabilized finite element method \cite{Whiting2001}.  As all \textit{a posteriori} simulations in this paper involve periodic boundary conditions, the following semi-discretization is employed:\\

\noindent Find $\bar{\bm{u}}^h : \overline{\Omega} \times [0,T] \rightarrow \mathbb{R}^3$ and $\bar{p}^h: \Omega \times (0,T) \rightarrow \mathbb{R}$ such that $\bar{\bm{u}}^h(\cdot, 0) = \bar{\bm{u}}_0^h$ and, for all $t \in (0,T)$, $\bar{\bm{u}}^h(\cdot,t) \in V^h$, $\bar{p}^h(\cdot,t) \in Q^h$, and
\begin{multline}
    \int_{\Omega} \left\{ \frac{\partial \bar{u}^h_i}{\partial x_i}(\cdot,t) \bar{q}^h  + \left(\rho \frac{\partial \bar{u}^h}{\partial t}(\cdot,t) + \rho \bar{u}^h_j(\cdot,t) \frac{\partial \bar{u}^h_i}{\partial x_j}(\cdot,t) - \bar{f}_i(\cdot,t)\right) \bar{w}^h_i + \tau^*_{ij}(\cdot,t) \frac{\partial \bar{w}^h_i}{\partial x_j} \right\} d\Omega  \\ + \sum_{e=1}^{n_{el}} \int_{\Omega^e} \left\{ \tau_m \mathcal{R}^m_i(\cdot,t) \left( \bar{u}^h_j(\cdot,t) \frac{\partial \bar{w}^h_i}{\partial x_j} + \frac{1}{\rho} \frac{\partial \bar{q}^h}{\partial x_i} \right) + \tau_c \mathcal{R}^c(\cdot,t) \frac{\partial \bar{w}^h_j}{\partial x_j} \right\} d\Omega^e = 0
\end{multline}

\noindent for all $\bar{\bm{w}}^h \in V^h$ and $\bar{q} \in Q^h$.\\

\noindent Above, $\Omega$ is the periodic tensor-product domain of interest, $\left\{ \Omega^e \right\}_{e=1}^{n_{el}}$ is the set of hexahedral elements for a tensor-product mesh of interest, $V^h$ is the finite element space of filtered velocity field approximations, $Q^h$ is the finite element space of filtered pressure field approximations, $\tau_{ij}^* = 2 \mu S_{ij} - \rho \tau_{ij} - \bar{p}^h \delta_{ij}$, $\tau_m$ and $\tau_c$ are the momentum and continuity stabilization parameters respectively, and
\begin{align}
    \mathcal{R}^m_i &= \rho \frac{\partial \bar{u}^h_i}{\partial t} + \rho \bar{u}^h_j \frac{\partial \bar{u}^h_i}{\partial x_j} + \frac{\partial \bar{p}^h}{\partial x_i} - \frac{\partial \tau^*_{ij}}{\partial x_j} - \bar{f}_i \\
    \mathcal{R}^c &= \frac{\partial \bar{u}^h_i}{\partial x_i}
\end{align}

\noindent are the momentum and continuity residuals respectively.  Superscript $h$ denotes a discretized quantity.  Trilinear finite element basis functions are used to discretize the filtered velocity and pressure fields, so
\begin{align}
    V^h &= \left\{ \bar{\bm{v}}^h \in \left(C^0_\text{per} (\Omega)\right)^3 : \bar{\bm{v}}^h\vert_{\Omega_e} \in \left(Q^1 (\Omega_e)\right)^3 \text{ for all } e = 1, \ldots, n_{el} \right\} \\
    Q^h &= \left\{ \bar{q}^h \in C^0_\text{per} (\Omega) : \bar{q}^h\vert_{\Omega_e} \in Q^1 (\Omega_e) \text{ for all } e = 1, \ldots, n_{el}\right\}
\end{align}
where $C^0_\text{per} (\Omega)$ is the space of continuous periodic functions over the domain $\Omega$ and $Q^1(\Omega_e)$ is the space of trilinear polynomials over element $\Omega_e$.  For the \textit{a posteriori} simulations in this paper, the momentum and continuity stabilization parameters are selected as
\begin{align}
    \tau_m &= \text{min} \left\{ \frac{\Delta t}{16}, \frac{h}{2| \bar{\bm{u}}^h|}, \frac{h^2}{12 \nu} \right\} \\
    \tau_c &= \frac{\rho |\bar{\bm{u}}^h| h }{20} \text{min} \left\{ 1, \frac{|\bar{\bm{u}}^h| h}{2 \nu} \right\}
\end{align}
where $\Delta t$ is the time step size and $h$ is the element size.  The semi-discrete equations of motion are discretized in time using the generalized-$\alpha$ method with $\rho_\infty = 0.5$ \cite{Jansen2000}.

\end{document}